\documentclass{aa}
\usepackage{natbib}
\bibpunct{(}{)}{;}{a}{}{,} % to follow the A&A 
\usepackage{hyperref}
\usepackage{aalongtable}
\usepackage{graphicx} 
\usepackage{lscape}
\usepackage{txfonts}
\usepackage{balance}
\usepackage{wrapfig}
\usepackage{latexsym}
\newcommand{\nc}{\newcommand}

\nc{\teff}{$T_{\rm eff}$\,}  
\nc{\teffns}{$T_{\rm eff}$}  
\nc{\glog}{log\,$g$\,}  
\nc{\kms}{\,${\rm km\,s}^{-1}$\,}  
\nc{\mic}{$\xi_{\rm t}$\,}
\nc{\nod}{--}
\nc{\ld}{\ldots}

\begin{document}

\title{{Silver and palladium help unveil the nature of a second
r-process}
\thanks {Based on observations made with the ESO Very Large Telescope 
at Paranal Observatory, Chile (ID 65.L-0507(A), 67.D-0439(A), 68.B-0475(A), 68.D-0094(A), 71.B-0529(A); P.I. F. Primas).}
}

\author{
C. J. Hansen\inst {1,2}\and F. Primas\inst {1}\and H. Hartman\inst{3,8} \and K.-L. Kratz\inst{4} \and S. Wanajo\inst{5,6,7} \and B. Leibundgut\inst{1} \and K. Farouqi\inst{4,2} \and O. Hallmann\inst{4} \and N. Christlieb\inst{2} \and H. Nilsson\inst{3} 
}

\institute{European Southern Observatory (ESO),
         Karl-Schwarschild-Str. 2, D-85748 Garching b. M\"unchen, 
Germany\\
   \email{cjhansen@lsw.uni-heidelberg.de,fprimas@eso.org,bleibund@eso.org}
        \and    
	Landessternwarte Heidelberg (LSW, ZAH),
	K\"onigstuhl 12, D-69117 Heidelberg, Germany\\
	\email{cjhansen@lsw.uni-heidelberg.de, N.Christlieb@lsw.uni-heidelberg.de} 
	\and
Lund Observatory, Department of Astronomy and Theoretical Physics, Lund University, Box 43, 22100 Lund, Sweden\\
        \email{Henrik.Hartman@astro.lu.se}
        \and
        Max-Planck-Institut f\"ur Chemie, Otto-Hahn-Institut, Joh.-J-Becherweg 27, D-55128 Mainz, Germany\\
        \email{klk@uni-mainz.de, k.farouqi@mpic.de, o.hallmann@mpic.de} 
        \and
        Technische Universit\"at M\"unchen,
        Excellence Cluster Universe,
        Boltzmannstr. 2, D-85748 Garching, Germany\\
        \and
        Max-Planck-Institut f\"ur Astrophysik, Karl-Schwarzschild-Str. 1, D-85748 Garching\\ 
	\and
	National Astronomical Observatory of Japan,
        2-21-1 Osawa, Mitaka, Tokyo 181-8588, Japan\\
	\email{shinya.wanajo@nao.ac.jp}
	\and
	Applied Mathematics, School of Technology, Malm\"o University, Sweden\\
}

\offprints{C. J. Hansen}

\date{Received 14 December 2011; Accepted 6 June 2012}

\abstract{The rapid neutron-capture process, which created about half of the heaviest elements in the solar system, is believed to have been unique. Many recent studies have shown that this uniqueness is not true for the formation of lighter elements, in particular those in the atomic number range 38 $< Z <$ 48. Among these, palladium (Pd) and especially silver (Ag) are expected to be key indicators of a possible second r-process, but until recently they have been studied only in a few stars. We therefore target Pd and Ag in a large sample of stars and compare these abundances to those of Sr, Y, Zr, Ba, and Eu produced by the slow (s-) and rapid (r-) neutron-capture processes. Hereby we investigate the nature of the formation process of Ag and Pd. 
}
{We study the abundances of seven elements (Sr, Y, Zr, Pd, Ag, Ba, and Eu) to gain insight into the formation process of the elements and explore in depth the nature of the second r-process. 
}
{By adopting a homogeneous one-dimensional local thermodynamic equilibrium (1D LTE) analysis of 71 stars, we derive stellar abundances using the spectral synthesis code MOOG, and the MARCS model atmospheres. We calculate abundance ratio trends and compare the derived abundances to site-dependent yield predictions (low-mass O-Ne-Mg core-collapse supernovae and parametrised high-entropy winds), to extract characteristics of the second r-process.}
{The seven elements are tracers of different (neutron-capture) processes, which in turn allows us to constrain the formation process(es) of Pd and Ag. The abundance ratios of the heavy elements are found to be 
correlated and anti-correlated. These trends lead to clear indications that a second/weak r-process, is responsible for the formation of Pd and Ag. On the basis of the comparison to the model predictions, we find that the conditions under which this process takes place differ from those for the main r-process in needing lower neutron number densities, lower neutron-to-seed ratios, and lower entropies, and/or higher electron abundances.}
{Our analysis confirms that Pd and Ag form via a rapid neutron-capture process that differs from the main r-process, the main and weak s-processes, and charged particle freeze-outs. We find that this process is efficiently working down to the lowest metallicity sampled by our analysis ([Fe/H] = $-$3.3).
Our results may indicate that a combination of these explosive sites is needed to explain the variety in the observationally derived abundance patterns.
}


\maketitle

\titlerunning{Silver and palladium help unveil the nature of a second
r-process}
\authorrunning{C. J. Hansen et al.}

%-----------------------------------------------------

\section{Introduction}
The heavy elements beyond the iron-peak are not created in the same way as the lighter elements, many of which form via hydrostatic core or shell burning in the star. These elements are generally created by various neutron-capture processes taking place as either the result of mixing in very evolved stars or explosions\footnote{We disregard proton processes here.}. 

Previous studies have shown that the slow neutron-capture (s-) process can be classified into two sub-processes, namely a weak s-process creating the lighter of the s-process isotopes \citep{prantzos,heil,pigna}, and a main s-process creating heavy isotopes, such as those of barium \citep{kaep89,busso,gallino,chrisrev}. The sites of the rapid neutron-capture (r-) processes remain unclear, and the exact conditions under which they operate continue to be investigated. Since the time of \citet{bbfh}, it has been evident that an explosive environment is needed to provide the proper conditions for an r-process to happen. After several attempts to make site-dependent predictions of the neutron-capture processes, \citet{klk93} provided a site-independent approach using the so-called waiting point approximation, which is based on the best available nuclear physics to shed light on the r-process. Nevertheless, the conditions are still poorly constrained. A number of sites have been suggested: neutron star mergers \citep{freiburg,GorJanka,goriely,wanBH}, massive core-collapse supernovae (SNe) \citep{wasserburg,argast}, neutrino-driven winds from core-collapse SNe \citep{duncan,meyerhot,taka,woosley,freineutron,wanajo,Fara,farou,arcones}, low-mass SNe from collapsing O--Ne--Mg cores (\citealt{Wanajo2003}; or iron cores \citealt{sumi}). However, no consensus on the formation site has been reached. 

Observationally, the discovery of r-process-rich stars which contain a factor of 20--100 more heavy elements than normal Population II halo stars; see \citealt{hill,sneden03,christlieb,barklem,frebel07,hayek,mpdw,cowan11} has offered an important opportunity to study in greater detail the r-process and its characteristics. By comparing light to heavy neutron-capture elements (i.e. 38 $<$ Z $<$ 50 vs Z $>$ 56), some of these studies \citep{sned2000U,westin,john02,christlieb,honda04,barklem,honda,honda07,francois,chrisrev,klk08,roed2010} have revealed a departure of the ``light'' neutron capture elements from the main solar-scaled r-process distribution curve, which was interpreted as an indication of an extra process. This suggests that the r-process may also split into two sub-channels, namely a 'weak' and main one \citep{cowan91,wan06,ott}, which are responsible for the production of the lighter and heavier r-process isotopes, respectively. The nomenclature is used to match the s-process \citep{kaep89}. 

The 'weak' r-process has received a lot of recent attention, but is still poorly constrained despite the many attempts that have been made to understand this process. Some of the proposed processes are the lighter element primary process (LEPP, \citealt{trav,arcones}), the weak r-process \citep{kratz,montes,Fara,wan11}, the $\nu$p-process \citep{frohlich}, and several more processes and comparisons, which can be found in \citealt{cowan01}, \citealt{qian01}, and \citealt{sneden03}. These processes can be considered when attempting to explain the abundances of the lighter heavy elements, which have been found to deviate from the solar-scaled r-process pattern\footnote{Solar-scaled r-process abundance: $N_r = N_{\odot} - N_s$}.
Palladium and silver are among these lighter heavy elements. Silver was studied for the first time by \citet{craw} more than a decade ago in a small sample of metal-poor stars. They applied a different hyperfine split oscillator strength from the one we adopt here, which together with the higher solar Ag abundance helps us to explain the low silver abundances they derive. A few years later, \citet{john02} studied both Pd and Ag in a sample that is the only other relatively large sample where both Pd and Ag were analysed. Hence, we compare our results to theirs.
\citet{cjhletter} presented the first results of an analysis of Ag and Pd in a large sample (55 stars) that demonstrated the need for an extra production channel. Here, we extend the study to the entire sample (71 stars) and compare our derived Ag and Pd abundances to those of five other heavy elements, namely Sr, Y, Zr, Ba, and Eu.
We furthermore wish to explore the nature of the second r-process in depth by investigating the trends of two particular tracer elements, palladium and silver. We characterise and constrain the fundamental parameters of the formation of these elements by means of a detailed comparison to yield predictions from several of the above-mentioned astrophysical sites and objects. Silver and palladium are important for two reasons. 
First, silver is predicted to be a good tracer of the weak r-process since nearly 80\% of its solar system abundance is predicted \citep{arland,chrisrev,loddersSun} to have come from the r-process, and more than 71\% of the r-process is estimated to originate from the weak r-process \citep{klk08,Fara,roed2010}. For comparison, only 54\% of palladium is created by the r-process \citep{arland}.
Second, these two elements had only been studied in a small number of stars ($<20$) until \citet{cjhletter}, whereas many other neutron-capture elements such as Ba have been studied in hundreds of stars \citep[e.g.][]{reddy,barklem,francois,roederer}.
A study of palladium and silver provides astrophysical information on a poorly studied part of the periodic table.

The paper is organised as follows: Sect. 2 describes the observations and data, Sect. 3 outlines the stellar parameter determination, Sect. 4 presents new atomic data and calibration of the line list, Sect. 5 presents the abundance analysis, and Sect. 6 and 7 provide the results and discussions of our abundance and model comparisons, respectively. Finally, our conclusions can be found in Sect. 8.  

\section{Observations and data reduction}
Our sample consists of a mixture of dwarf and giant stars, which were observed at high resolution ($R > 40,000$). The dwarfs were observed in the years 2000 -- 2002 with the UltraViolet Echelle Spectrograph at the Very Large Telescope, UVES/VLT, \citet[][]{dekker} for a project targeting beryllium, which requires high signal-to-noise (S/N) data of the near-ultraviolet (near-UV) particularly the Be doublet at 313\,nm \citep{fp2010}. Similarly high quality data are also needed to detect silver and palladium (328.6, 338.3\,nm, and 340.4\,nm, respectively). The spectra cover the wavelength ranges $\sim 305$ -- 680\,nm (in some cases up to 1000\,nm), including the wavelength gaps between the CCD detectors. All of our UVES spectra have a S/N $>$ 100 per pixel at 320\,nm. The dwarf spectra were reduced with the UVES pipeline (v. 4.3.0). The pipeline performs a standard echelle spectrum data reduction. It starts with bias subtraction, removes bad pixels due to e.g. cosmic ray hits, and locates the orders. Then a background subtraction is followed by flat field division, order extraction, and wavelength calibration, and finally the orders are merged. We tested the quality of the data products against a manual data reduction carried out in IRAF\footnote{IRAF is distributed by the National Optical Astronomy Observatory, which is operated by the Association of Universities of Research in Astronomy, Inc., under contract with the National Science Foundation.} because previous versions of the pipelines had problems with the order merging. However, this pipeline performs very well and the reduced data were compatible with manually reduced data. Finally, the reduced spectra were radial velocity corrected/shifted via cross correlation, coadded, and had their continua normalised (in IRAF).

The spectra of the giants were instead extracted from public data archives of the Very Large Telescope (VLT) and the Keck telescopes. In both cases, the spectra were observed with the high-resolution spectrographs available on both sites, i.e. UVES \citep{dekker} on the VLT and {HIgh REsolution Spectrometer} HIRES \citep{vogt94} on Keck. The wavelength coverage of HIRES spans 300 - 1000nm, which is very similar to the wavelength range of UVES but might have gaps above 620nm. Only spectra of high and comparable (to the dwarfs') quality were added to the sample. The giant spectra extracted from the respective archives had already been reduced, and were carefully inspected, radial velocity shifted, coadded, and continua normalised in IRAF. 

\subsection*{Sample}\label{sample}
The final stellar sample consists of 42 dwarf and 29 giant field stars, belonging to the Galactic halo, the thick, and the thin disks. The sample spans a broad parameter range exceeding 2000 K in temperature, 4 dex in gravity, and 2.5 dex in metallicity. Such a sample composition allows us to explore the chemical evolution of the Galaxy, as well as test the different chemical signatures of different stellar evolutionary stages. This in turn can shed light on the importance of mixing and non-local thermodynamic equilibrium (NLTE) effects. 

Our sample includes some of the most well-known r-process enhanced giant stars including CS~31082--001 \citep[][]{hill}, which we compare to CS~22892--052 \citep[][]{sneden03}, and BD +17 3248 \citep{cowan11}. We note that only one r-process enhanced metal-poor dwarf star has been found and observed so far \citep[][]{mpdw}, which is not included here. %This may introduce a small sample bias towards metal-poor r-process enhances giants. 
Furthermore, silver lines can be detected in giants of all the metallicities studied here, but can only be detected in dwarfs with [Fe/H]$> \sim -2.0$. This may introduce a small sample bias towards metal-poor r-process enhanced giants. %together with the fact that silver lines can only be detected in giants, and not in dwarfs below a metallicity of [Fe/H] $\simeq -$2.0. 
No carbon-enhanced stars were included in our sample. 

\section{Stellar parameters}
\begin{figure}[h!]
\begin{center}
\includegraphics[width=0.48\textwidth]{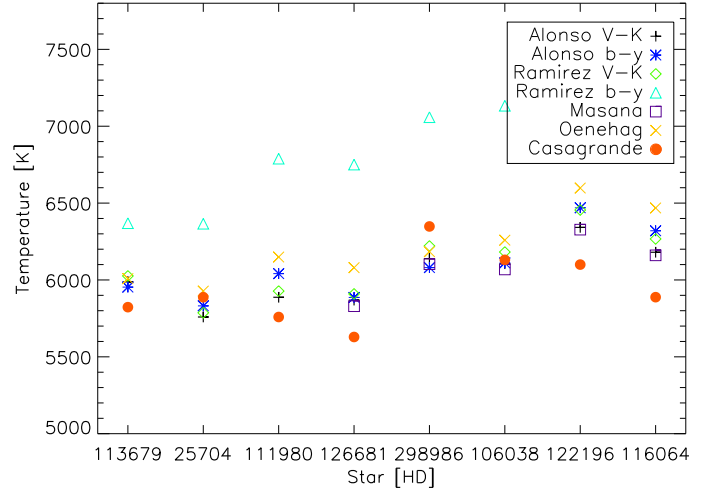}
\caption{Effective temperatures derived for eight stars (of different metallicity, from higher to lower as one moves from left to right along the x-axis) with seven different colour-$T_{\rm eff}$ calibrations (see figure legend).}
\label{figTscale}
\end{center}
\end{figure}

We followed different methods to determine the optimal set of stellar parameters. With such a large sample, we faced some difficulties in applying the same method to the determination of the stellar parameters for the entire data-set. The effective temperature of most of our stars was derived from colour-$T_{\rm{eff}}$ calibrations to which we applied the necessary band-filter and colour corrections. In this respect, we tested several different colour calibrations from \citet{alondw}, \citet{AlonsoG}, \citet{ramirez}, \citet{masana}, \citet{oenehag}, and \citet{casagra}, who make use of both ($V-K$) and ($b-y$) colour indices. In the end, we chose the calibration of \cite{alondw,AlonsoG} because these lead to temperature predictions that generally fall in the middle of the range shown in Fig. \ref{figTscale}. The temperature has a large influence on the derived stellar abundances. Hence, we wished to avoid systematic effects in the abundances by over-/under-estimating the temperature, and therefore selected an intermediate temperature scale. The photometry was from 2MASS (K) and Johnson V \citep[the $V - K$ was taken from][]{2MASSref} and the parallax was taken from the Hipparcos catalogue \citep{hipcat}. 

Our final effective temperatures are based only on the ($V-K$) colour index. Among the indices, we considered it to be the most metallicity-independent one \citep{AlonsoG}, since infra-red magnitudes are less affected by reddening (K is the only infra-red magnitude that is available for all our sample stars). Additionally, the temperatures derived for the dwarfs based on this colour are in good agreement with those determined via H$_{\beta}$ line fitting \citep{nissen07}. We note, however, that the ($b-y$) colour tends to predict slightly higher temperature values than ($V-K$).

The reddening corrections, $E(B-V)$, were mostly derived from the Schlegel dust maps\footnote{http://spider.ipac.caltech.edu/staff/jarrett/irsa/dust.html} \citep{schlegel} and corrected according to \citet{bonEBV} if they exceeded 0.1\,mag. 
For a few stars, we took the corresponding $E(B-V)$ values from the literature \citep{nissen02,nissen04,nissen07}. We applied the formula of \citet[][]{alondw} of $E(V$-$K)=2.72E(B$-$V)$, which corresponds to the average of those of \citet{ramirez}, \citet{kinman}, and \citet{nissen02}. A filter conversion of $-$0.04 from \citet[][2MASS to Johnson]{bessel} transformed the K magnitudes from the 2MASS to the Johnson system, and brought both magnitudes to the Johnson scale leading to:\\

$V-K_{\rm 0,Johnson} = V_{\rm Johnson}-K_{\rm 2MASS}  - 0.04 - 2.72 E(B-V)$. \\

\noindent Having both magnitudes on the Johnson scale, we converted $V-K$ from Johnson to TCS (Observatorio del Teide), which can be done by applying the following relation from \citet{alonsofilter}\\

$(V-K)_{\rm TCS} = 0.05 + 0.994(V-K)_{\rm Johnson}$.\\

\noindent This last part of the filter conversion -- Johnson to TCS -- corresponds on average to +0.04 mag. We keep all transformations for the sake of accuracy.

In the case of stars (typically, from the disk) found to have unrealistically large $E(B-V)$ values we decided to derive their temperatures spectroscopically. The gravity was calculated from Hipparcos parallaxes by applying the classical formula \\

$\log \frac{g}{g_{\odot}} = \log \frac{M}{M_{\odot}} + 4 \log \frac{T_{\mathrm{eff}}}{T_{\mathrm{eff\odot}}} - 4V_0 +0.4BC +2\log \pi +0.12 $,\\

\noindent where $M$ is the mass, $V_{0}$ is the dereddened apparent magnitude, $BC$ is the bolometric correction\footnote{Adopted from \citet{nissengrav}}, and $\pi$ is the parallax. Stellar masses were taken from the literature \citep{nissen02,nissen04,nissen07}. On the basis of \citet{alonsobc}, we calculated the BC for each of our stars.
For the few stars for which no parallax was available, we constrained their gravities by enforcing ionisation equilibrium between Fe I and Fe II\footnote{In total, we have 13 stars for which no reliable information on either their ($V - K$) colour, parallax, or reddening correction, $E(B-V)$, is available. Hence, we resort to spectroscopically derived stellar parameters, i.e. excitation temperatures and gravities constrained via Fe I/Fe II ionisation equilibrium (see also letter 'a' and 'c' in online material B).}. In general, the Fe I and II abundances are in good agreement, although we note that for stars labelled with an 'a' or 'c' in online Table \ref{stelpar} the parallax was neglected (either owing to their large uncertainties or wide-ranging Fe I and II abundances when the gravity was derived from the parallax). 
The metallicity was derived from Fe I equivalent widths (EWs) and is in good agreement with previous studies.  
The microturbulence was determined by requiring that all Fe I lines yield the same abundance regardless of their EW. The final values and adopted methods are presented in the online material \ref{stelpar}.

\subsection{Error estimates for the stellar parameters}

The largest source of uncertainty in estimating the temperature is the dereddening of the colour indices, e.g. applying overestimated reddening values from the Schlegel dust maps to stars close to the Galactic plane. These can easily translate into uncertainties of several hundred Kelvin in the derived temperature. Disregarding these extreme cases, we found that the general uncertainty in the reddening values is usually 0.05 mag. Combining these values of 0.05 mag with the uncertainty due to the Johnson--2MASS transformation led to typical uncertainties of the order of 100 -- 150\,K. A slight magnitude--temperature offset was found between giants and dwarfs owing to the stronger colour dependence of the dwarfs' temperature compared to that of the giants. Similar uncertainties were found for the excitation temperatures. 

Since all stellar parameters to some extent are inter-dependent, we also tested the influence of gravity and metallicity on the temperature. For instance, an uncertainty of $\pm$0.15 dex in metallicity has a negligible effect on the temperature (the uncertainty is usually a few Kelvin). An uncertainty of $\pm$ 0.2 dex in gravity causes an uncertainty in the temperature of $< \pm$1 -- 10 K. Finally, the microturbulent velocity is found to have a negligible impact on the temperature. 

The main uncertainty in the gravity comes from the uncertainty in the parallax, which is on average $\pm$1.0" \citep{hipcat}. This translates into $\lesssim$ 0.2 dex in \glog. 
A change of $\pm$100 K in temperature only causes a gravity uncertainty of $\pm$0.04 dex. 
By altering the gravity by $\pm$ 0.25 dex, the Fe II abundance is affected by $\pm$0.15 dex, whereas the Fe I abundance remains basically the same. 

The metallicity is based on EW measurements for which Fe I and Fe II lines provided consistent results, usually agreeing to within 0.1 dex. Since our derived metallicities closely match those found in the literature (most of our stars are well-studied Galactic halo and disk stars), our typical adopted uncertainty in the metallicity is $\pm$0.1 dex ($\pm$0.15 dex in only a few cases).

For the microturbulence velocity, we estimated uncertainties of the order of 0.15 km/s, stemming from the uncertainty in the [Fe/H] and the uncertainty in the Fe EW measurements (which is of the order of $\pm$2 m\AA\,, as tested via repeated independent measurements).

\section{Atomic data and line lists}
This section is divided into two. The first part presents the newly
calculated $\log gf$ values of silver, and the second part describes the
adjustments and calibrations carried out on the line lists.
We first note that similar calculations are not necessary for palladium. This element has six naturally occurring stable isotopes (102, 104, 105, 106, 108, 110), of which only four are accessible to the r-process. $^{105}$Pd is the only odd-mass isotope with nuclear spin 5/2 for which hyperfine splitting exists. The effect on the oscillator strength is, however, minor, since this isotope only contributes 22.33\%\footnote{http://www.tracesciences.com/pd.htm} of its solar elemental abundance. Hence, we continue focusing only on the hyperfine structure (hfs) of silver.

\subsection{Atomic data}
This section focuses on the derivation of the hfs of the
resonance lines in Ag I.

Silver has two stable isotopes with mass numbers 107 and 109, respectively. The nuclear spin is $I$=1/2 for each of the isotopes. As a consequence, each fine structure level is split into two hyperfine levels. The resonance lines in Ag I connect the lower 5s level to the 5p levels.

The isotopic and hyperfine structures commonly used in abundance studies of the Ag resonance lines are those given in \citet{RossAller}. They derived $\log gf$ values for the different hyperfine and isotopic components using the experimental studies of the relative hfs pattern conducted by \citet{jackson} and \citet{crawphys}. These are intensity measurements of different components studied by interferometric experiments. \citet{RossAller} label four components, i.e. two hyperfine components for each isotope. The expected number of components are three for each of the isotopes 107 and 109 (see Table \ref{table:isotopic} and \ref{table:nonisotopic}). 
The uncertainty in the old intensity measurements resulted in a misinterpretation and misidentification of the components.

We derive new hyperfine transition components based on several experimental measurements of the hfs from more recent studies, using the theory of the addition of angular momenta to derive the hyperfine components. We also derive experimental oscillator strengths, $\log gf$ values, for the different components. The transition energies are derived from unresolved Fourier transform spectroscopy (FTS) measurements. 

\subsection*{Hyperfine structure components}
The splitting due to the hfs of a level is given by

\[ \Delta E_{\mathrm{hfs}} = \frac{1}{2}A_{\mathrm{hfs}}[F(F+1)-J(J+1)-I(I+1)],
\]
where $A_{\mathrm{hfs}}$ is the hyperfine magnetic dipole constant. For nuclei with larger spin, the electric quadrupole moment can be significant, but for nuclei of spin I=1/2, as for Ag, only the magnetic dipole is non-zero \citep{cowanphys}. The quantum numbers $I, J,$ and $F$ are those related to the nuclear spin, total angular momentum of the electrons, and the total angular momentum with the nuclear spin taken into account, respectively. This expression assumes that the coupling among the electrons, resulting in a total angular momentum $J$, is much stronger than the coupling to the nuclear angular momentum $I$. The interaction between \textbf{\textit{I}} and \textbf{\textit{J}} are coupled to a moment \textbf{\textit{F}}. 

The energy splitting for a given level can thus be derived from the hyperfine constant. The hyperfine constants $A_{\mathrm{hfs}}$ for the 5p levels were measured by \citet{carlsson} by observing quantum beats. The splitting of the 5s level is an order of magnitude larger and was measured by \citet{dahmen}. From the energy splittings, the relative wavelengths for the transitions can be derived.

The intensity ratios for the transitions between the different hyperfine components can be derived using the expressions for the addition of angular momenta (e.g. \citealt{cowanphys}), where the decay in each channel is proportional to 
\[ A \propto (2F+1)(2F'+1) \left\{ \begin{array}{ccc}
J & I & F \\
F' & 1 & J' \end{array} \right\}^2, \] 
and the prime is for the lower level. From the hyperfine constants of the 5s and 5p levels, the hyperfine pattern with relative intensities and splitting can be derived. This gives the relative intensities and positions of the hyperfine components for one isotope, but not the relative shift between the isotopes.

\begin{figure}[!h]
\begin{center}
\includegraphics[width=0.5\textwidth]{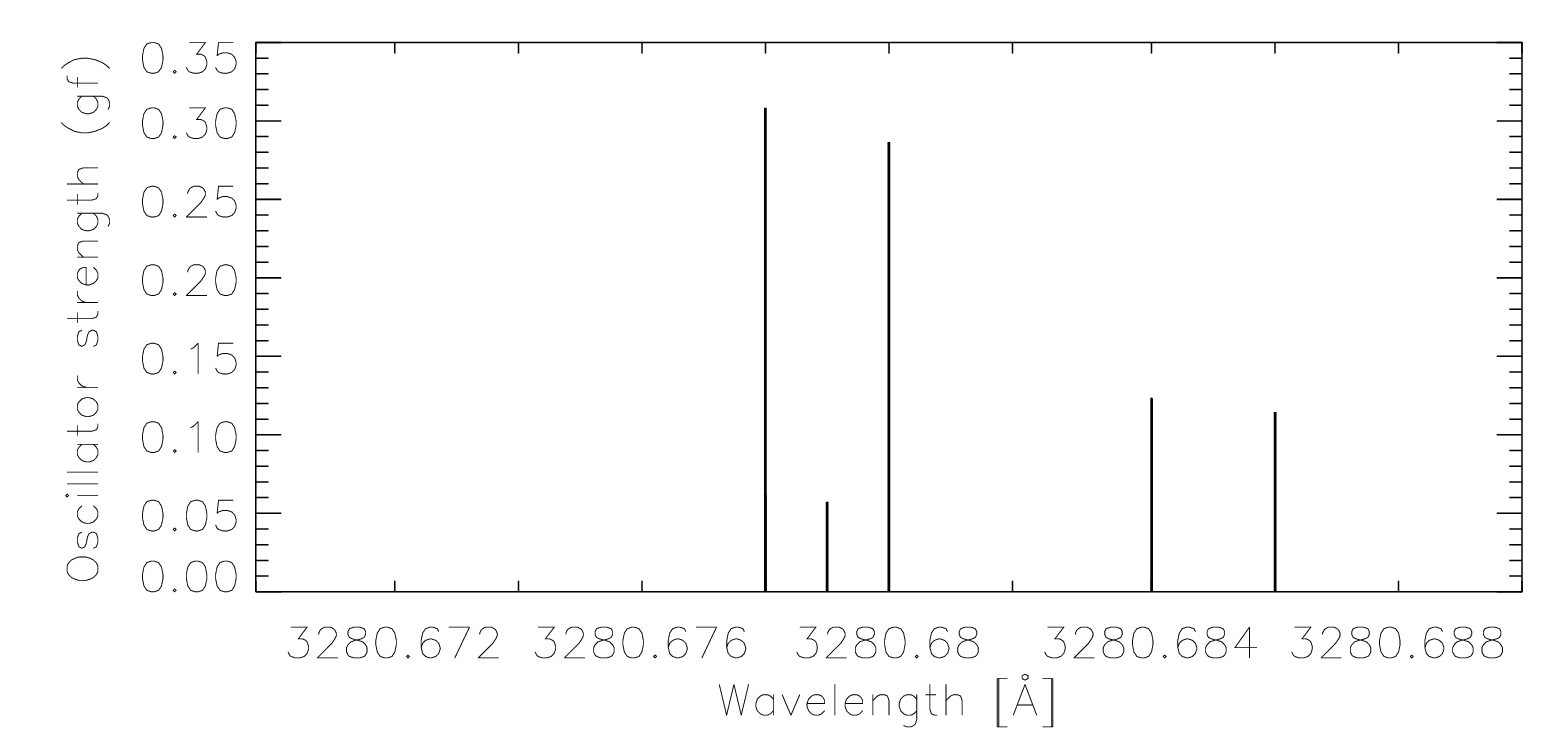}
\includegraphics[width=0.5\textwidth]{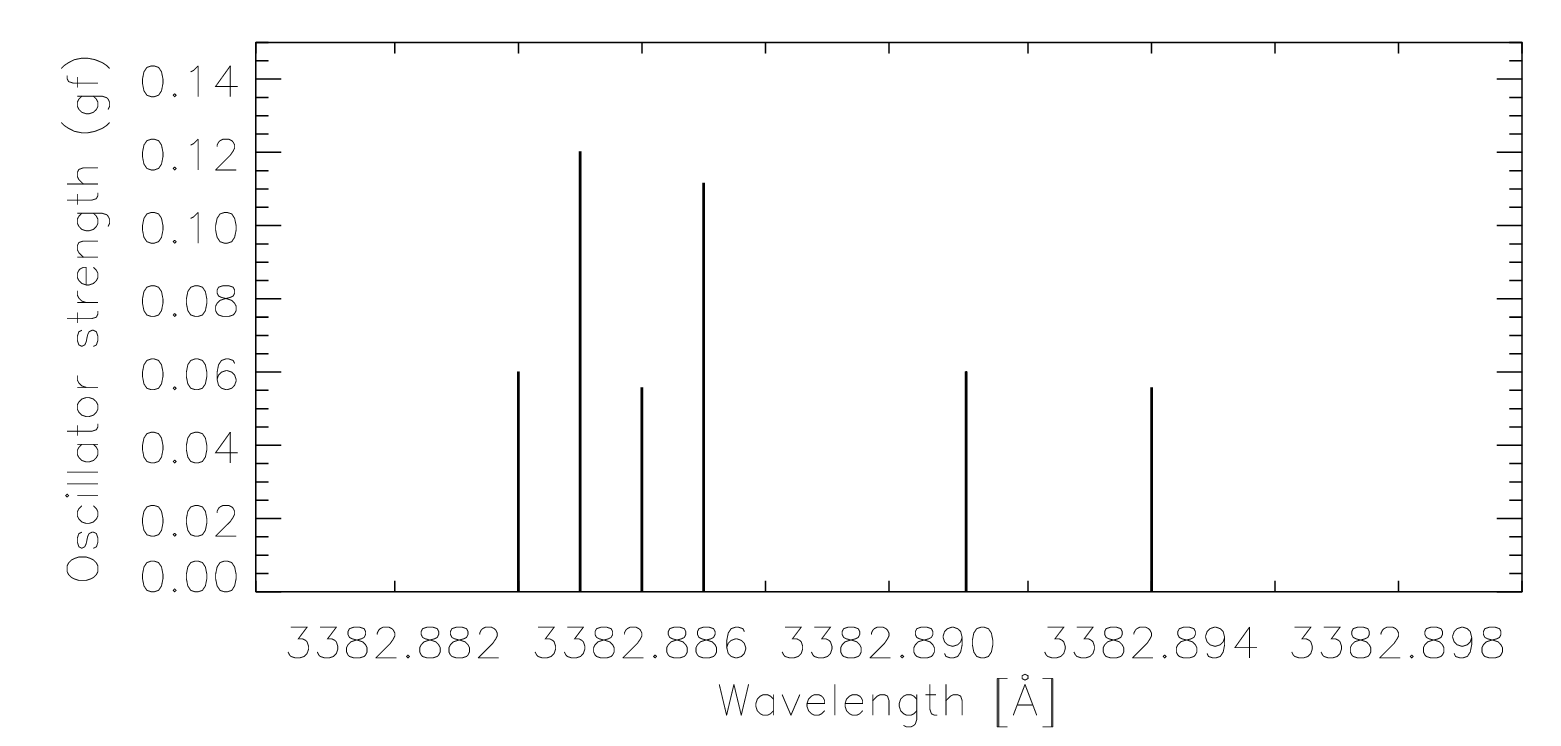}
\caption{The hyperfine and isotopic structure of the resonance lines of 
Ag, calculated using a natural isotopic abundance.}
\label{hfsstructure}
\end{center}
\end{figure}

We used the interferometric observations of \citet{jackson} to derive the shift between the two isotopes. The resolved components in their measurements were, with the aid of the predicted hfs for each isotope, used to derive the isotopic shift. We used the resolved components ($F_u-F_l$: 1--0) to establish the isotopic shift, which are 0.026\,cm$^{-1}$ and 0.022\,cm$^{-1}$ for the 5s~$^2$S$_{1/2}$ -- 5p~$^2$P$_{1/2}$ and 5s~$^2$S$_{1/2}$ -- 5p~$^2$P$_{3/2}$, respectively. 
The resulting structure for the two resonance lines are shown in Fig. \ref{hfsstructure}.

The absolute wavelengths of the different components were derived from the centre of gravity of the resonance lines measured by \citet{pickering}, who used a hollow cathode discharge and Fourier Transform Spectrometer. The hyperfine and isotopic structure are too small to be resolved in the Doppler broadened line profiles.

\subsection*{Transition strengths}
The derivation of the line structure due to isotopic and hyperfine structure above give the relative intensities. To use the transitions for quantitative studies, we need the absolute values, i.e. the oscillator strengths ($\log gf$), which can be derived from the radiative lifetime of the upper levels. 

The lifetimes for the upper levels of the resonance transitions, 5p $^2$P$_{1/2, 3/2}$ were measured using a laser induced fluorescence technique by \citet{carlsson}. Since there is only one decay channel per level, the transition rates ($A$) are simply given by the inverse of the lifetime as $A=1/\tau$. 

The absolute transition rates can, combined with the relative intensities of the hyperfine components for a given fine structure transition as discussed above, give the $\log gf$ value for the individual hyperfine components according to 
\[ gf=1.499 \cdot 10^{-14}\lambda^2 gA, \] 
where $\lambda$ is given in nm and $g$ is the statistical weight. These are reported in Table \ref{table:nonisotopic}.

The hyperfine and isotopic structures of Ag is rather small and cannot be resolved in the stellar spectrum. The contribution from the different isotopes can thus rarely be measured. To handle the different isotopes in the stellar spectrum, it is usually assumed that the isotope ratio is the same as the natural abundance: 51.84\% for isotope 107 and 48.16\% for isotope 109.  
It is convenient to derive the contribution to the Ag absorption lines from the different isotopes, normalising to the isotopic ratio. The line parameters for a natural abundance mix of isotopes is given in Table \ref{table:isotopic}. It should be noted, however, that the true $\log gf$ is an atomic parameter for each isotope, which is independent of the isotopic ratio, and the values in Table \ref{table:isotopic} are to be used only with a fixed isotopic ratio and for a total Ag abundance. 
\begin{figure}[!h]
\begin{center}
\includegraphics[width=0.45\textwidth]{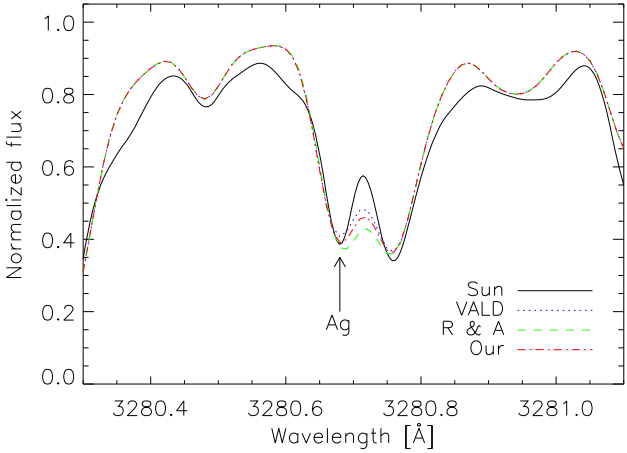}
\caption{The Kitt Peak solar spectrum with spectrum synthesis computed with different line lists overplotted: VALD's $\log gf$ without hfs (dotted blue line); our most recent $\log gf$ values (dash-dotted red line); and the old hfs (only two levels) values from \citealt{RossAller} (R \& A, dashed green line).}
\label{hfs}
\end{center}
\end{figure}
For a strict treatment of the individual abundances for the two isotopes, the values in Table \ref{table:nonisotopic} should be used. 
\begin{table*}[!ht]
\caption{Model parameters for the silver resonance lines, assuming an isotopic ratio of 51.84\% for isotope 107 and 48.16\% for isotope 109. \label{table:isotopic}}
\begin{center}
\begin{tabular}{cccccc}\hline \hline
Isotope	& Lower level	& Upper level	& Flow--Fup & $\lambda_{\mathrm{air}}$	& Reduced $\log gf$ \\ 
 & & & & [\AA] & \\%[3mm]
\hline					
107	& $^2$S$_{1/2}$	& $^2$P$_{1/2}$	& 0--1	& 3382.891	& $-$1.221 \\
107	& $^2$S$_{1/2}$	& $^2$P$_{1/2}$	& 1--0	& 3382.884	& $-$1.221 \\
107	& $^2$S$_{1/2}$	& $^2$P$_{1/2}$	& 1--1	& 3382.885	& $-$0.920 \\
109	& $^2$S$_{1/2}$	& $^2$P$_{1/2}$	& 0--1	& 3382.894	& $-$1.253 \\
109	& $^2$S$_{1/2}$	& $^2$P$_{1/2}$	& 1--0	& 3382.886	& $-$1.253 \\
109	& $^2$S$_{1/2}$	& $^2$P$_{1/2}$	& 1--1	& 3382.887	& $-$0.952 \\
\hline
&&&&total $\log gf$ & $-0.334$\\ 
\hline
&&&&&\\[-2mm]
107	& $^2$S$_{1/2}$	& $^2$P$_{3/2}$	& 0--1	& 3280.684	& $-$0.909 \\
107	& $^2$S$_{1/2}$	& $^2$P$_{3/2}$	& 1--1	& 3280.678	& $-$1.210 \\
107	& $^2$S$_{1/2}$	& $^2$P$_{3/2}$	& 1--2	& 3280.678	& $-$0.511 \\
109	& $^2$S$_{1/2}$	& $^2$P$_{3/2}$	& 0--1	& 3280.686	& $-$0.941 \\
109	& $^2$S$_{1/2}$	& $^2$P$_{3/2}$	& 1--1	& 3280.679	& $-$1.242 \\
109	& $^2$S$_{1/2}$	& $^2$P$_{3/2}$	& 1--2	& 3280.680	& $-$0.543 \\ 
\hline
&&&&total $\log gf$ & $-0.022$\\ 
\hline
\end{tabular}
\tablefoot{
The transition strengths (reduced $\log gf$) given are not the true $\log gf$, but were adjusted for the natural isotopic ratio. For studies treating the isotopes using individual abundances, the data in Table \ref{table:nonisotopic} should be used.} 
\end{center}
\end{table*}

Figure \ref{hfs} shows the effect of including hyperfine splitting with zero, two or three hfs levels. If we had adopted the $\log gf$ value available from VALD (the Vienna Atomic Line Database\footnote{http://vald.astro.univie.ac.at/$\sim$vald/php/vald.php}, \citealt{VALD}) without hfs, all the Ag abundances would have been overestimated. This effect is even more pronounced in the cool metal-rich stars, where the silver lines are stronger. In dwarf stars such as the Sun, the new hfs predicted $\log gf$ values can lead to a difference of $\lesssim$+0.2 dex in the derived silver abundances, compared to the results based on \citet{RossAller} values (see Fig. \ref{hfs}).  
Hence, neglecting hfs would lead to overestimated silver abundances.

\subsection*{Silver isotopes}
Based on measurements of the visual and near-infrared Ag I and II lines \citep{tysk}, silver is predicted to show a relatively small isotopic shift, which would barely affect the spectral line at our spectral resolution. We carried out a test for the near-UV lines with natural isotopic abundance (which is $\sim 48/52$\% for 109/107 Ag) and compared this to two other test cases with ratios of 25/75 \% and 1/99\% for the 109/107 Ag isotopes, respectively. The actual change in the synthetic spectrum was less than the width of the plotted line. Hence, the change in isotopic fraction could be seen in neither our high quality spectra nor the high-resolution Kitt Peak spectrum of the Sun.  

\subsection{Line list}
We now focus on the silver and palladium lines and their atomic data, since these elements are the ones that have been studied the least. The line list for the Sr, Y, Zr, Ba, and Eu lines is not reported here. They include the most commonly used transitions of these elements, and can be found in the online material (Table \ref{heavylist}).

In general, all atomic data were taken from VALD \citep{VALD}, and we cross-checked excitation potentials and oscillator strengths ($\log gf$) against the NIST\footnote{http://physics.nist.gov/PhysRefData/ASD/lines\_form.html}  (National Institute of Standards and Technology) compilation and recent literature, in order to get the most up-to-date line list and best possible abundances. 

From VALD, we excluded all weak lines\footnote{By adjusting the VALD 'extract stellar' search the minimum $\log gf$ around the silver lines found is $-3.4$\,dex, whereas using the VALD 'extract all' yields a factor of five more lines reaching minimum $\log gf$ values of $-$9.7. This large number of weak lines evidently affects the continuum placement.}, i.e. lines with excitation potential higher than 4\,eV and $\log gf$ values smaller than $-4$ dex. These weak lines have no significant influence on the continuum, thus do not affect the derivation of the Ag abundances. We note that the same approach was followed by \citet{john02}, which we adopted to be able to make a direct comparison to their (the only other) large available sample.

The silver lines are situated at 3280.7 \AA\, and 3382.9 \AA\, and the palladium line used in this study falls at 3404.58 \AA\,. In this near-UV region, the molecular lines (OH and especially NH) make a significant contribution to the spectrum, and all molecular line information was taken from Kurucz's database\footnote{http://kurucz.harvard.edu/molecules.html}. In addition, we note that this wavelength region suffers from unidentified transitions. Therefore, one predicted line from Kurucz -- the 3382.96 \AA, Fe I line -- was included in our final list in order to produce a satisfactory synthetic spectrum. 

\begin{figure}[h!] 
\begin{center}
\includegraphics[width=0.45\textwidth]{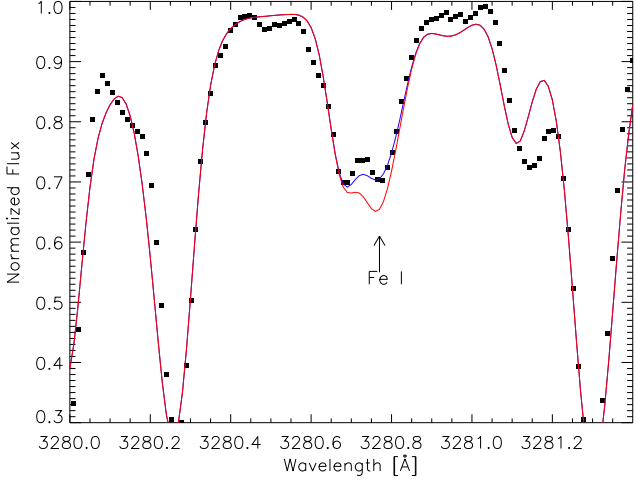}
\caption{The effect of a wrong $\log gf$ of the blending Fe line (marked by an arrow) shown for HD\,121004. The $\log gf$ of Fe I transition (red line) over-predicts the Fe line strength, resulting in an underestimation of the Ag abundance. The synthesis using our adjusted Fe I $\log gf$ value is shown in blue.
}
\label{GFfigure}
\end{center}
\end{figure}
For the 3280.7 \AA\, line, the red wing is severely affected by blends from the Zr II and Fe I transitions. By synthesising the region around the blue silver line using the derived metallicities of the stars, we found that the blending Fe line (3280.76 \AA\,) in most cases is overpredicted (red line in Fig. \ref{GFfigure}). Because our sample covers a large range of stellar parameters, we ran several syntheses, for a large number of stars spanning our entire parameter space with different $\log gf$ values for this line. In the end, we constrained the value of its transition probability so that it gives a reasonable fit to the entire sample. We thus altered the Fe I line $\log gf$ value from $-$2.231 dex to $-$2.528 dex.
An example of this procedure is provided in Fig. \ref{GFfigure} for the star HD\,121004. The value listed in the VALD database ($-$2.231) could be found in neither NIST nor the Fe line list of \citet{fuhr}.  

Furthermore, we note that with this change we were also able to derive
consistent solar abundances from both silver lines. Both solar spectra, the one
observed with UVES\footnote{$R \sim 85,000$,
http://www.eso.org/observing/dfo/quality/UVES\\/pipeline/solar\_spectrum.html}
and the Kurucz Solar Flux Atlas\footnote{R $\sim$ 500,000, 
http://kurucz.harvard.edu/sun.html} yielded silver abundances that differed by
$\sim$ 0.3 dex, with the bluer of the two lines giving the lowest silver abundance. The
Kitt Peak solar spectrum\footnote{ftp://nsokp.nso.edu/pub/atlas/fluxatl/}, which has the highest resolution ($R \sim$ 840,000),
also yields different abundances, of the order of 0.19\,dex. The alteration
of the Fe $\log gf$ to $-$2.528 dex led to an agreement between the two Ag lines/abundances within 0.04 dex of the two solar silver abundances and yielded a value of 0.93 $\pm$ 0.02 dex. This is in good agreement with the previous solar photospheric abundances summarised in \citet[][where log $\epsilon (Ag)_{\odot}$ = 0.94 dex]{asp09}.

\begin{figure}[h!] 
\begin{center}
\includegraphics[width=0.45\textwidth]{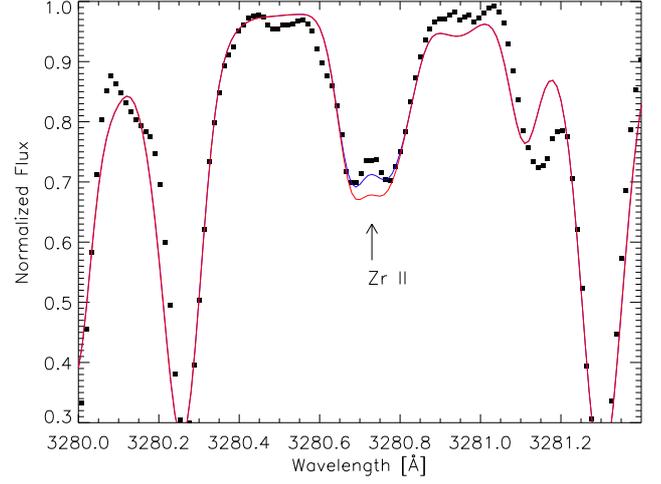}
\caption{A spectrum of HD121004 with the results of two spectral syntheses with different $\log gf$ values for Zr; in blue we plot the results for $-$1.5 dex and in red for $-$1.1 dex. This demonstrates that a reduction in this zirconium line's $\log gf$ value was necessary to obtain a better synthesis and the correct silver abundances. \label{FigZrgf}}
\end{center}
\end{figure}

The synthesis of this region requires one more change to provide an acceptable fit. Based on equivalent width measurements of Zr II lines in the optical (see Sect. \ref{abun}), we first determined the Zr abundance of each sample star, and used those values when synthesising the Ag line at 3280\,\AA. We noticed a similar feature as for the above-mentioned Fe line: the Zr abundance derived from the Zr II line in the red wing of the Ag line was always overestimated by $\sim$ 0.4 dex (in all sample stars) when using the Zr abundance derived from the Zr optical lines. We then reduced the Zr $\log gf$ of the 3280.735\,\AA\, by 0.4 dex and obtained an overall much better fit (see the blue line in Fig. \ref{FigZrgf}).   

There are two additional important blends that contribute to the region around 3280.7\,\AA\,, namely that of Mn I and NH; however, for neither of these lines are changes needed to their atomic data, but they can be properly synthesised once we determined their abundances from other spectral lines/regions. 
\begin{figure}[h!] 
\begin{center}
\label{feup}
\includegraphics[width=0.45\textwidth]{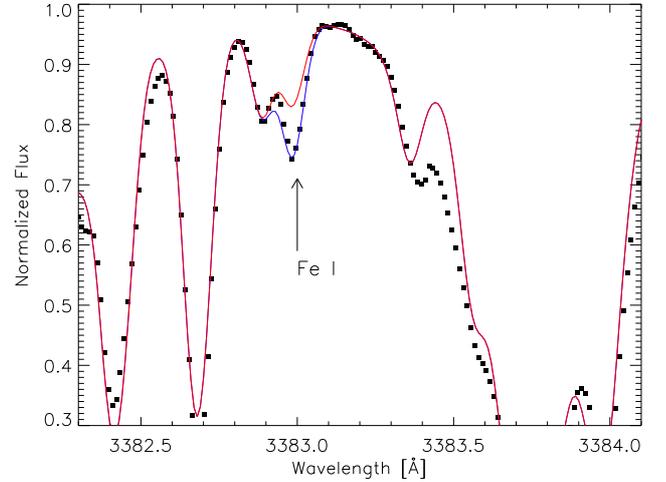}
\caption{A spectrum of HD121004 (dots) to which two syntheses are fitted. The red indicates that the $\log gf$ value is too low, while the blue shows the properly adjusted $\log gf$ for the blending Fe I line.}
\end{center}
\end{figure}

The 3382.9\,\AA\, silver line has a strong Fe blend in its red wing (3382.985\,\AA). This line is taken from the line list of \citet{moore}, because it was not found in either VALD or NIST. However, \citet{moore} only provide the excitation potential of this line, and we had to adjust the $\log gf$ empirically to obtain acceptable fits for this wavelength region. We adopted a $\log gf$ value of $-$3.28 $\pm$0.1 dex, which provides a good fit to the vast majority of our 71 sample stars.

The palladium line list was partially based on the line list published in \citet{john02} and partly on VALD. The list required few (negligible) empirical adjustments and the solar value obtained from synthesising the line in the Kitt Peak solar spectrum was log $\epsilon$ (Pd)$_{\odot}$ = 1.52 dex. As previously noted in \citet{cjhletter}, this value compares very well to the solar photospheric abundance of Pd summarised in \citet{asp09}, log $\epsilon$ (Pd)$_{\odot}$ = 1.57 dex. 

For Ba and Eu, we used the hfs calculated relative oscillator strengths from \citet{mcwil98} and \citet{ivans}, respectively. To derive accurate abundances, we applied a weighting to the lines from which we synthesised the abundances. For barium, we assigned the 5853\,\AA\, line the highest weight (3) since this line is clean, and the 4554\,\AA\, line has an intermediate weight (2) owing to the weak blends. Only when neither of the two aforementioned lines were detectable was the 4934\,\AA\, line used (with weight 1 --- otherwise it was given a weight 0) owing to the severe blends, yielding consistently lower abundances. Furthermore, we note that the 4554\,\AA\, line tends to yield higher abundances ($\sim 0.1-0.15$dex) than the 5853\,\AA\, line due to the presence of blends. Similarly, we assign weights to the Eu lines: 4129\,\AA\, was given the highest weight (3) since it is clean, 4205\,\AA\, an intermediate weight (2) owing to the weak blends, and the 6645\,\AA\, line (weight 1 or 0) is only used when the two blue lines are neither detectable nor observed. The 4205\,\AA\, Eu line yields abundances that on average 0.1 dex higher than those of the 4129\,\AA\, line, while the abundances of the 6645\,\AA\, line agree with the 4129\,\AA\, derived ones. However, the 6645\,\AA\, line is weak and generally only provides upper limits for our stars.

\section{Abundance analysis}
The abundances were calculated based on MARCS model atmospheres\footnote{See http://www.marcs.astro.uu.se/ for model atmospheres in radiative and convective scheme (MARCS models).} \citep{Gus08}, which were interpolated to match the stellar parameters derived for our stars using the code written by \citet{masseron}. Additionally, the 1D LTE synthetic spectrum code MOOG \citep[][version 2009 including treatment of scattering]{snedenphd} was applied to derive the stellar abundances. To date, neither NLTE corrections nor three-dimensional (3D) model effects have been studied for Ag or Pd. However, NLTE corrections can be found in the literature for Sr, Zr, Ba, and Eu and we briefly comment on these when we discuss our results. 

Owing to the severe line blanketing affecting the near-UV/blue part of the spectra of all stars, blending plays a major role, thus spectrum synthesis is required to derive accurate abundances of Ag and Pd. Since hfs is substantial for the Ba and Eu abundances, we also derived their abundances via spectrum synthesis. For the other elements that we studied (Sr, Y, Zr, and Fe), we measured equivalent widths mostly in the redder parts of the spectra to avoid line blends. We measured most equivalent widths manually, by fitting Gaussian line profiles in IRAF ({\it splot} task), except for iron for which we used {\tt Fitline} \citep{fitline}, due to the large number of Fe lines available in our spectra\footnote{The abundances are calculated as:\\
${\rm [A/B]} = {\rm \log(A/B)} - {\rm \log(A/B)_{\odot}},  {\rm where} \log \epsilon (A) = \log (\frac{N_{A}}{N_H}) + 12,$

where $N_A$ and $N_H$ are the number densities of absorbing atoms of element A and hydrogen, respectively. We adopted a scale where the number of H atoms is set to 10$^{12}$.}.

\subsection{Correlation with stellar parameters?}
To ensure that our abundances are pure tracers of formation and evolution processes, and unaffected by spurious analytical effects and method biases, it is important to carefully investigate the trends of the derived abundances with temperature, gravity, and microturbulence.

\begin{figure}[!th]
\begin{center}
\includegraphics[width=0.49\textwidth]{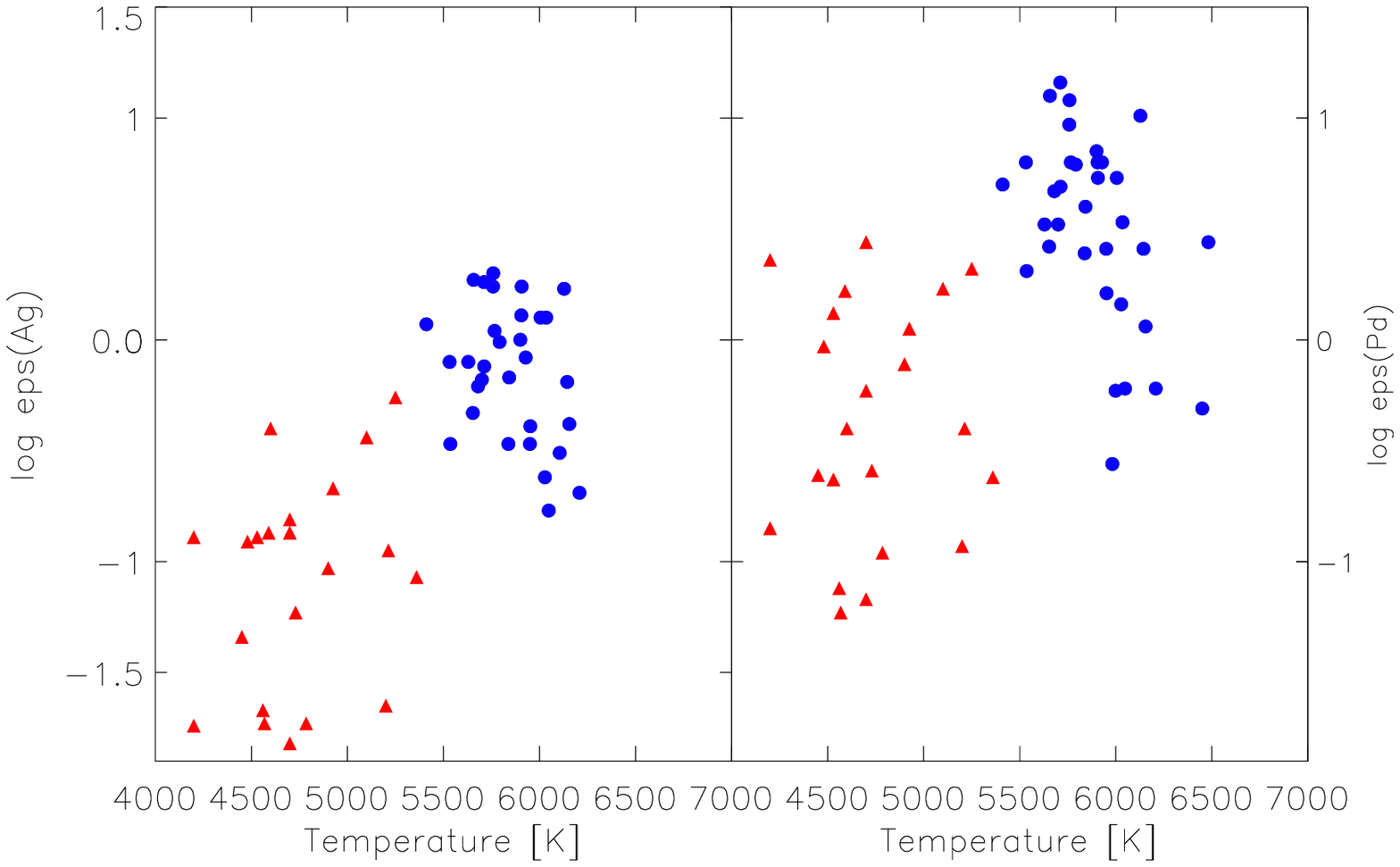}
\includegraphics[width=0.49\textwidth]{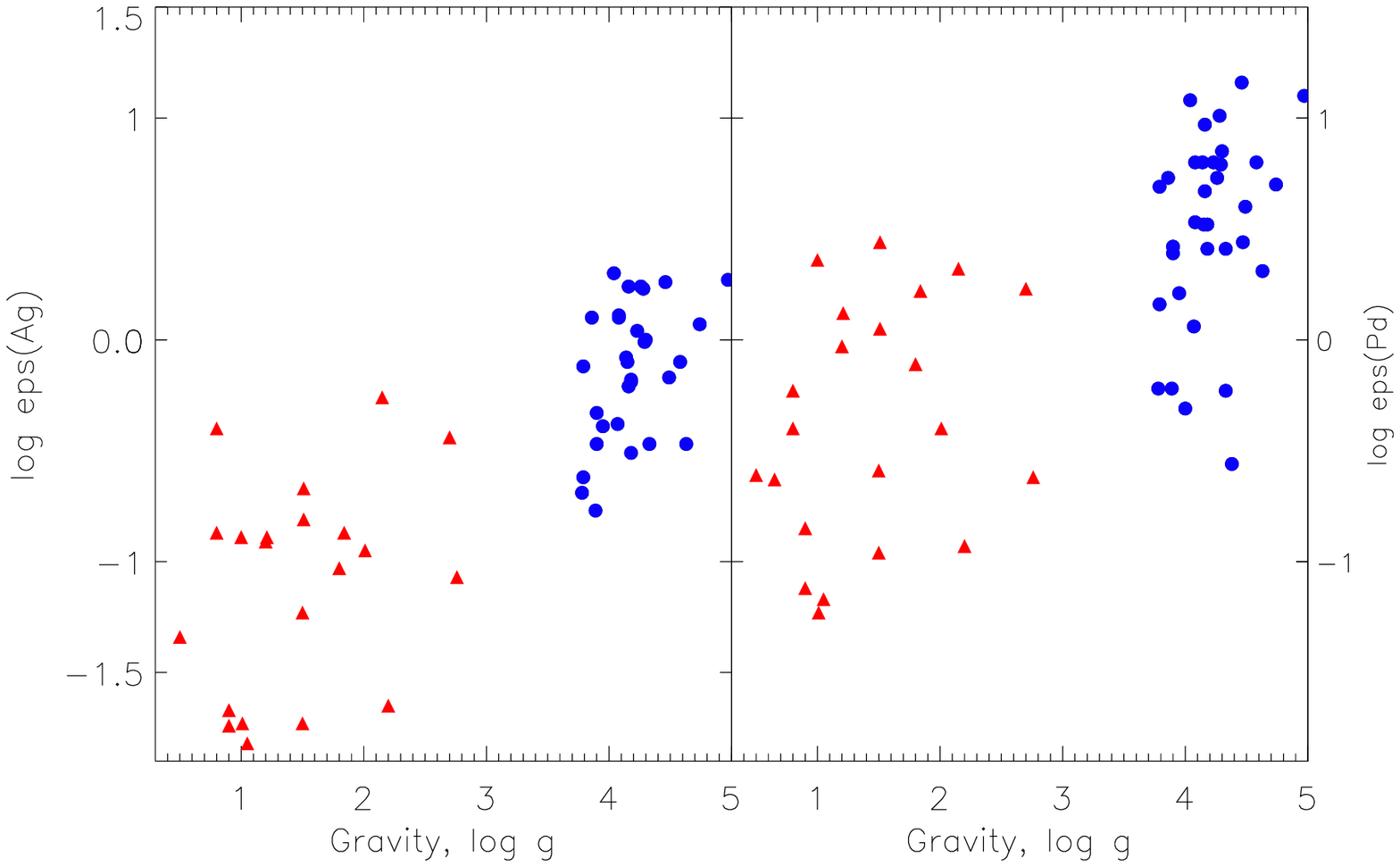}
\includegraphics[width=0.49\textwidth]{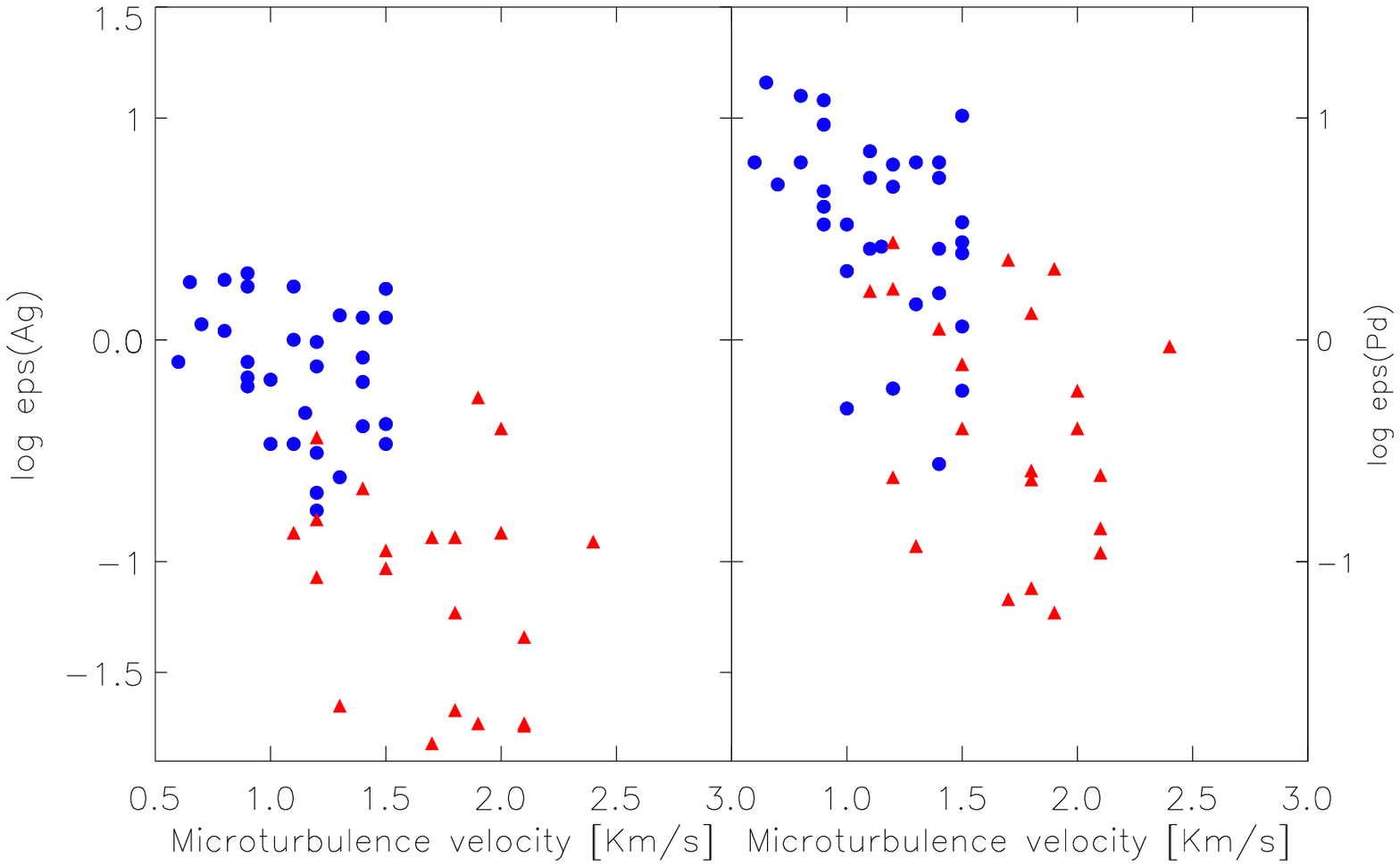}
\caption{ Abundances of Ag (left) and Pd (right) compared to stellar parameters. They show a clear division between the dwarfs and the giants. No trends could be fitted owing to the very large $\chi^2$.
}
\label{epstrends}
\end{center}
\end{figure}

Figure \ref{epstrends} shows that no trend with any of the three parameters is found, but it is evident that there is an abundance difference between dwarfs and giants. Non-local thermodynamic equilibrium effects could be one possible explanation of this difference; other possibilities could be mixing effects \citep{salaris,korn,lind}, microturbulent velocity, an incorrect treatment of the $T-\tau$ relation in the model atmospheres of giants, or unknown line blends in the spectra \citep{lai}. This abundance difference cannot be explained by differences in the stellar evolutionary stages (cf. \citealt{preston}).

The comparison of the Pd and Ag abundances to [Fe/H] can be found in \citet{cjhletter}, where flat trends with metallicity were found. This means that the abundances are not biased by the stellar parameters or the methods applied to determine these, and our abundances can be seen as pure tracers of the formation processes. This allows us to apply the abundances as direct indicators of the chemical evolution of the Galaxy. 

\subsection{Error estimation}
The final error in the derived abundances stems from uncertainties in the stellar parameters, the synthesis/equivalent width measurements, and the continuum placement.
The stellar parameter uncertainties are ($T_{\mathrm{eff}}$/log g/[Fe/H]/$\xi$): $\pm$100K/0.2--0.25dex/0.15dex/0.15km/s (cf. Sect. 3.1). Their effect on the abundances was constrained by running different models in which each parameter was varied by its corresponding uncertainty, one at a time. 

Furthermore, since we synthesised both Pd and Ag transitions, we needed to include the uncertainty in the continuum placement (about $\pm$0.05 dex) and the possible incompleteness of stellar model atmospheres, the synthetic code, and the line list (i.e. missing atomic data), which all together sums up to an uncertainty of $\pm$0.1 dex. Adding all three contributions in quadrature yields uncertainties of the order of $\pm$0.2 dex and $\pm$0.25 dex in the Pd and Ag abundances, respectively.
The average error in the equivalent width measurements of Sr and Y is around 2.5\,m\AA\, and slightly larger for Zr, Ba, and Eu ($\sim$ 4 m\AA\,). These errors were incorporated into the total uncertainty in the abundances shown in the figures in Sect. \ref{abun}.

Propagating the uncertainties in the heavy element abundances derived from equivalent width measurements and stellar parameters resulted in abundance errors of 0.1 -- 0.3 dex. Details can be found in Table \ref{DWallabun} and \ref{Gallabun}.

\section{Indications of a second r-process}\label{abun}
To characterise the formation process of Pd and Ag, we compare their abundances to those of various different elements that trace the weak/main s-process and the main r-process. This comparison allows us to detect either similarities or differences between the yet unidentified formation process of Pd and Ag and the known formation processes of the elements we compare to.
For this purpose, we selected the following tracer elements, which at solar metallicity are created by the process we have listed in Table \ref{tracer}. 
\begin{table*}[!ht]
 \caption{We list the elements and the process that they trace at solar metallicity. }
\label{tracer}
\begin{center}
\begin{tabular}{lccc}\hline \hline
Elements & Formation process [1] & Reference & r-process fraction [8]  \\
 \hline
Sr & 85\% s-process (weak s-process) & [1,3,6] & 2.7\% \\ 
Y & 92\% s-process (in part weak s-process) & [1,3,6] & 5\%\\
Zr & 83\% s-process (low weak s-process) & [1,6] & 9.2\%\\ 
Pd & 54\% r-process (some ($<\sim67$\%) weak r-process) & [1,2] & 46.9\% \\ 
Ag & 80\% r-process (mainly ($>\sim71$\%) weak r-process) & [1,2,5]& 77.9\%\\ 
Ba & 81\% s-process, (main s-process) & [1,4,7] & 11.3\% \\
Eu & 94.2\% r-process (main r-process) & [1] & 94\% \\
\hline \hline
\end{tabular}
\tablefoot{The references refer to the following papers: [1] \citet{arland}, [2] \citet{Fara}, [3] \citet{heil}, [4] \citet{loddersSun}, [5] \citet{montes}, [6] \citet{pigna}, [7] \citet{chrisrev}, and [8] \citet{bister} for a comparison to more recent r-process fractions. }
\end{center}
\end{table*}
This means that a correlation of Ag with Ba around solar metallicity would indicate that Ag had a common formation process to Ba, which in this case would be the main s-process.
However, at low metallicity this picture changes: Sr, Y (and Zr) could be created by charged particle freeze-outs \citep{klk08,Fara}, and Ba mainly by the main r-process. We find indications that Zr also receives weak r-process contributions at low ([Fe/H] $< -$2.5) metallicities, which agrees with \citealt{Fara} (see also Sect. \ref{models}).  

\subsection{Chemical evolution trends of Sr -- Eu}
We first compare the elemental abundances of Sr -- Eu with Fe\footnote{All abundances are available online --- see Table \ref{DWallabun}--\ref{Gallabun}} to follow the chemical evolution of these elements, and detect the onset of the various formation processes. 
We also compare our derived abundances to other studies from the literature, which include measurements for some or all of the elements studied here. 
The following five large abundance studies were chosen:
\citet[][J02]{john02}, \citet[][B05]{barklem}, \citet[][F07]{francois}, \citet[][B09]{bonifaciodw}, and \citet[][R09*]{roederer}. The last (R09*) is a compilation of previous studies by \citet{edvard}, \citet{fulbright}, \citet{nis97}, and \citet{stephens}.
As mentioned in Sect. \ref{sample}, we include and compare with some r-process enhanced stars. These are: BD+17$^{\circ}$3248 \citep{cowan02}, CS~22892--052 \citep{sneden03}, and CS~31082--001 \citep[][included in our sample]{hill}. These are clearly labelled in the figures. 
\begin{figure}[ht!]
\begin{center}
\includegraphics[width=0.5\textwidth]{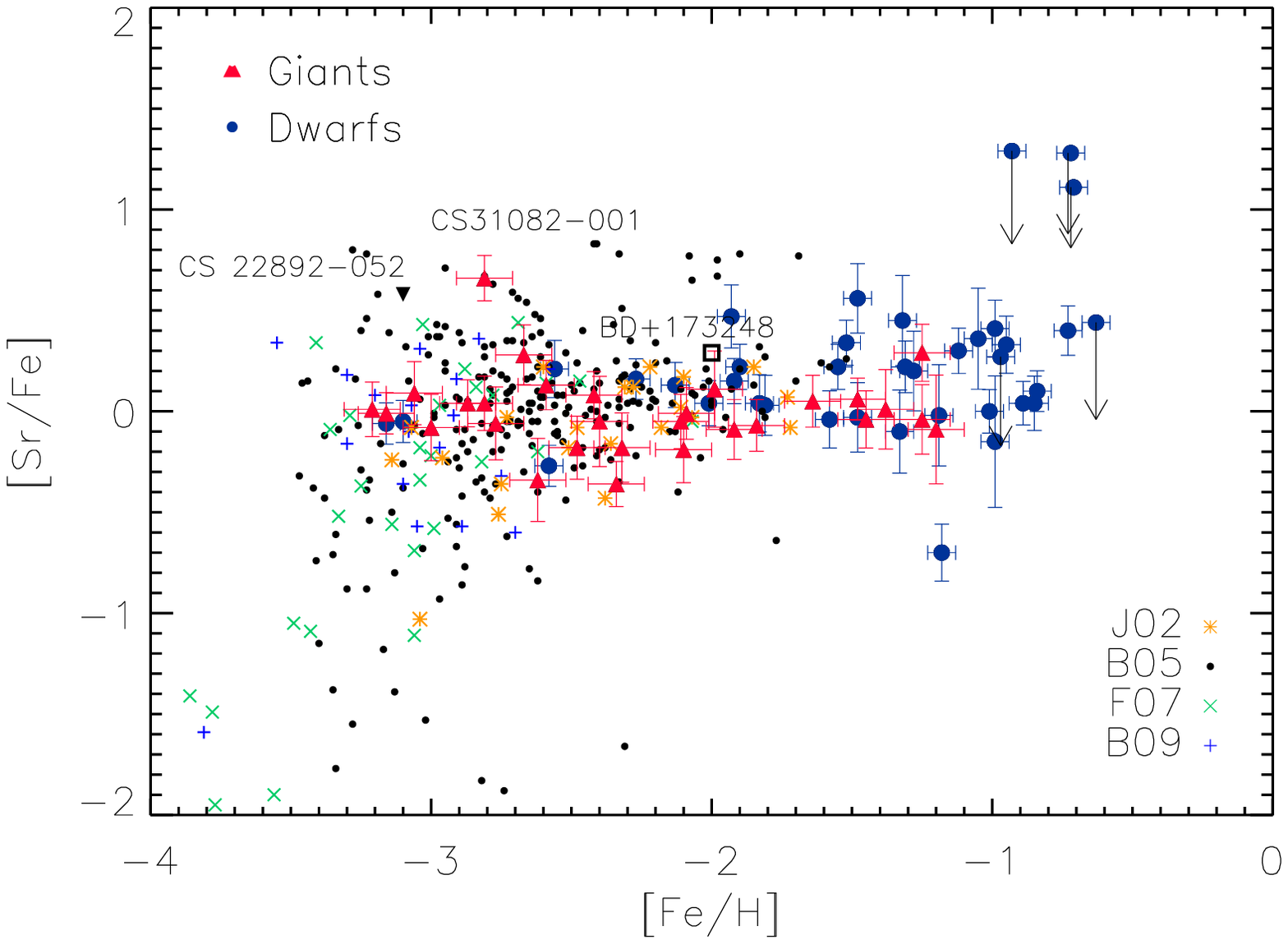}
\caption{[Sr/Fe] as a function of [Fe/H] for the entire sample, compared to \citet[][J02 --- orange asterisk]{john02}, \citet[][B05 --- black dots]{barklem}, the 'First Stars' giants \citet[][F07 --- green $\times$]{francois}, and dwarfs \citet[][B09 --- purple $+$]{bonifaciodw}, respectively. The dwarfs from our sample are shown as filled blue circles, while filled red triangles represent our giants. Three very enhanced stars are shown and labelled in this and the following figures:  BD+17$^{\circ}$3248 \citep[][open black square]{cowan02}, CS 22892--052 \citep[][filled black triangle]{sneden03}, and CS 31082--001 \citep[][also analysed in this study, hence the red triangle]{hill}. Arrows indicate upper limits to the abundances. A flat trend of [Sr/Fe] is seen down to [Fe/H] $\sim -$2.5, below which the scatter becomes dominant. 
}
\label{SrFe}
\end{center}
\end{figure}

Starting with the lightest element, Sr, we see that down to [Fe/H] $= -$2.5, [Sr/Fe] presents a relatively clean and flat trend with a mean value around 0.14 dex (see Fig. \ref{SrFe}). Below this metallicity, the scatter becomes dominant. Only three stars deviate from this picture (HD175179, HD195633, and G005--040), for which only upper limits were attainable from near-UV lines (no spectra covering the wavelength range 3800 -- 4800\,\AA\, were available in the ESO archive). 
\begin{figure}[!ht]
\begin{center}
\includegraphics[width=0.5\textwidth]{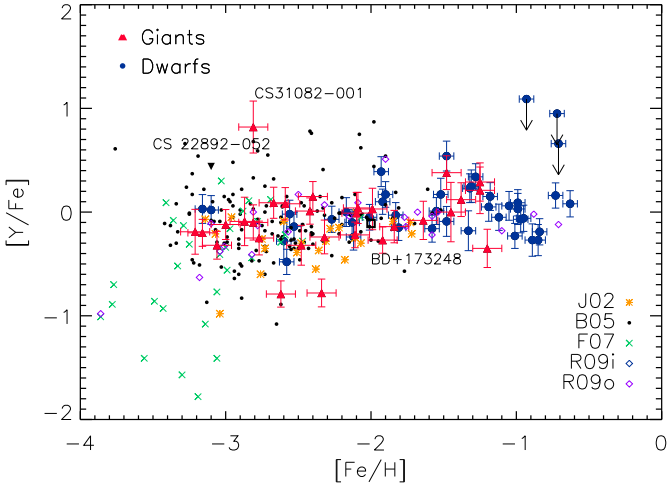}
\caption{[Y/Fe] vs [Fe/H] similar comparison samples as in Fig. \ref{SrFe}, but also including a fourth sample --- \citet[][R09i/o]{roederer} --- shown as blue/purple open diamonds indicating stars belonging to the inner/outer halo, respectively. The enhanced stars agree with the other comparison samples as well as our sample. However, CS 31082--001 is seen to be particularly enhanced in Y. [Y/Fe] shows almost no variation with metallicity down to [Fe/H] $\sim -$2.5 dex.}
\label{YFe}
\end{center}
\end{figure}

The trend for yttrium is also seen to be flat down to [Fe/H] = $-$2.5 dex (Fig. \ref{YFe}). We find the same increase in star-to-star scatter of [Y/Fe] with decreasing [Fe/H] as detected in \citet{roed2010}. However, the average Y abundance is sub-solar. In general the abundance predictions of the Sr/Y--ratio from SN models are found to be very high, most likely due to incorrect solar scaled residuals\footnote{The Sr/Y--ratio can be correctly predicted by the high-entropy wind models \citep{Fara}, where these residual assumptions are not considered.}.  
A too-low solar abundance of Y could have explained this, but this does not seem to be the case, since the solar photospheric and meteoritic Y abundance agree to within 0.04 dex \citep{asp09}, making this a trustworthy value. 
\begin{figure}[!hb]
\begin{center}
\includegraphics[width=0.5\textwidth]{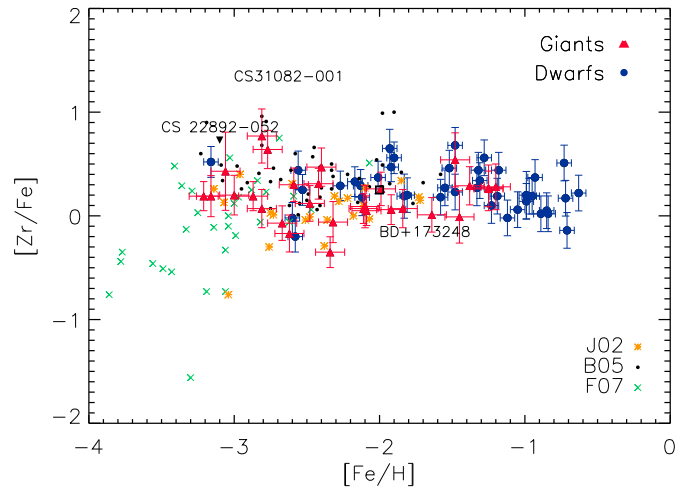}
\caption{[Zr/Fe] as a function of [Fe/H]. Zr does not vary much with metallicity. Symbols and colour are the same as in Fig. \ref{SrFe}.}
\label{ZrFe}
\end{center}
\end{figure}
\begin{figure}[!ht]
\begin{center}
\includegraphics[width=0.5\textwidth]{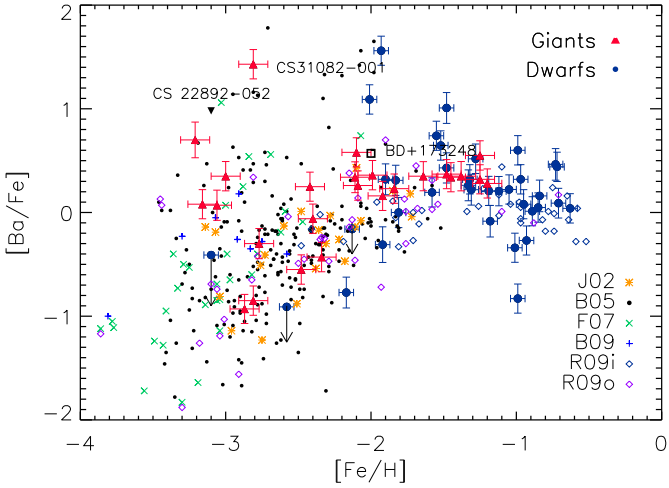}
\caption{[Ba/Fe] plotted vs [Fe/H]. Below [Fe/H] $\sim -2.0$, a very large scatter in all samples is seen. The very large scatter is indicative of a poorly mixed ISM. Symbols and colour coding as in Fig. \ref{SrFe}.
}
\label{BaFe}
\end{center}
\end{figure}

The zirconium abundance distribution is also flat and found to have a mean value of 0.2\,dex down to a metallicity of at least $-2.5$ dex (see Fig. \ref{ZrFe}). The scatter in [Zr/Fe] below [Fe/H] $= -2.5$ is less pronounced than for [Sr/Fe], which may be due to there being fewer Zr abundance determinations at low metallicities compared to, e.g., Sr. One can see from Table \ref{heavylist}, that the Zr lines are intrinsically much weaker than, e.g., the Sr and Ba resonance lines. 

Figure \ref{BaFe} shows the evolutionary trend of [Ba/Fe] vs [Fe/H] which is characterised by a large scatter ($>$ 2 dex) below a metallicity of [Fe/H] $= -$2.0. The large scatter can be interpreted as an indication of different yields from one enrichment event to another, creating an inhomogeneous interstellar medium (ISM). However, it could also point towards several formation processes being at work and releasing very different elemental ratios into the ISM. Even when correcting the derived Ba abundances for NLTE effects \citep[see][]{andrievsky09}, the scatter is far in excess of any possible uncertainty stemming from observations and model assumptions. It is therefore a possible indication that different formation processes are at play. 
\begin{figure}[!ht]
\begin{center}
\includegraphics[width=0.5\textwidth]{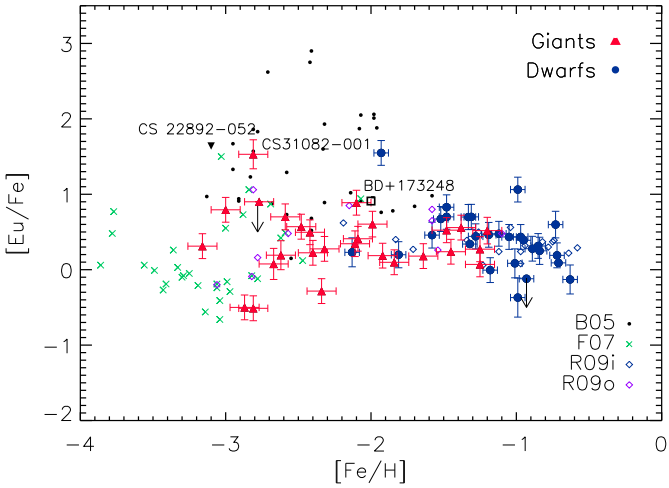}
\caption{[Eu/Fe] as a function of metallicity. A very large scatter is seen at all metallicities (also within the various samples). Symbols and colour coding as in Fig. \ref{SrFe}.
}
\label{EuFe}
\end{center}
\end{figure}
Figure \ref{EuFe} shows a large spread in the europium abundances.

The evolutionary trends of both [Pd/Fe] and [Ag/Fe] relative to [Fe/H] were previously presented in \citet{cjhletter} and were found to be flat and scattered, similarly to the other five elements discussed above. Here, we thus decided to show new plots of Pd and Ag abundances, relative to their neighbouring elements (see following sub-sections). 

We note that, in general, the r-process enhanced stars follow the overall trends, but fall on the upper abundance envelope as one would expect from their enhancements. 
For CS~31082--001, we re-derived all abundances and found them to agree very well with the results of \citet{hill}. The only exception is yttrium, which we propose is caused by uncertainties in the continuum placement ($\pm 0.1$dex) and the profile fitted. The Y lines to which we fit Gaussian profiles are very sensitive to the exact shape and broadening of the profile, and we can only reproduce the observed spectral line by fitting much broader line profiles to the Y lines than the surrounding spectral lines. The offset in line profile between the Y lines and the nearby other spectral lines introduces an $0.3$dex abundance offset in our Y abundance. We can attribute our higher Y abundance compared to that derived in \citet{hill} to a combination of uncertainties and offsets. 

The star-to-star abundance scatter revealed by all the elemental trends discussed here points to a rather inhomogeneous ISM below a metallicity of $-2.5$ (see Sect. \ref{discus} for further discussion). Below this metallicity, the varying abundances indicate that the stars have been affected by different productions (or processes) from various nucleosynthetic events. The main contribution at these low metallicities must come from primary processes, since the sites of the secondary processes (the s-processes) have not yet had enough time to both reach the evolutionary stages where they yield s-process contributions and have their yields incorporated into later generations of stars. 
This is why any monitoring of the r-process is carried out most efficiently below [Fe/H] $= -$2.5. 
From Fig. \ref{SrFe} -- \ref{BaFe}, the s-process might start around [Fe/H] = $-2.5$ dex, since we see a change in the abundance behaviour (trend flattening/lower scatter) at this metallicity. Unfortunately, our data do not allow us to identify the metallicity for the onset of the weak s-process, a problem that we discuss further in Sect. \ref{zrtrans}.

\subsection{Correlations and anti-correlations}
We now turn to a different set of abundance plots, of the type [A/B] vs [B/H] (where A and B are two of the neutron-capture elements under investigation), to see whether and how they (anti-) correlate with each other. This is determined by the abundance trends to which we fit lines. The slopes determine the anti-/correlation. The fitting of linear trends has been made to all points (stars) taking their uncertainties into consideration, and the uncertainties in the fits are expressed in the figures in parentheses. 
These plots are powerful diagnostics for constraining formation processes and can help us to identify similarities and differences among the neutron-capture elements. If A and B correlate (i.e. the [A/B] ratio is flat across the spanned values of [B/H]), it means that they grow in the same way (constant ratio) and that they are most likely created by the same process. If they anti-correlate (e.g. [A/B] decreases with
increasing [B/H]), this is usually interpreted in terms of their having different amounts of A and B, hence different processes being responsible for their formation.
To define our terminology, the strengths of the correlations can be described as follows; a weak/mild anti-correlation is stated for slopes between $-0.25$ and $-0.5$ and a strong anti-correlation is assigned to negative slopes around or steeper than $-0.5$.  
We choose hydrogen (H) as our reference element because we wish to focus only on the formation processes of elements A and B. Had we selected iron instead, the interpretation of the plots would have become more complex because of the different sites contributing to the formation of iron.

In the following, there are two important factors to bear in mind, namely the difference between dwarfs and giants and that below [Fe/H] $< -$2 dex the silver lines could only be detected in giant stars. The giants might have been affected by NLTE or mixing effects, whereas the inclusion of the dwarfs may affect our constraints on the formation processes. The giants could be affected by almost pure r-process yields, whereas the dwarfs might carry a mixture of r- and s-process yields. Therefore, we need to test the purity of the r-process as we do in Sect. \ref{models}. Furthermore, it is very important to look for differences in the behaviour of the Ag and Pd abundance ratios in dwarf and giant stars (see Sect. \ref{discus}). 

Now focusing on the formation process of Pd and Ag, we start by comparing these two elements to Sr, Y, and Zr, which may be formed by the weak s-process elements or charged particle freeze-out (depending on metallicity).

\begin{figure*}[!htbp]
\begin{center}
\includegraphics[width=0.495\textwidth]{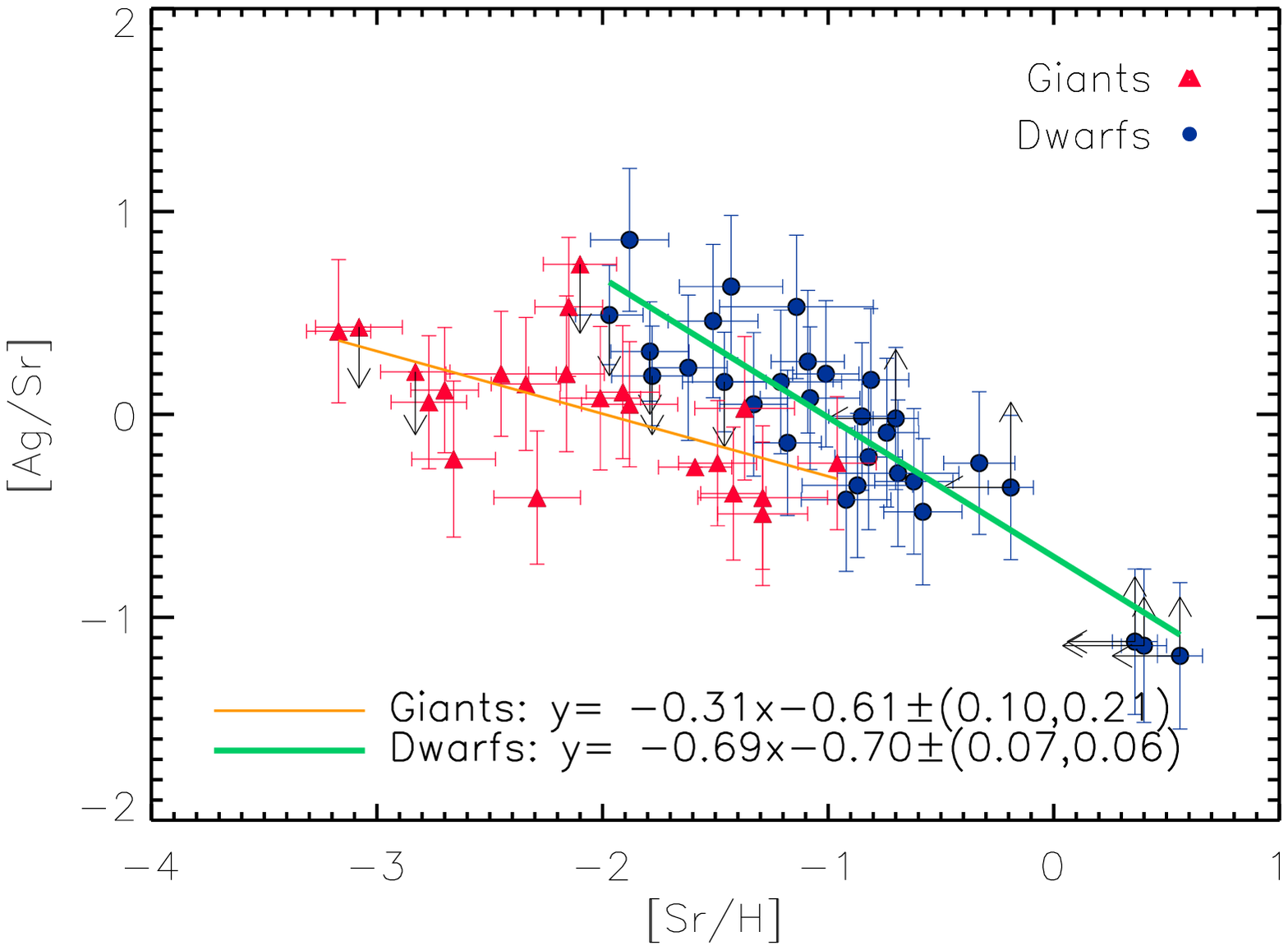}
\includegraphics[width=0.495\textwidth]{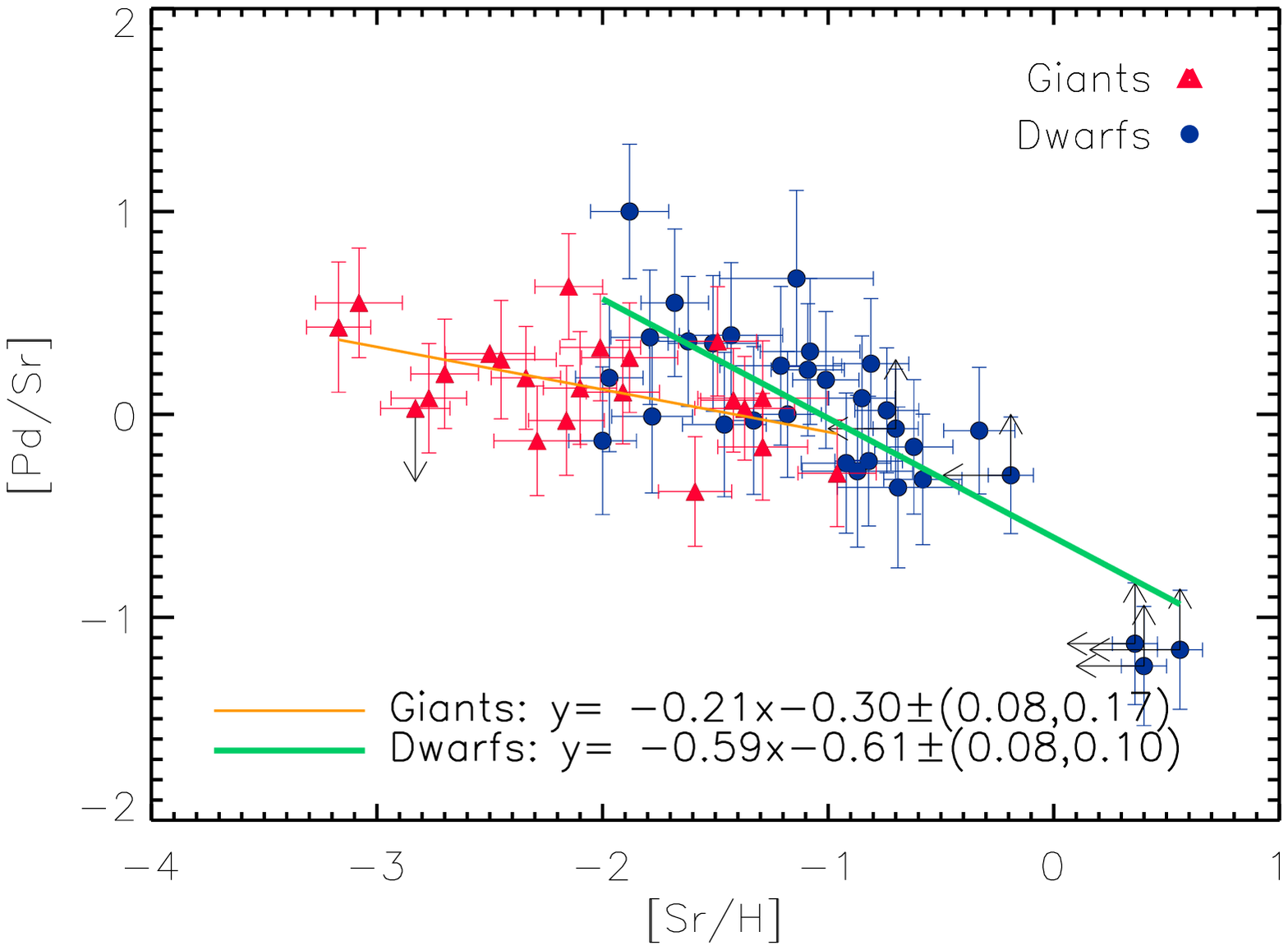}
\caption{[Ag/Sr](left) and [Pd/Sr] (right) as a function of [Sr/H] is shown here for both dwarfs (filled blue circles) and giants (filled red triangles). An anti-correlation is seen in this figure, which is strongest for the dwarfs (see the slopes in the figure). The values given in parenthesis are the uncertainties in the linear fits: the first number is the error in the slope, the second number is the uncertainty in the intersection with the y-axis. }
\label{figAgSr}
\includegraphics[width=0.495\textwidth]{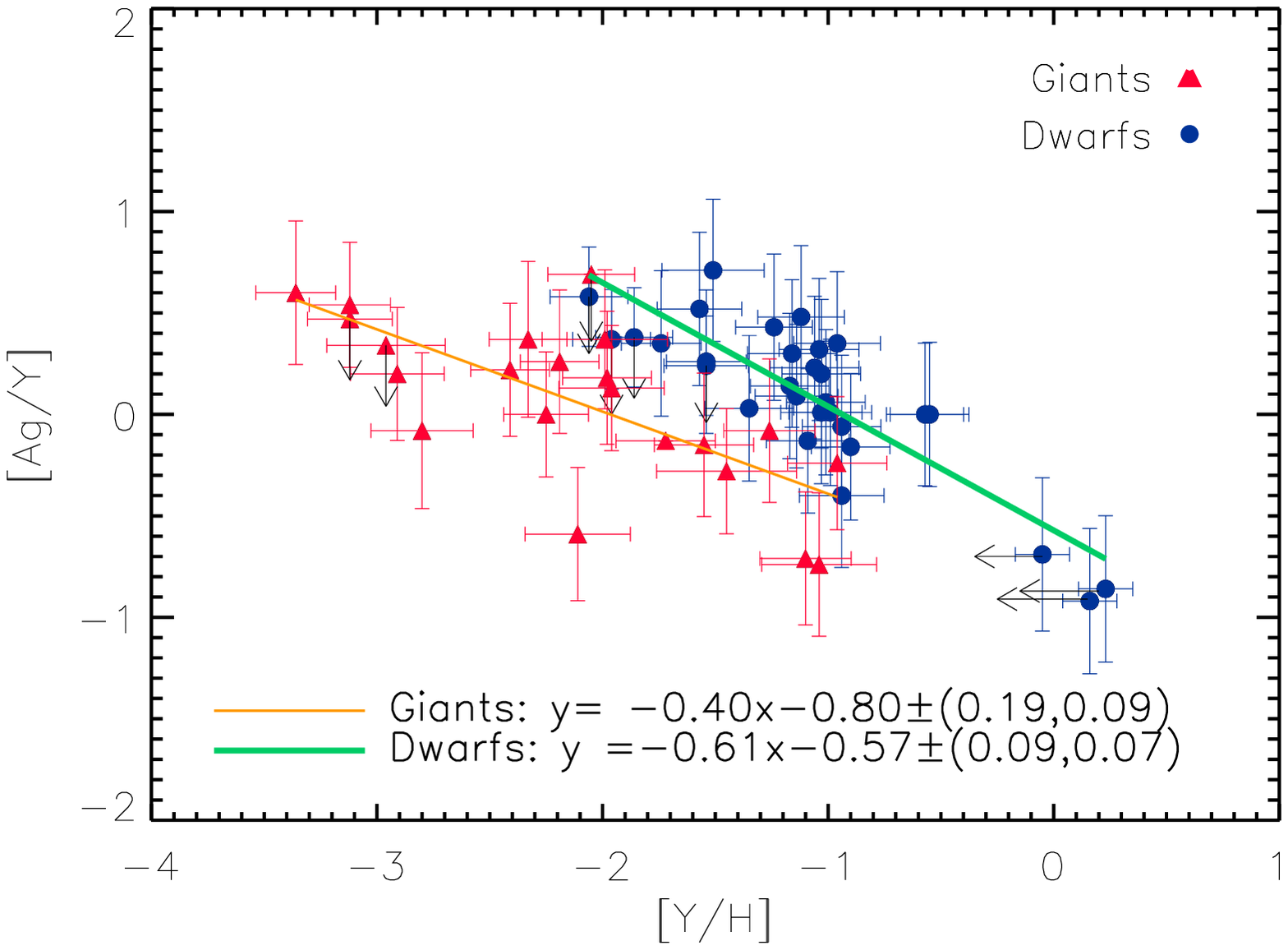}
\includegraphics[width=0.495\textwidth]{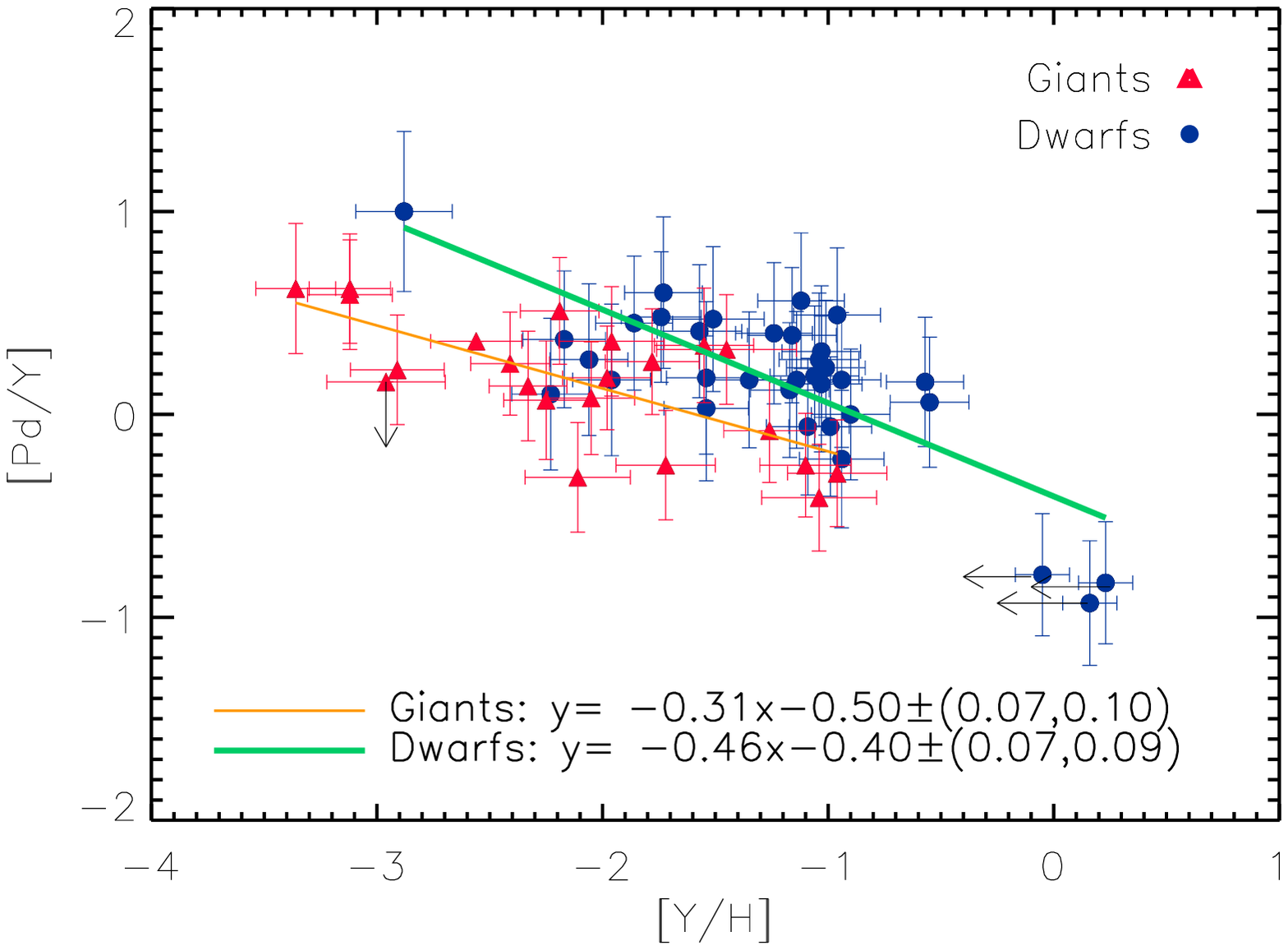}
\caption{ Left: [Ag/Y] as a function of [Y/H]. Right: [Pd/Y] vs [Y/H]. Legend is described in Fig. \ref{figAgSr} and shown in the figure. Anti-correlations between the weak s-process element Y and Ag and Pd are seen in this figure.}
\label{AgY}
\includegraphics[width=0.495\textwidth]{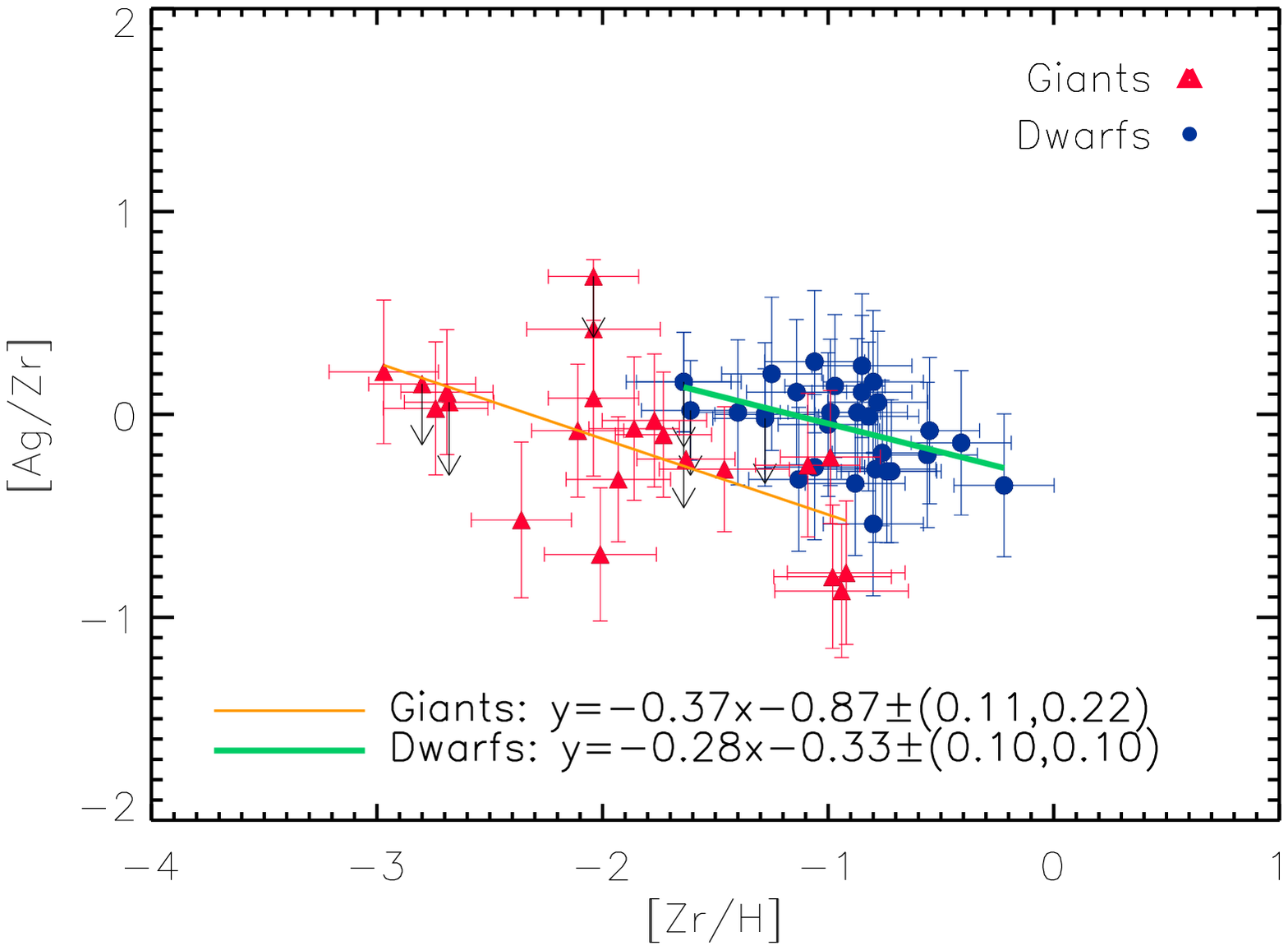}
\includegraphics[width=0.495\textwidth]{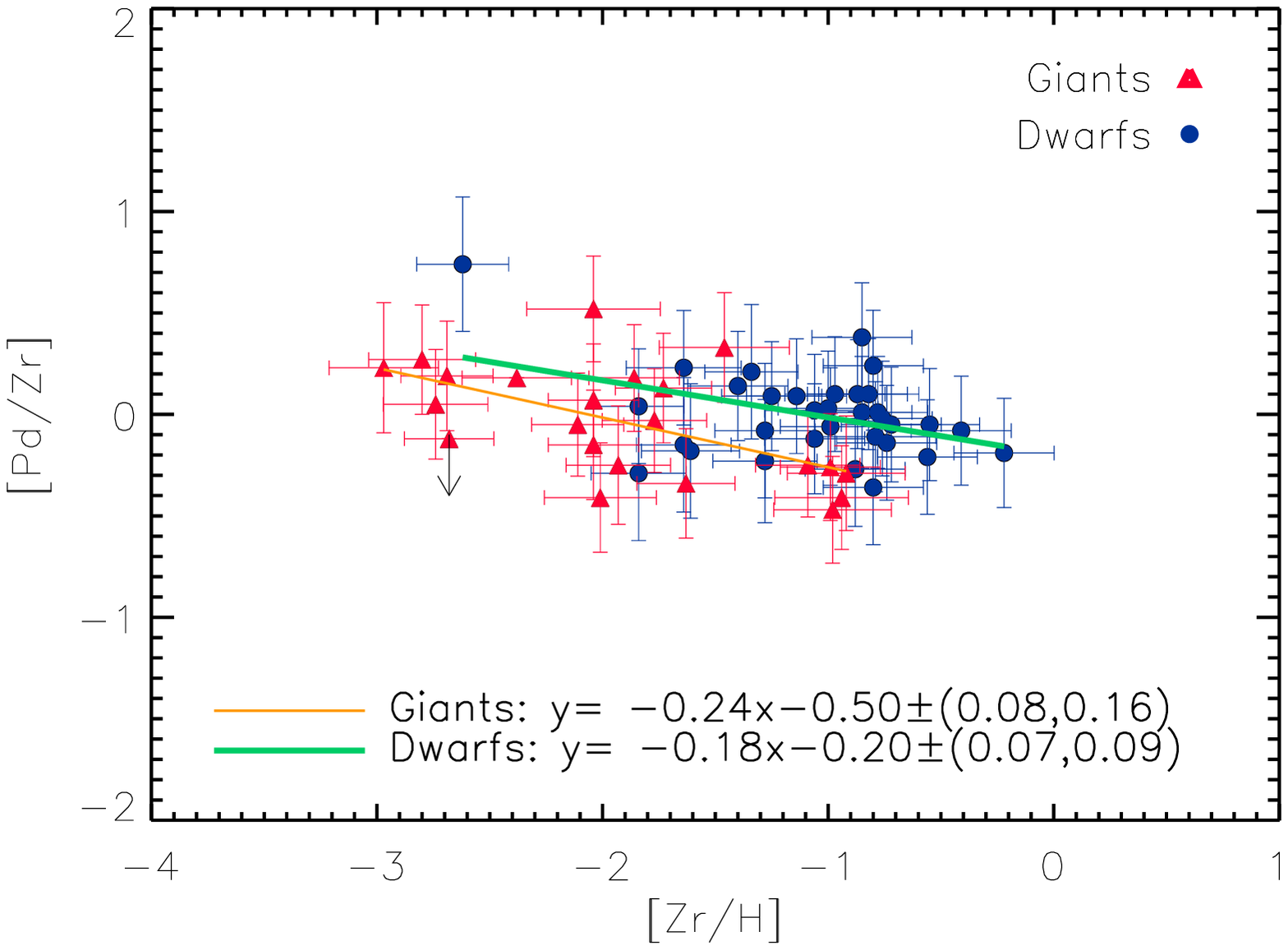}
\caption{ [Ag/Zr] and [Pd/Zr] vs [Zr/H] to the left and right, respectively. The trend of the fitted line is only slightly negative, which could be interpreted as a slight correlation, but the abundances clump. Upper limits to the abundances are indicated by arrows. The formulas of the lines fitted are given in the lower left corner for giants and dwarfs, respectively.}
\label{AgZr}
\end{center}
\end{figure*}
\begin{figure*}[!htbp]
\begin{center}
\includegraphics[width=0.6\textwidth]{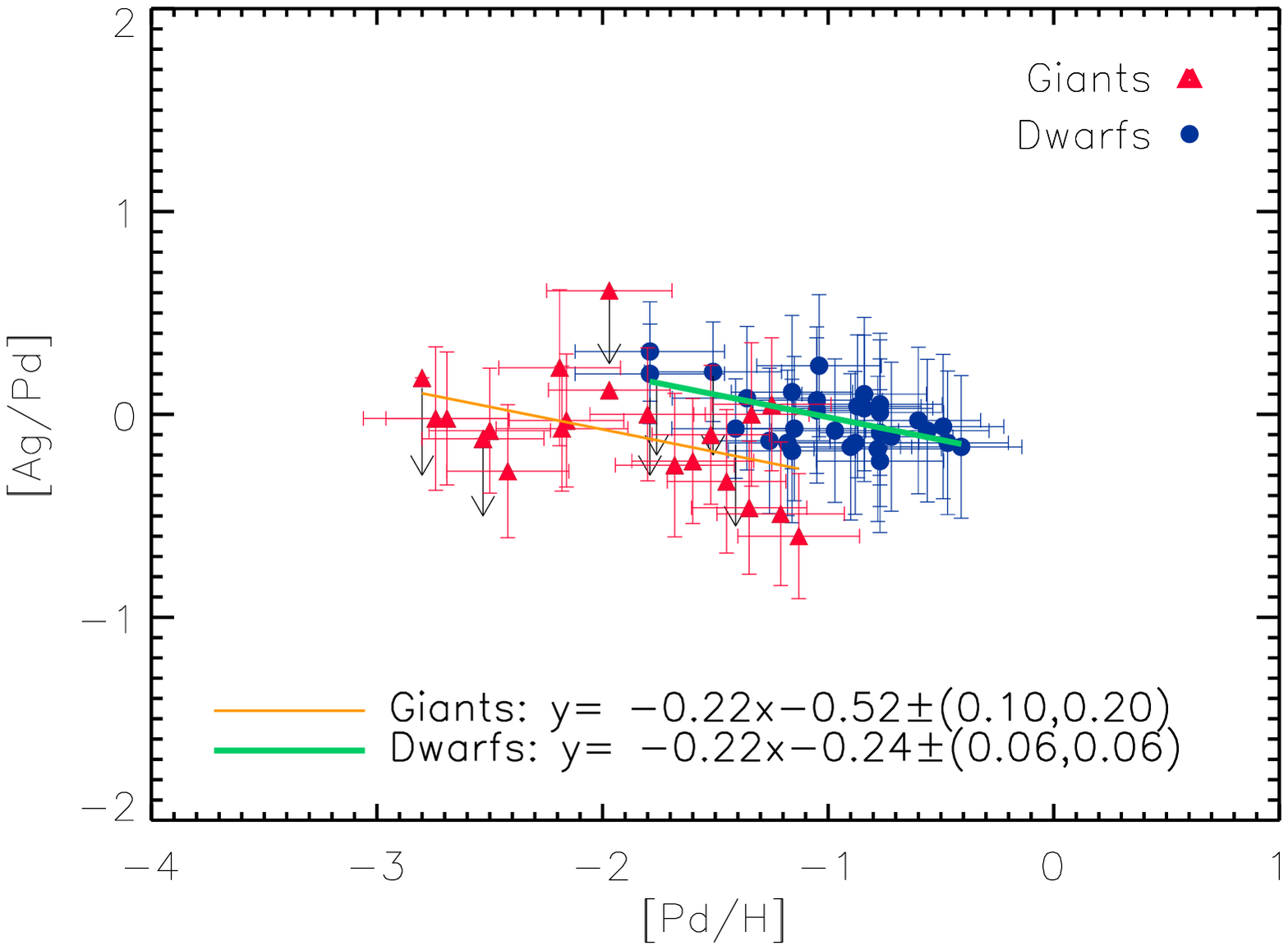}
\caption{An almost flat trend (correlation) is seen in the figure showing [Ag/Pd] as a function of [Pd/H], which is indicative of a similar origin of Ag and Pd.}
\label{AgPdfig}
\includegraphics[width=0.495\textwidth]{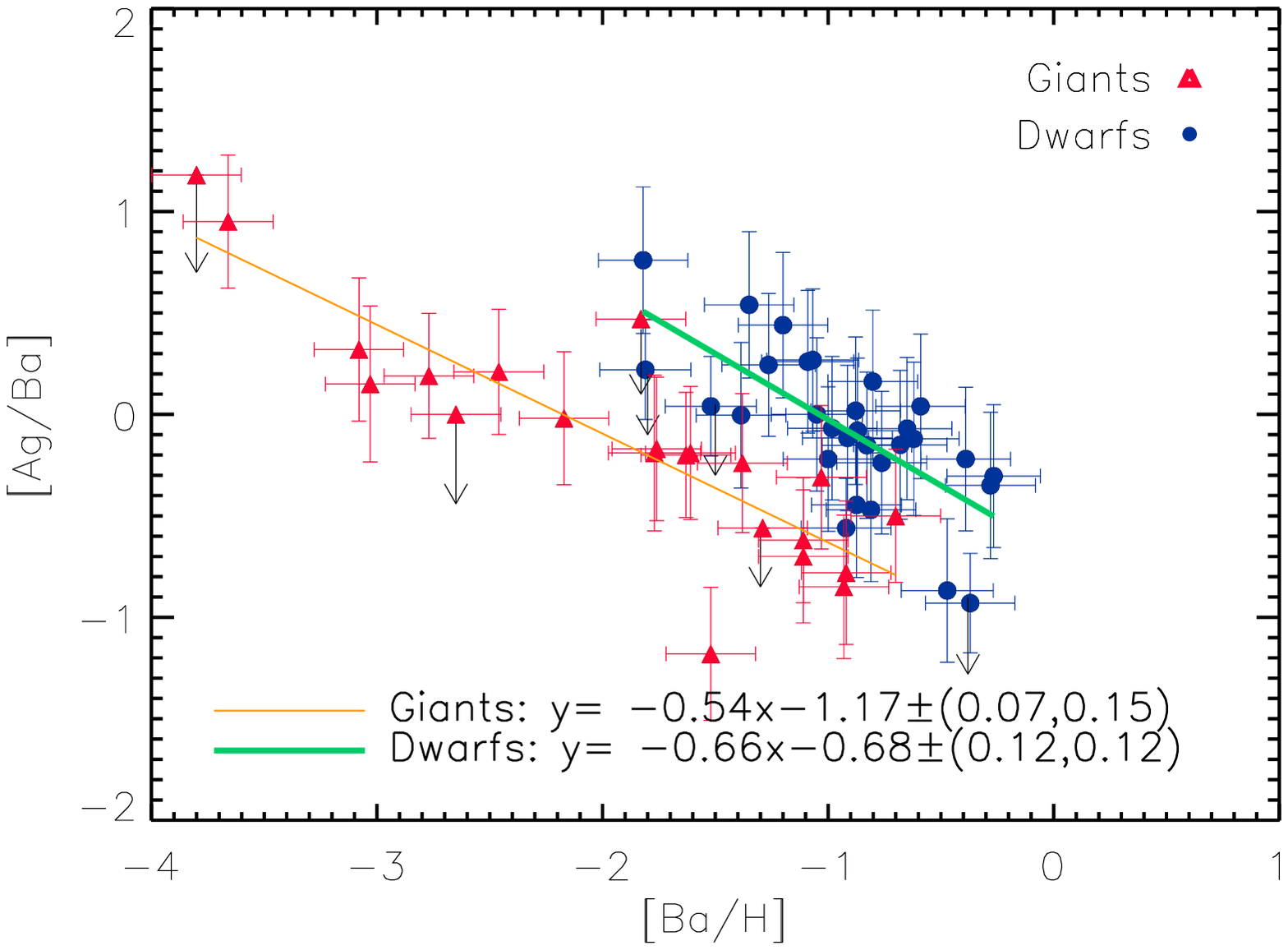}
\includegraphics[width=0.495\textwidth]{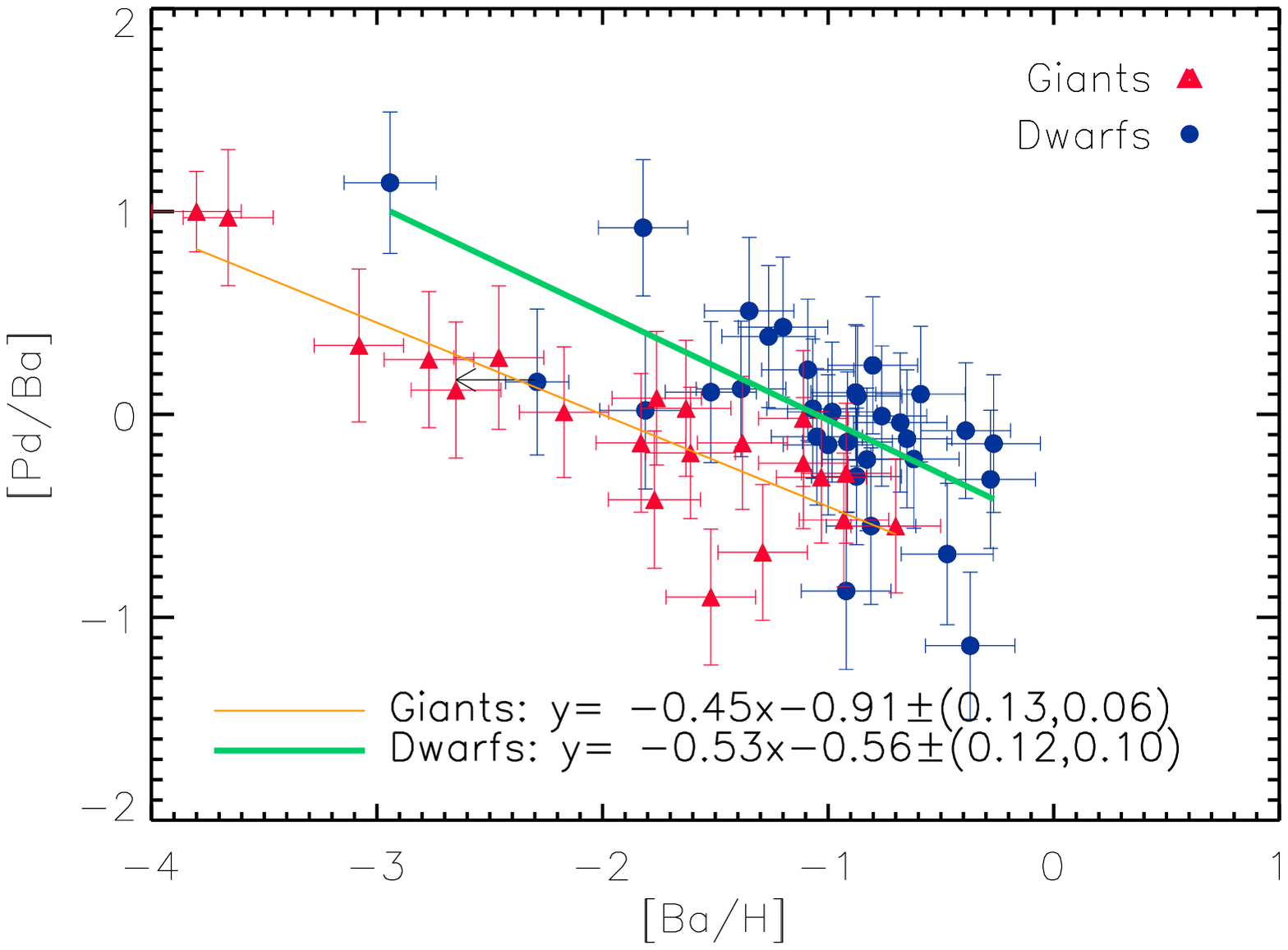}
\caption{A strong anti-correlation is seen in this plot of [Ag/Ba] vs [Ba/H] and [Pd/Ba] vs [Ba/H]. Silver and palladium are therefore not main s-process elements.}
\label{AgBa}
\includegraphics[width=0.495\textwidth]{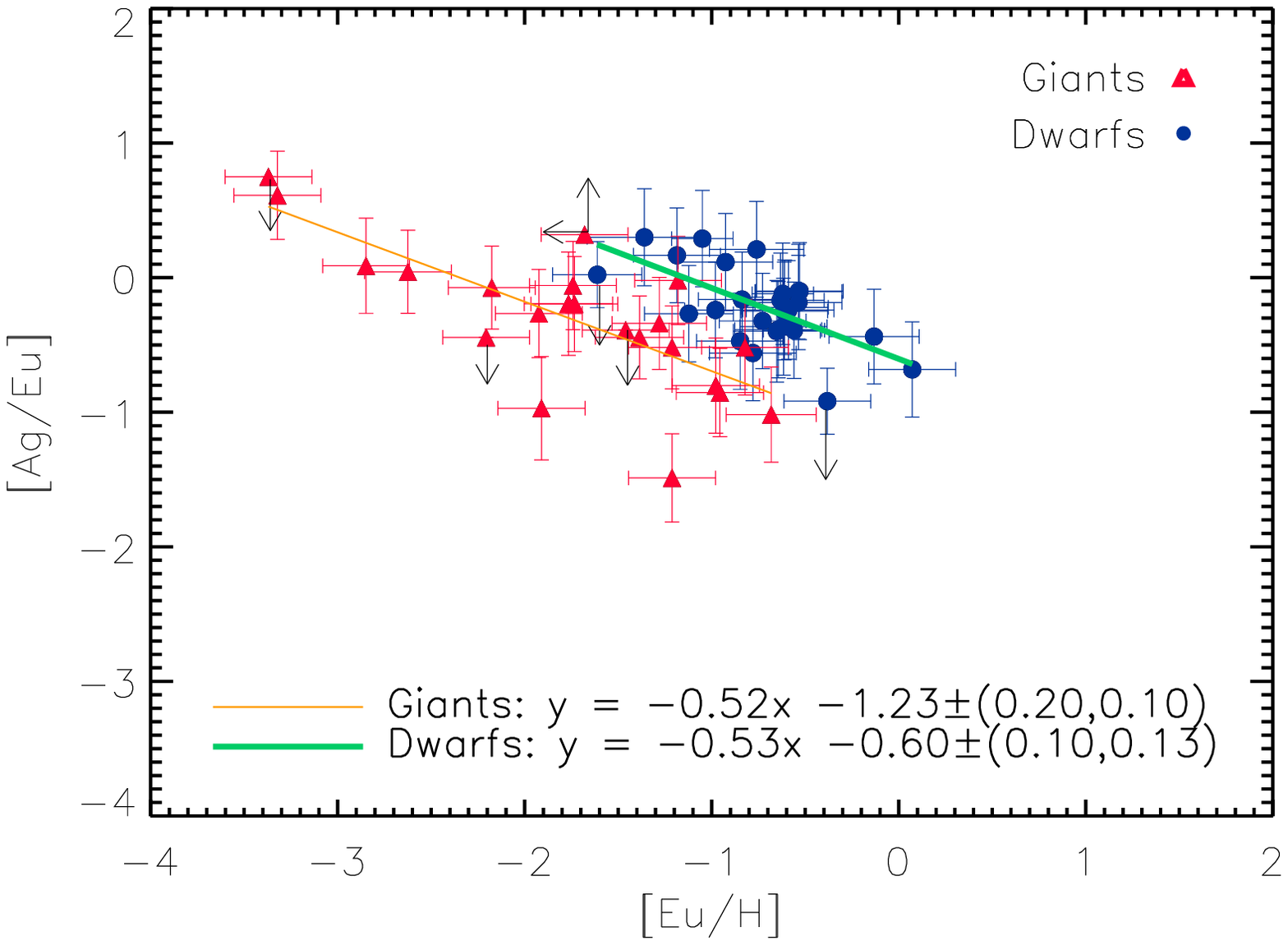}
\includegraphics[width=0.495\textwidth]{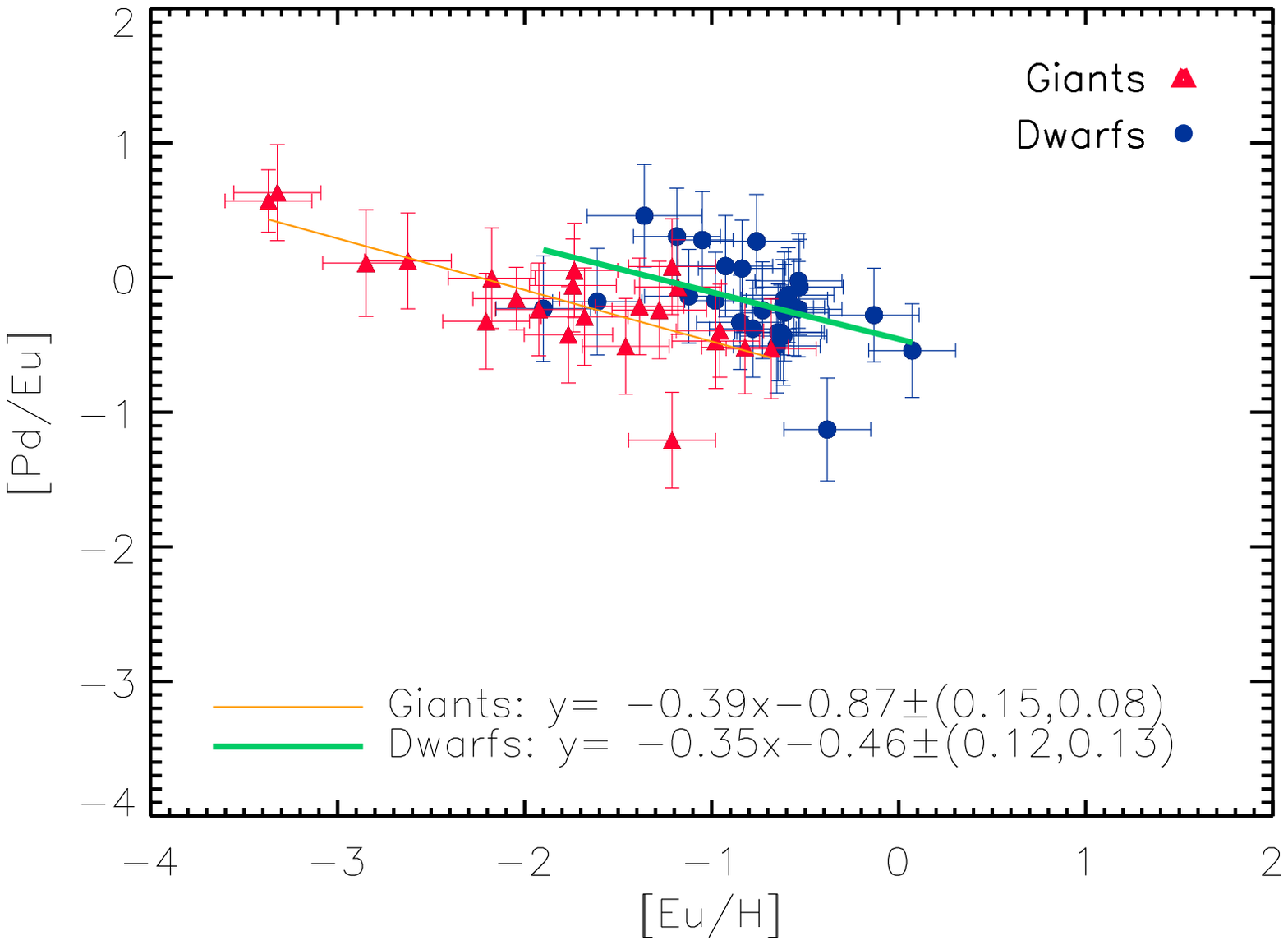}
\caption{To the left: [Ag/Eu] plotted as a function of [Eu/H], showing a clear and strong anti-correlation. To the right: [Pd/Eu] vs [Eu/H]. This means that Ag and Eu are not synthesised by the same process, nor are Pd and Eu. Silver and palladium are therefore not produced by the main r-process.}
\label{AgEu}
\end{center}
\end{figure*}
In general, Fig. \ref{figAgSr}, \ref{AgY}, and \ref{AgZr} have one common feature, i.e. they all clearly show that the elements plotted in each graph anti-correlate. Although these anti-correlations are characterised by slightly different (negative) slopes, all of these plots agree that neither Pd nor Ag are formed by the same mechanism that produced Sr, Y, or Zr (i.e. weak s-process or charged particle freeze-outs).
However, these negative slopes do not merely differ randomly between the elements, but there seems to be a clear decreasing trend (i.e. the slopes become shallower) going from Sr to Y and then to Zr. The slopes derived by fitting the data-points in [Ag, Pd/Zr] are between $-0.37$ and $-0.18 \pm0.07$, which thus indicate that there is only a mild anti-correlation. We interpret this as an indication that Zr may be produced (at least in part) by the same formation process producing Pd and Ag.

When comparing Ag to Pd (see Fig. \ref{AgPdfig}), it becomes difficult to draw a firm conclusion about the exact trend of their abundance ratio [Ag/Pd] as a function of [Pd/H]. Despite the slopes overplotted on the graph being indicative of a very mild anti-correlation, they may be misleading especially since they take into account giants and dwarfs separately. If one were to ignore these slopes and consider the entire sample as a whole, we could argue that we find a flat [Ag/Pd] trend, especially when considering the associated error-bars and excluding upper limits. The latter is also supported by our earlier finding of an almost 1:1 linear slope between [Ag/H] vs [Pd/H] (\citealt{cjhletter}), which strongly indicates a common origin for these two elements.

If we now consider how Ag and Pd compare to Ba (Fig. \ref{AgBa}), which is the most representative tracer of the main s-process, we see that both Ag and Pd strongly anti-correlate with Ba, which excludes the main s-process as one of the possible production channels responsible for the formation of Ag and Pd. At low metallicity ([Fe/H] $< -2.5$ dex), Ba is created by the main r-process, which indicates that Pd and Ag are also not created by the main r-process, although we compare them to Eu to confirm this finding. Finally, Fig. \ref{AgEu} shows that strong anti-correlations of Ag and Pd are found with Eu, which means that the process forming Pd and Ag cannot be the main r-process. We cannot, however, exclude that Ag and Pd are partly produced by the main r-process. 

Therefore, the formation process of Pd and Ag is neither a charged particle freeze-out, a weak, main s-process, nor a main r-process. Both Ag and Pd are seen to form at extremely low metallicity ([Fe/H] $< -3$). These results, combined with the predictions of \citet{montes}, \citet{kratznewrev}, and \citet{Fara}, indicate that their formation process must be of primary and likely r-process nature, but we need to resort to model comparisons in order to characterise this second r-process.

As mentioned at the beginning of this sub-section, one needs to keep in mind two caveats when discussing these abundances: i) we derived all abundances based on 1D LTE model atmospheres and spectral syntheses; ii) we were able to track the evolution of Ag down to the lowest metallicities only with giant stars. We adopted the former approach because NLTE corrections are available for only some of the elements investigated here, namely Sr \citep[e.g.][and Bergemann et al, 2012 submitted]{srnlte,andrSr}, Zr, Ba \citep[e.g.][]{andrievsky09}, and to some extent Eu. However, no NLTE corrections have been calculated for our two key elements Pd and Ag, and only a few for Y and Zr \citep{Zrnlte}. Because we use primarily [A/B] ratios (where A can be either Ag or Pd, and B is one of the other neutron-capture elements), we decided to keep a 1D LTE consistency across all ratios, instead of correcting only some elements. We are, however, fully aware of the importance of NLTE corrections, and that would ideally be a better way to proceed, were NLTE corrections to become available for all elements. As for the latter, dwarfs and giants show in general very similar trends (see Fig. \ref{figAgSr} - \ref{AgEu}), with the dwarfs having higher abundance values than the giants at similar metallicities. However, the overall good agreement between dwarfs and giants suggests that the process creating Ag and Pd is likely to be the same at all metallicites.

\subsection{Formation processes and transitions around Zr}\label{zrtrans}
\begin{figure}[!h]
\begin{center}
\includegraphics[width=0.495\textwidth]{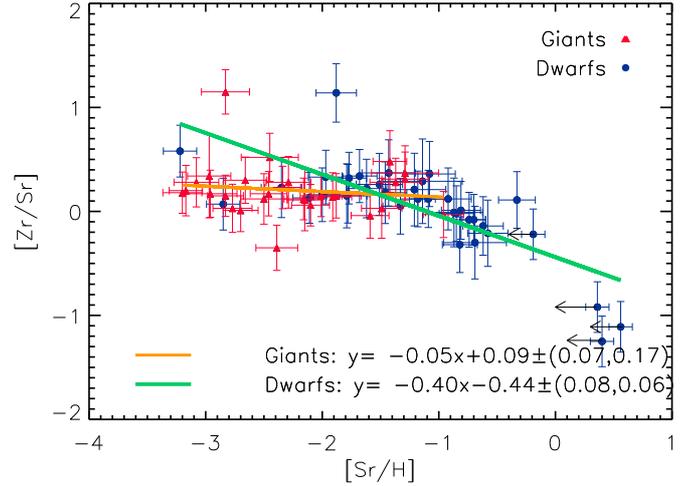}
\caption{Zr and Sr correlate in the metal-poor giants which indicates a similar formation process of these two elements. This is in agreement with the findings of \citet{Fara} and \citet{klk08}. At higher metallicities ([Sr/H] $> -1$) the formations of Sr and Zr differ. The upper limits are due to the before mentioned lacking visual spectra of the three stars (see text).}
\label{Zrweakcor}
\end{center}
\end{figure}

Zirconium and strontium clearly share a common formation process at low metallicities down to and even slightly below [Zr/H] $= -3$ (see the flat correlation for giants in Fig. \ref{Zrweakcor}). A similar trend is found when comparing yttrium to zirconium and yttrium to strontium. However, at higher [Fe/H] and [Sr/H] abundances above $-1$ dex, we find an anti-correlation between Sr and Zr for the dwarfs. At higher metallicities, this can indicate differences in the formation process --- or a difference between the process primarily responsible for the formation of the two elements.

\begin{figure}[h!]
\begin{center}
\includegraphics[width=0.495\textwidth]{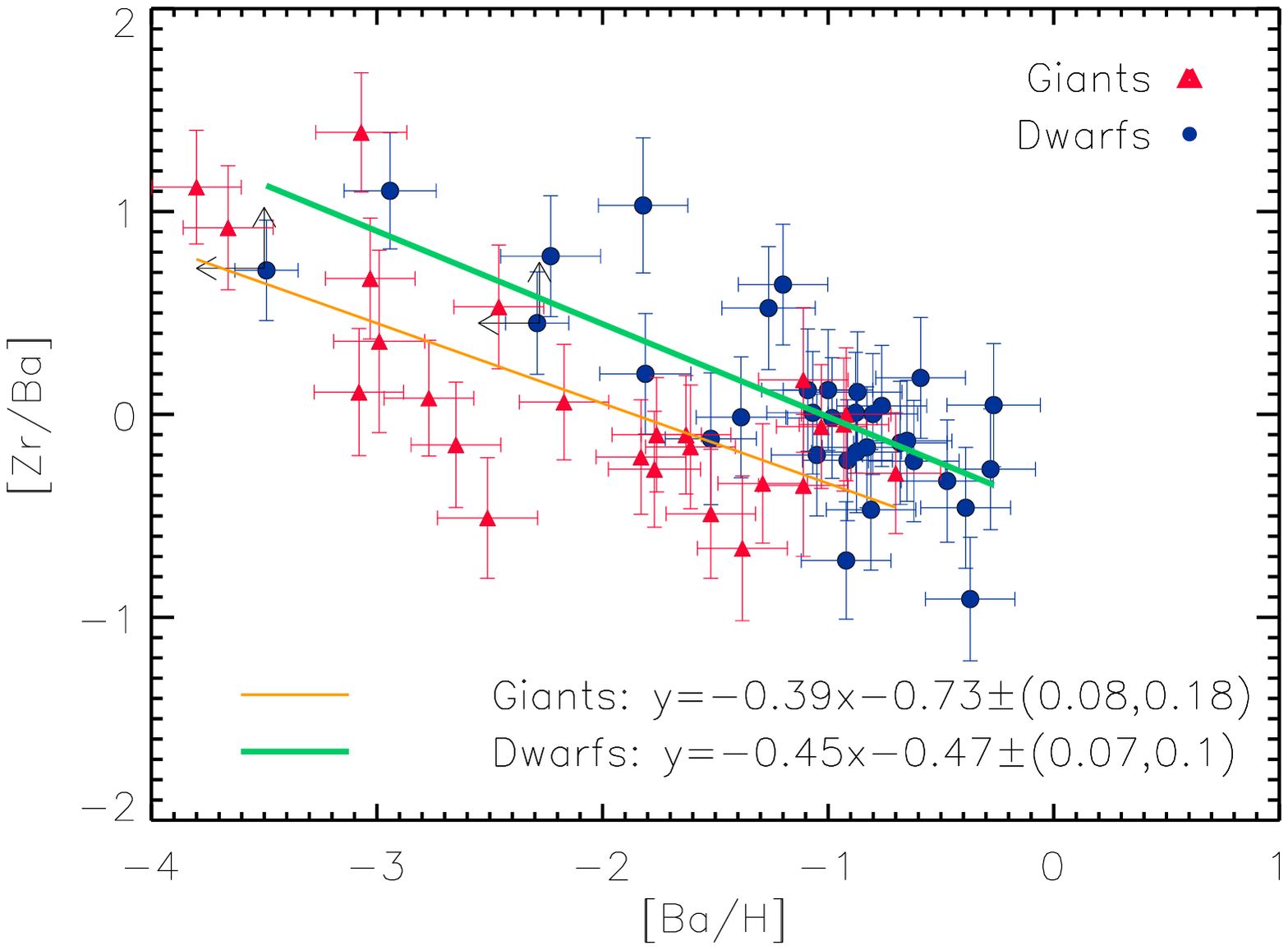}
\includegraphics[width=0.495\textwidth]{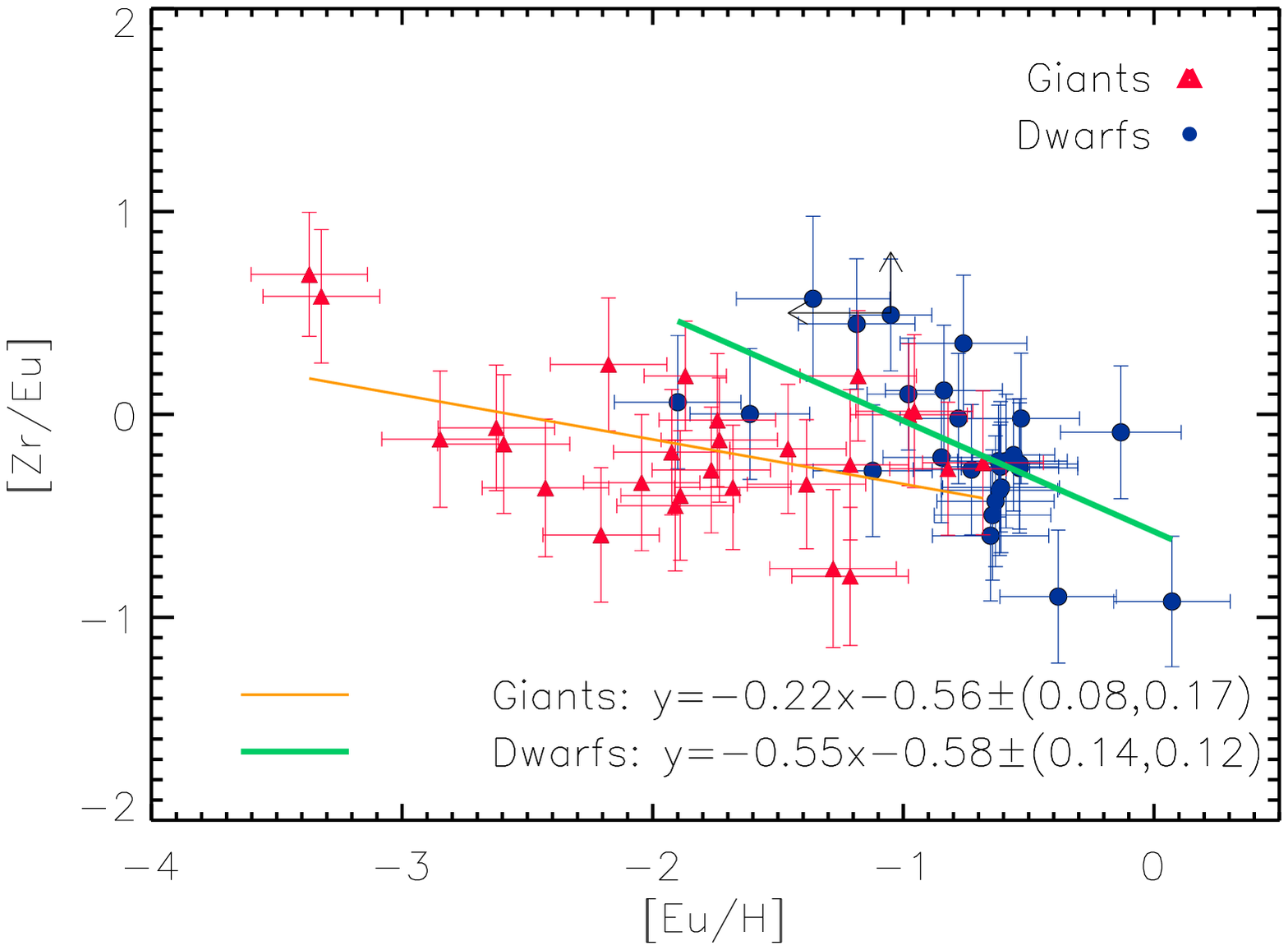}
\caption{
Top: [Zr/Ba] vs [Ba/H] showing anti-correlations. The clumping visible at higher [Ba/H] abundances may be indicative of some common formation (s-) process for Zr and Ba. Bottom: [Zr/Eu] vs [Eu/H], showing a clear, strong anti-correlation over the entire range of [Eu/H] values for the dwarfs. This resembles the behaviour seen for [Ag/Eu] vs [Eu/H]. 
}
\label{Bacor}
\end{center}
\end{figure}

Zirconium and barium seem to have different origins, as shown in Fig. \ref{Bacor} (Zr; e.g. charged particle freeze-out or weak r-process vs Ba; main r-process origin at low metallicities). These findings confirms those of \citealt{Fara} and \citet[][see their Fig. 4]{kratznewrev}, who found a low-entropy charged-particle freeze-out process to be the primary formation process of Sr, Y, and Zr at low metallicity. Here, we find indications of Zr being created in a slightly different way from Sr and Y. Similar trends are also seen for [Sr/Ba] and [Y/Ba] ratios, where the giants show clear anti-correlations. The trends for giants were already reported by e.g. \citet[][]{francois}. For the dwarfs, this trend is less pronounced and they have a greater scatter in the abundances. From the dwarfs' trends, we might conclude that around [Ba/H] = $-2$ the s-process yields from asymptotic giant branch stars are no longer negligible formation sites of Ba, and that the larger scatter is evidence of multiple formation sources. Comparing the giant abundances of Zr to Eu shows that like Pd and Ag, Zr is not produced by the main r-process at higher metallicites (see Fig. \ref{Bacor}), although we note that Zr and Pd follow a weaker anti-correlation with Eu than Ag does. 

In the solar system, Zr appears to have been partly produced by the weak and main s-processes (as well as there being a minor contribution from the weak/second r-process), owing to the correlations (and only mild anti-correlation) of Zr with Sr, Pd, Ag, and Ba. At low metallicities, the s-process contribution to Sr, Y (and Zr) is substituted with a charged particle freeze-out creation. These statements are confirmed in Sect. \ref{models}. This means that Zr may represent a transition in the periodic table around atomic number 40 from the weak s-process/charged particle freeze-out process (depending on metallicity) to the weak r-process. This second r-process could be responsible for the creation of elements in the atomic number range 40 -- 50. However, this process would cease to create elements somewhat below barium. This upper limit is uncertain owing to the lack of elements investigated (observationally in large samples) in the range 48 -- 55. We note that a natural end to the weak r-process from a nuclear physics point of view would be around the element tin because of the bottle neck occurring at N = 82, beyond which many more neutrons are needed to continue the fusion.
 
\subsection{Discussion}\label{discus}
This section highlights our findings and addresses key points mentioned in the previous sections, namely, scatter and inhomogeneities, the presented abundance trends, and differences between dwarfs and giants (possibly NLTE effects).

The consistently large scatter or ISM inhomogeneity seen at metallicities below [Fe/H] $< -$2.5 dex is found in the majority of the abundance trends. Many of the large abundance studies have found similar large star-to-star scatters at these low metallicities \citep[e.g.][]{barklem,preston,francois,bonifaciodw}. A NLTE study of the latter carried out by \citet{andrievsky09}, confirmed that the scatter in Ba was so large even after applying the NLTE corrections to the abundances, that they could not assume that the ISM is homogeneous. However, the very low star-to-star scatter of $\alpha$- and iron-peak element abundances provides a counter argument to this statement \citep{firstV,preston}, since these elements would suggest that the ISM is very well mixed.

Our findings seem to favour an inhomogeneous early ([Fe/H] $< -2.5$) ISM for the reasons that follow. Considering all these (alpha, iron-peak, and neutron-capture) abundances above [Fe/H] = $-2.5$, all star-to-star scatters are much smaller and the ISM seems to be well-mixed. 
This implies that single (or a few) nucleosynthetic events
such as SNe exhibit smaller effects on the stellar abundances at
higher metallicity \citep{ishi}.
However, this is not the case below $-2.5$ dex in metallicity, where we may be witnessing the effects of very different (single?) exploding SNe \citep[this was also suggested by][]{john02}. Owing to the different supernova features their yields will vary: we refer to \citealt{hegerNOr,Wanajo2003,galchem,Izu,Fara}, and \citealt{wan11} who discuss the impact that various parameters such as peak temperature, mass-cut, and entropy have on the SN yields. The $\alpha-$elements are mainly yielded by type II SNe and produced in one process only; they do not show this kind of scatter in their abundance pattern. 
The neutron-capture elements, on the contrary, seem to have several underlying formation processes, even for the same element, which may help explain the variations in the star-to-star scatters.
The exact site of the neutron-capture elements is yet not known, as we have seen in the previous sections, different neutron-capture elements might be created via different channels \citep{john02, Fara}. Hence, the lack of one dominating source could cause a larger scatter compared to that of the $\alpha$-elements. Furthermore, the different supernovae that create the neutron-capture elements could, due to their differing nature, lead to different neutron-capture processes, i.e. a main and a second r-process, which would help us to explain the scatter. Simply put, the inhomogeneity could in part be explained by several sources/sites yielding different amounts of the neutron-capture elements, whereas the alpha-elements are dominated by SNe II which yield relatively similar amounts of these elements. In contrast to the suggested range of one single r-process \citep{klk08,roed2010}, which is needed to explain the different abundance patterns of HD122563 and CS22892--052, we confirm that no other group of elements be it $\alpha$, odd-Z, or Fe-peak show this kind of scatter in abundances when these originate from only one process. Furthermore, on the basis of our findings, we see that two r-processes (or primary processes) are likely needed to fully explain the correlations and the scatter.   

The differences between these processes are clearly evident in Fig. \ref{AgEu}, where the strong anti-correlation between Ag and Eu, as well as that between Pd and Eu, can be seen. Europium is created by the main r-process, a process that requires very high neutron number densities to produce Eu \citep[typically around $10^{26-28} \rm{cm}^{-3}$,][]{kratz}, whereas the lighter isotopes of e.g. Pd can be created in environments with neutron number densities that are lower by several orders of magnitude. It is impossible to create Eu in environments with such low neutron densities \citep{kratz,Fara,wan11}. This suggests that the properties of the formation sites for the heavy and the light r-process elements differ, or that the processes themselves are different. We note that Fig. \ref{AgPdfig} indicates that the process creates both Ag and Pd at almost the same rate \citep[see also][]{cjhletter}. 
The second r-process seems to operate effectively at all metallicities down to [Fe/H] = $-3.3$. This process, or its production site, must be less efficient than the main r-process. For [Eu/H] $> -3$, the [Ag/Eu] is below zero and rapidly decreases with increasing Eu (see Fig. \ref{AgEu}). However, at the lowest metallicities and europium abundances ([Eu/H] $< -3$) the amount of Ag is at the same level or slightly higher than the Eu abundance, as can be seen from the [Ag/Eu] abundance being larger than zero. This could indicate that the second r-process is more efficient at low [Eu/H]. 
We cannot rule out that Ag and Pd both receive small contributions from the main r-process, since this process is generally ([Eu/H] $>-3$) predominant in the ISM gas.

Figures \ref{figAgSr} and \ref{AgBa} show anti-correlations of Ag and Pd compared to Sr and Ba. At high metallicities, [Fe/H] $\sim > -1$, the s-process is far more prevalent in the ISM than the second (weak) r-process (e.g. [Ag/Ba] $<$ 0). However, the same figures show abundance ratios of around 0 in the range [Fe/H] $= -2.5$ to $-1.0$. This could indicate that the s-process and the second r-process have some features or sites in common (e.g. super AGB stars), but this has yet to be confirmed.  

Another important outcome of this study is the discovery of Zr as a 'transition' element. Figures \ref{figAgSr} to \ref{AgZr} show a gradual increase in the slopes of Ag and Pd compared to Sr, Y, and Zr; i.e. an indication of the growing similarities in their formation processes. Within the uncertainties in the slopes, Ag and Pd almost correlate/show a very weak anti-correlation with Zr. When Ag and Pd are compared to each other (Fig. \ref{AgPdfig}), an almost 1:1 correlation is seen. This could be the first observational evidence that at higher metallicities ([Fe/H] $> -2.5$), Sr and Y are weak s-process products, as claimed by \citet{heil} and \citet{pigna} (at lower metallicities Sr and Y might be created by charged particle freeze-outs; \citealt{farou} and \citealt{wan11}). Zirconium should mainly be an s-process element (in the solar inventory), which also receives considerable contributions from a type of weak r-process. This r-process is responsible for the main production of Pd and Ag. The transition from charged-particle freeze-out or weak s- (Sr, Y) to 'weak' r-process (Pd, Ag) takes place around Zr (Z = 40), hence the name transition element. 
Moreover, the figures showing [Ag/Ba] and [Ag/Eu] yield anti-correlations (both strong, see Fig. \ref{AgBa} - \ref{AgEu}), suggesting that the formation processes differ. The strong anti-correlation with Ba shows that this process is not a main s-process and the anti-correlation with Eu demonstrates the differences between the main and the second r-process.

Finally, we turn the discussion to the differences between dwarfs and giants. Unfortunately, a full NLTE analysis is not yet available, owing to incomplete and complicated model atoms of these heavy elements. However, on the basis of previous studies of some of the heavy elements such as Sr and Ba \citep{srnlte,andrSr,andrievsky09}, the NLTE corrections can be relatively large for low-gravity metal-poor stars. The Sr II abundance may need a correction of the order of $\sim<0.2>$\,dex \citep{andrSr}, while the Zr II abundance corrections are lower and generally between 0.08 dex and 0.17 dex according to \citet{Zrnlte}. These corrections are very dependent on the spectral line, the stellar parameters, and therefore vary from star to star. Additionally, it is insufficient to correct only one of the elemental abundances in the abundance ratios we have discussed so far. A detailed NLTE study would be needed, but is beyond the scope of this paper. Any estimate of the behaviour of the NLTE corrections of e.g. silver would be very speculative at this point, although we note from Fig. \ref{AgBa} that the [Ag/Ba] ratio of the giants would need an NLTE correction of $\sim$ +0.5 dex, as estimated from the offset in the figure between the giants and the dwarfs. We note that a fraction of this estimated value would be due to the NLTE correction of Ba.

\section{A comparison to supernova yields}\label{models}
To gain information on the formation site and process of our sample's abundance patterns, we compare these to two different models. The first model ({\it model Ia+b}) we focus on is that of high-entropy winds (HEW) \citep{Fara,farou} triggered by type II SN explosions, whereas the second model ({\it model II}) is tied to low-mass electron-capture SNe \citep[arising from collapsing O-Ne-Mg cores,][]{wan11}. In the last case, the neutrino interactions are taken into account.

\begin{figure}[h!]
\begin{center}
\includegraphics[width=0.495\textwidth]{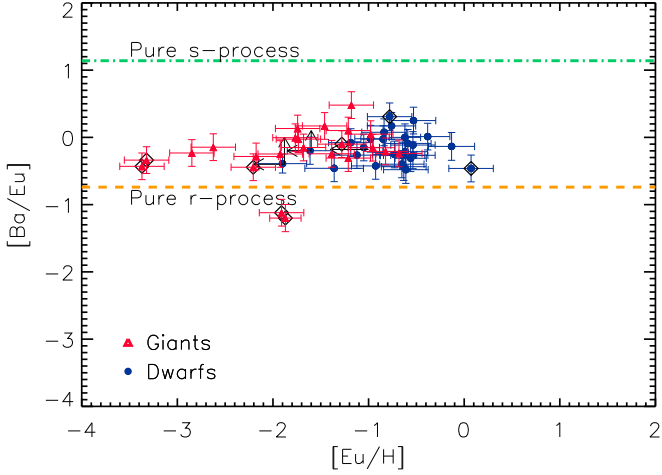}
\caption{Barium vs Europium. The purity of the process is estimated based on the numbers from \citet{arland}. The black diamonds indicate the [Ba/Eu] abundance of the eight stars shown in Fig. \ref{hewdw} and \ref{wanafig}.}
\label{BaEupure}
\end{center}
\end{figure}

To ensure that the abundance--to--model prediction comparison is as informative and complete as possible, we selected eight stars distributed in the following way.
To probe abundance patterns that include Ag, two dwarf and six giant stars were singled out, where the giant sub-sample includes stars with special patterns such as r-rich stars. Furthermore, the selection was carried out, so that the stars cover a wide range of stellar parameters, especially metallicity. By considering stars with a wide range of [Ba/Eu] ratios, we include stars with mixed as well as pure r-process abundance patterns (see the black diamonds in Fig. \ref{BaEupure}). 
The stars selected are: CD--453283 and HD106038 (dwarfs with mixed r- and s-process patterns), BD+42621 and CS~22890-024 (giants; pure r-process tracers), HD122563 and HD88609 (r-poor giants), and CS~31082-001 and HD115444 (r-rich giants). The dwarf star CD--453283 has a very high europium abundance ([Eu/Fe] =  0.78), which is the highest Eu abundance measured for the dwarf stars in our sample. Over all, this star is overabundant in elements heavier than Zr.

\citet{Fara,farou} explored a large parameter space especially in entropy, where the values were varied between 20 and 275 ${k_B/\rm baryon}$, and the electron fractions, $Y_e$, cover the range from 0.4 to 0.49. The wind velocity adopted is 7500 km/s. The output is neutron-to-seed ratios and corresponding yields/summed abundances. For further information we refer to \citet{Fara,farou}. Owing to the lack of well-constrained (3D) supernova explosion parameter output, it remains unknown whether a high entropy or a low $Y_e$ is more likely to happen in an actual explosion. Therefore, we carry out two different comparisons when contrasting the HEW model predictions. In model Ia) $Y_e$ is fixed and chosen so that the value reproduces the observationally derived abundances, while the entropy, $S$, is varied. In model Ib) the entropy is fixed, while $Y_e$ is varied. The latter case enables a more direct comparison to the yield calculations of \citet{wan11}.

\subsection*{Model Ia}
\begin{figure}[h!] 
\begin{center}
\includegraphics[width=0.5\textwidth]{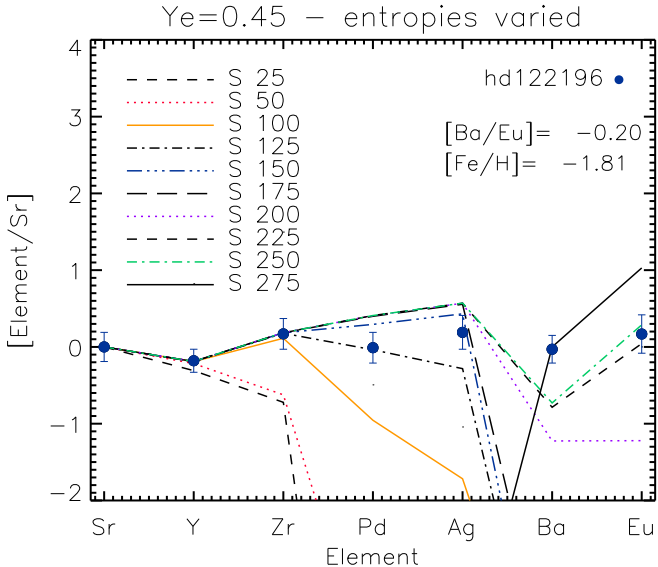}
\caption{A comparison of the abundances of light neutron-capture elements as derived in HD 122196 to the yields [X/Sr] produced by HEW models. The latter are computed with $Y_e$=0.45 and varying entropies $S$ (all values listed are given in $k_{B}/\mathrm{baryon}$ units. The metallicity ([Fe/H]), and [Ba/Eu] ratio for the star are shown as well. }
\label{entropies}
\end{center}
\end{figure}

The free parameters of the HEW models are entropy, $S \sim T^3/\rho$, $Y_e$, and V$_{exp}$. All parameters are correlated and define the free neutrons per heavy seed nucleus ($Y_n/Y_{\mathrm{seed}}$). Neutrino interactions have only been taken into consideration in terms of their estimated impact on the final value of $Y_e$. 
The model predictions with a fixed $Y_e$ of 0.45 are in good agreement with the derived abundances, for both the intermediate (125 $k_B/{\rm baryon}$) and high (200 $\rm{k_B/baryon}$) entropy (see Fig. \ref{entropies}).
Different values of $Y_e$ were tested in addition to 0.45. We note that for $Y_e$ = 0.48, the abundances of elements heavier than Zr are underestimated, whereas $Y_e$ = 0.4 predicts too high abundances for these elements. The estimates of 0.42 closely reproduce the observed abundances, but only the intermediate values of entropies agree well with observations --- not the high entropies. 

Figure \ref{entropies} helps us to constrain the entropy value and/or intervals that provide enough neutron-captures to activate an r-process.
Our empirical comparison to the abundances derived for HD 106038 confirms the findings of \citet{Fara}, who found two entropy intervals 110 $< S <$ 150 and $150 < S <$ 287 to provide enough neutrons for a weak and a main r-process, respectively. Figure \ref{entropies} shows indeed that the entropies needed to create Pd and Ag are likely between 100 and 150 $\rm{k_B/baryon}$. At very high entropies ($S \sim$ 275), no lighter elements (Sr -- Ag) are efficiently produced, since the fusion continues far past these elements owing to the high neutron flux. 

In Fig. \ref{hewdw}, we extend the comparison between HEW model predictions and derived stellar abundances for eight stars. For simplicity, we perform a comparison for only four entropies $\geq 125 k_B/{\rm baryon}$.
Additionally, we provide in all graphs the [Fe/H] and the [Ba/Eu] ratios we derived for each star. The [Ba/Eu] ratio is especially important, because it indicates the purity of the r-process (see Fig. \ref{BaEupure}). According to \citet{arland}, Ba is a 81\% s-process product, while Eu is a 94\% r-process product (both in the Sun), hence, the smaller the ratio, the stronger the r-process influence is. The r-process is accepted as being pure if [Ba/Eu] $< -$0.74 dex \citet{arland}, which agrees with the value (-0.738) from \citealt{arndt}. However, stars such as HD122563 \citep{honda} provide observationally derived upper limits to this range ($\sim-$0.2 dex), while a pure s-process would be found above 1.14 dex. Furthermore, the metallicity is also an important indicator of the predominant formation process, and is therefore included in the figures. Below [Fe/H] = $-$2.5 dex, we generally expect to see r-process yields. In Fig. \ref{hewdw} - \ref{wanafig} we have scaled all our derived abundances to Sr, since we detect this element in most stars and the element is produced/predicted at most of the entropies and electron fractions considered here\footnote{Unfortunately, Fe is not predicted in these models.}. We note that plotting [element/Sr] ensures that [Sr/Sr] corresponds to zero for all lines.

Within the error bars, the observationally derived abundances agree well with the model predictions calculated with $S = 125$ and $S = 175 k_B/{\rm baryon}$, although, in a few cases a model with $S = 150$ would have provided closer agreement (see Fig. \ref{hewdw}).
\begin{figure*}[!ht] 
\begin{center}
\includegraphics[width=0.6\textwidth]{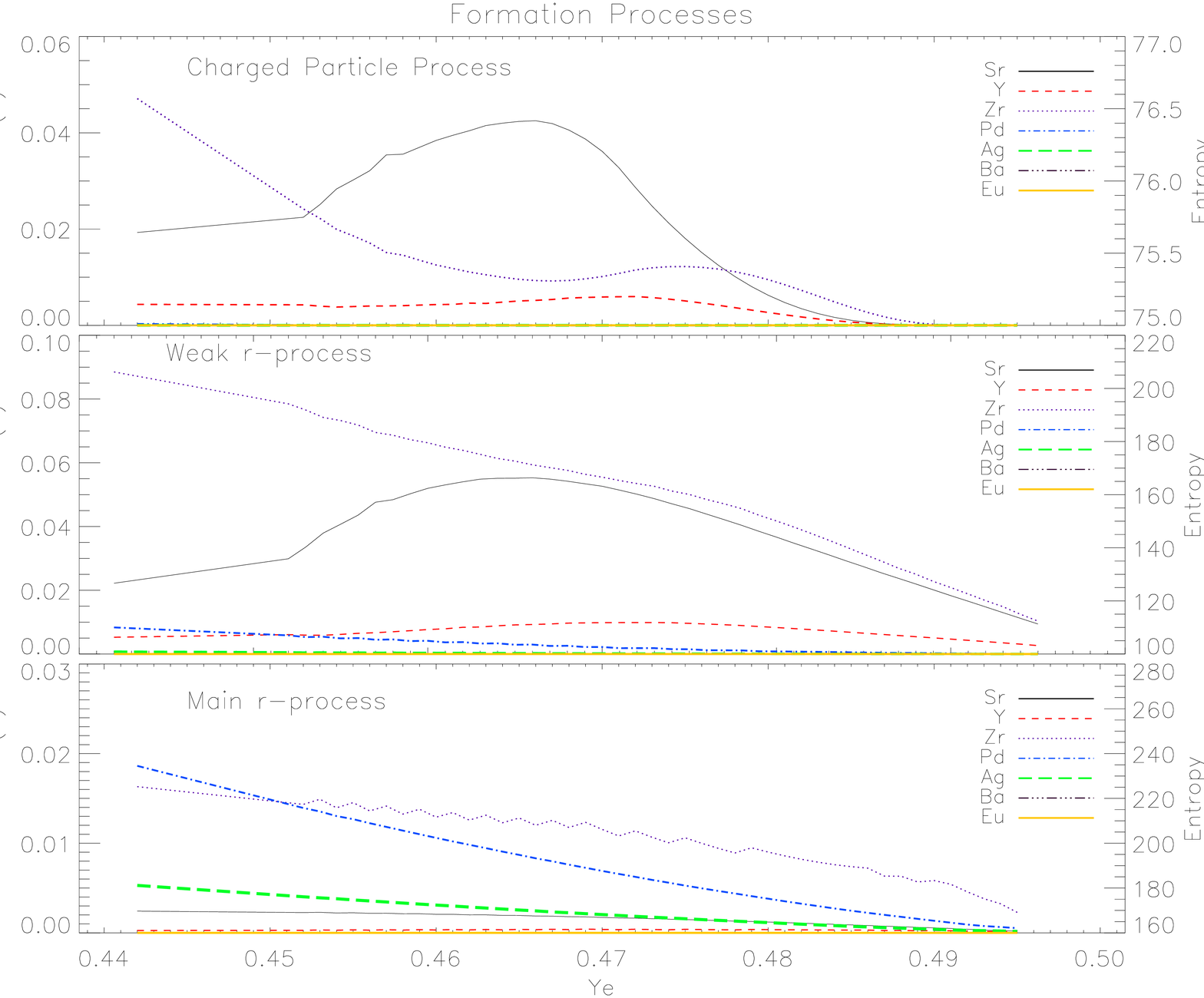}
\caption{HEW abundance, Y(Z), predictions as a function of $Y_e$ for the three different processes. These predicted abundances are the accumulated abundances summed over entropy for a given nucleus (see \citealt{farou} for further description). Note that the y-axis on the right-side of all three panels indicate the entropy intervals for each of the three processes. Every element is identified by a different colour -- see legend.}
\label{varye}
\end{center}
\end{figure*}
The neutron-to-seed ratio related to these models are in the range $Y_n/Y_{\mathrm{seed}}$ = 4 -- $24 k_B/{\rm baryon}$. From the same figure, it is furthermore evident that the heavy elements (Ba, Eu) need much higher entropies to be produced. In general, we find that the entropy interval facilitating the occurrence of the main r-process is 200 -- 275 $k_B/{\rm baryon}$, which is in good agreement with \citet{Fara}. However, we find a slight increase in the difference between the weak (125 $< S <$ 175) and the main (200 $< S <$ 275) r-process.

Additionally, these two different processes must be r-processes since they are observed in very metal-poor stars and show patterns similar to those in the pure r-process stars. 

\subsection*{Model Ib}
If we now vary the electron fractions, $Y_e$, in the HEW model predictions, allowing these to run from S = 2 $k_B/{\rm baryon}$ to a differing final entropy, we see as shown in Fig. \ref{varye} that the charged particle freeze-out, both the weak and main r-processes relate to different entropy ranges in the following way: $50 < S < 100$ corresponds to a charged particle freeze-out for a $Y_e$ of 0.45, or generally speaking, this process takes place when the neutron-to-seed ratio is less than one. For the representation of this process, we adopted the mid-range value S = 75 $k_B/{\rm baryon}$ --- a value that always falls below a neutron-to-seed ratio of one. The weak r-process exists at $125 < S < 175$ (for $Y_e$ = 0.45) or in general in the neutron-to-seed ratio from 1 to 15 (where 15 is reached at $S = 155$), the predictions shown in Fig. \ref{varye} are made with a neutron-to-seed ratio of 5. Finally, the main r-process operates when the neutron-to-seed ratio is above 16, and here we have shown a ratio of 30 as the representation of the main r-process (for a $Y_e = 0.45$, this corresponds to an entropy of $\sim$ 200 --- similar to what we found in the previous section). The yields in percentage for two different electron fractions can be found in Table \ref{HEWpercent}.

\begin{table}[!h]
\caption{Percentage of each element created, according to the HEW predictions by three different processes.  \label{HEWpercent}}
\begin{center}
\begin{tabular}{cccc}\hline \hline
Element	& Ch. part.	&  weak r  & main r \\
   &  Y$_n$/Y$_{\mathrm{seed}} < 1$ & $1 < $Y$_n$/Y$_{\mathrm{seed}} < 15$ &  Y$_n$/Y$_{\mathrm{seed}} > 15$ \\
\hline	
$Y_e$ = 0.442 &&& \\				
Sr& 97.6& 2.2& 0.1  \\
Y& 96.7& 3.3& 0.3  \\
Zr& 81.6& 18.2& 0.2 \\
Pd& 4.6& 85.3& 10.2 \\
Ag& 0.7& 82.5& 16.8 \\
Ba& 0& 0& 100 \\
Eu& 0& 0& 100\\
$Y_e$ = 0.493&&&\\
Sr&97.6& 2.3& 0.1 \\
Y&94.4& 5.5& 0.1 \\
Zr&72.2& 27.6& 0.2\\
Pd&0.4& 72.1& 27.5\\
Ag&0& 63.4& 36.6\\
Ba&0& 1.7& 98.3\\ 
Eu&0& 0& 100\\
\hline\hline
\end{tabular}
\tablefoot{The listed processes are: charged particle freeze-out process (Ch. part.), weak r-process (weak r), and main r-process (main r). These fractional yields (abundances) are functions of the electron fractions, $Y_e$.}
\end{center}
\end{table}
In the HEW predictions with $Y_e = 0.45$ and low entropy ($S < 50$), mainly iron group elements are produced owing to a very low neutron-capture efficiency at these low entropies. Therefore, we disregard this part of the entropy range to ensure that we produce and consider only heavy elements. Furthermore, not all material will necessarily be ejected in the explosion, and some fall back is to be expected. 

In the uppermost panel of Fig. \ref{varye} (the charged particle plot), we see that Sr peaks at a $Y_e$ of 0.47, i.e. an environment that is not very neutron-rich, whereas the Zr yield rapidly increases in a more neutron-rich environment and receives contributions from both the charged particle process, the weak r-process, and a smaller contribution from the main r-process (note the different y-axes in Fig. \ref{varye}). This was also seen in Sect. \ref{abun}. However, the contribution from the main r-process was too small to be detected in the abundances. Palladium is according to the HEW predictions created by both the weak and the main r-process, but as for Zr the contribution from the main r-process is difficult to see in the observationally derived abundances (from which we found weak anti-correlations between Pd and Eu and Zr and Eu). Silver needs considerably more neutron-enriched environments 
\begin{figure*}[!ht] 
\vspace{-0.0cm}
\begin{center}
\includegraphics[width=0.4\textwidth]{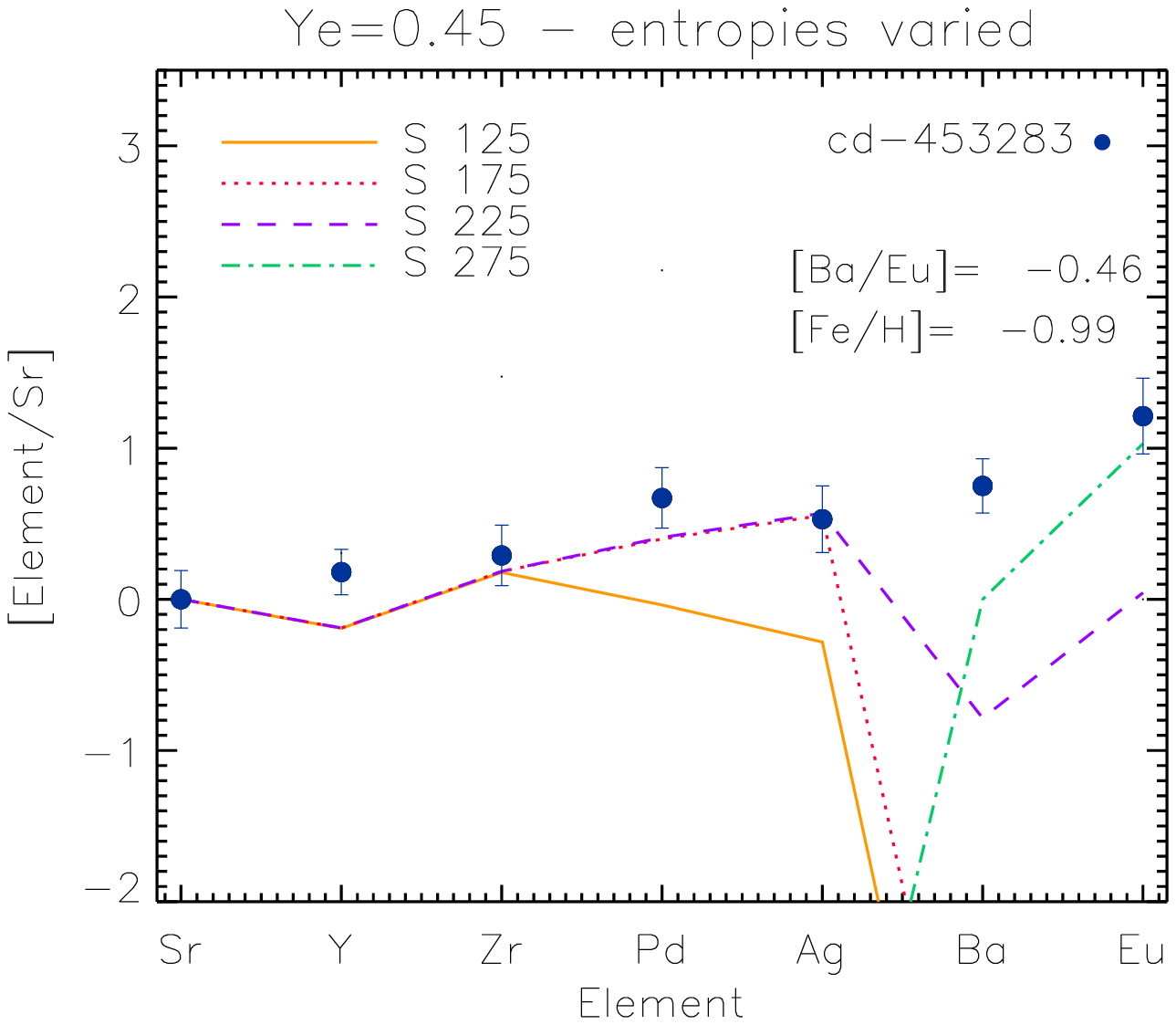}
\includegraphics[width=0.4\textwidth]{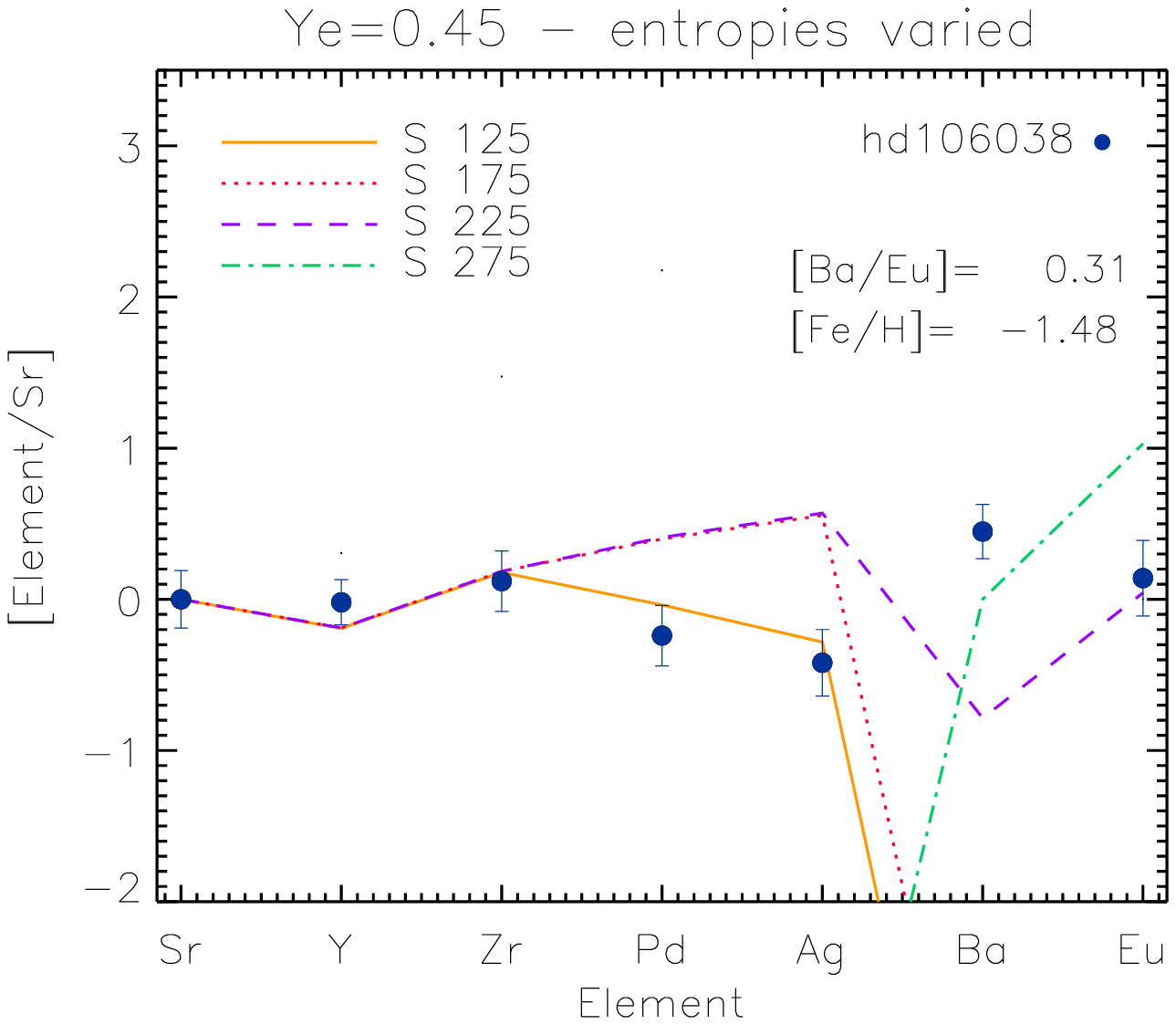}
\includegraphics[width=0.4\textwidth]{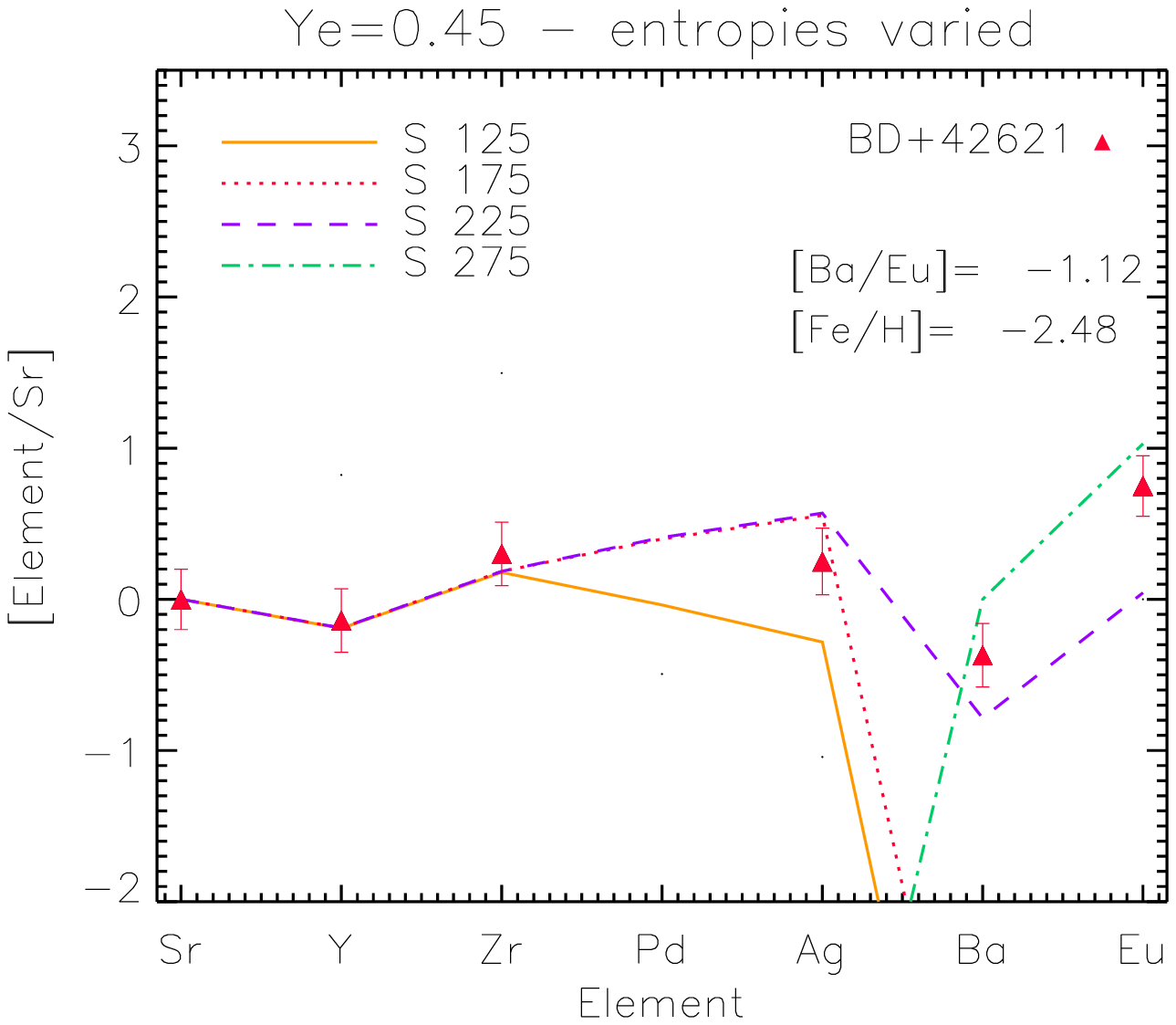}
\includegraphics[width=0.4\textwidth]{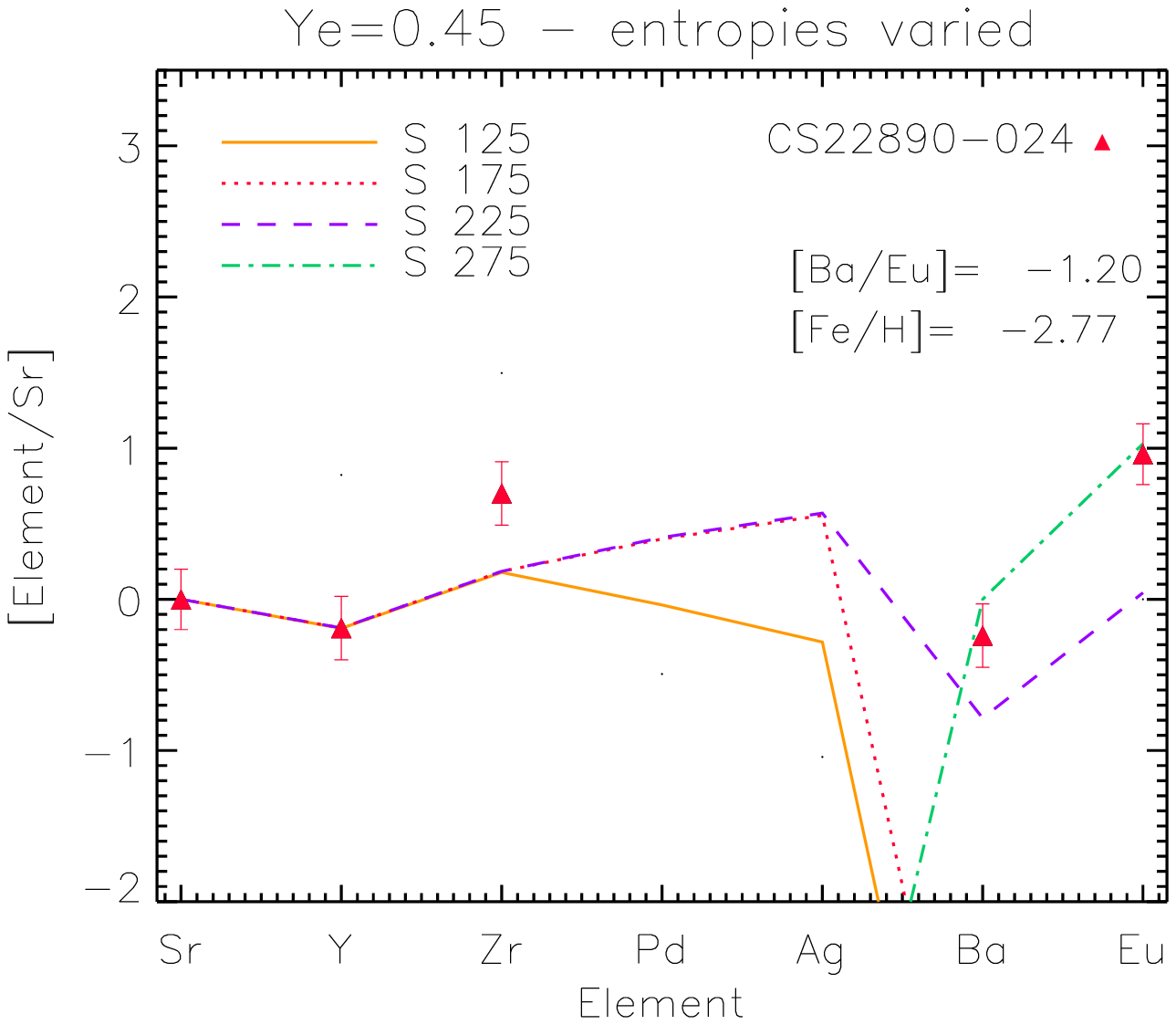}
\includegraphics[width=0.4\textwidth]{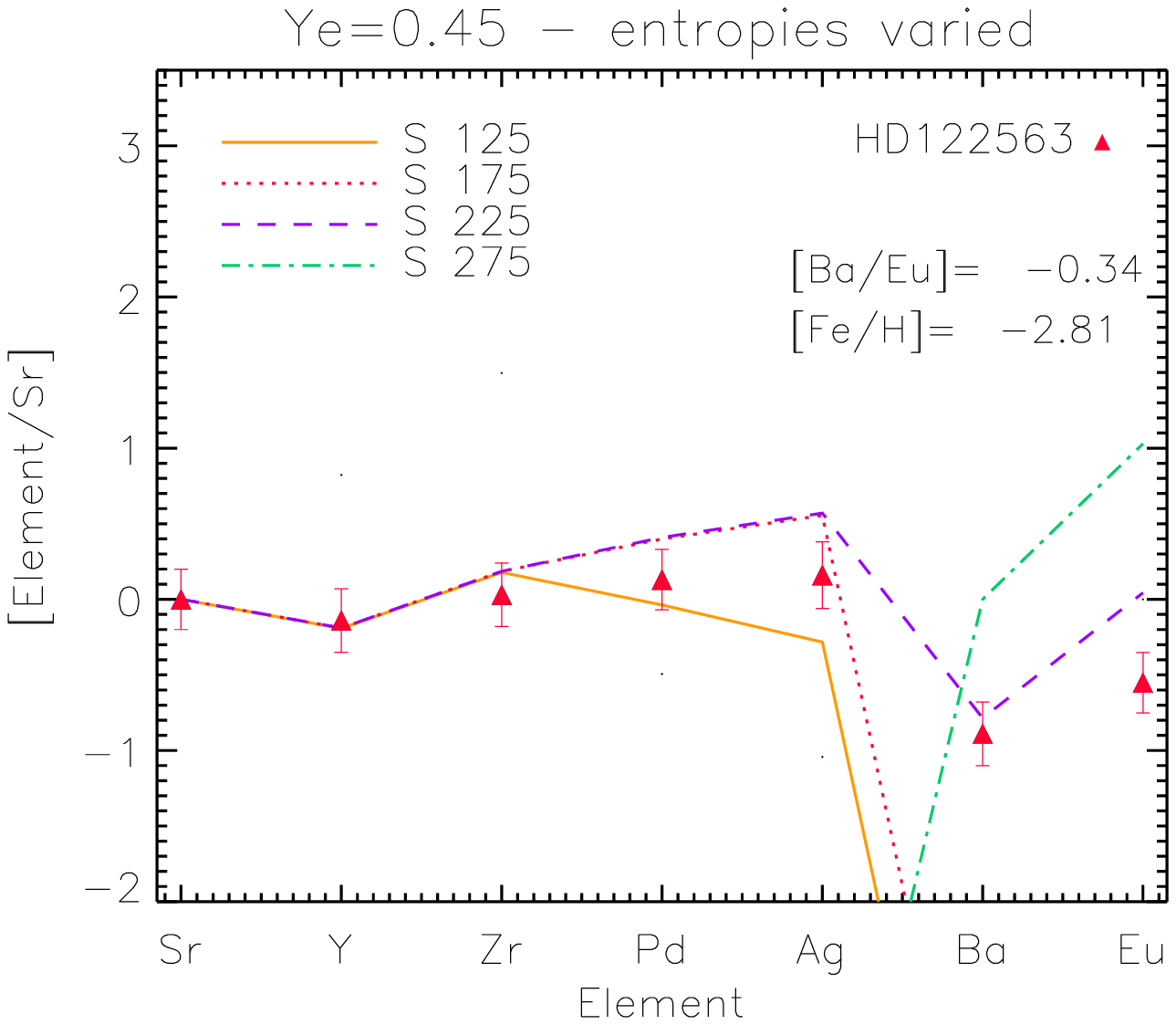}
\includegraphics[width=0.4\textwidth]{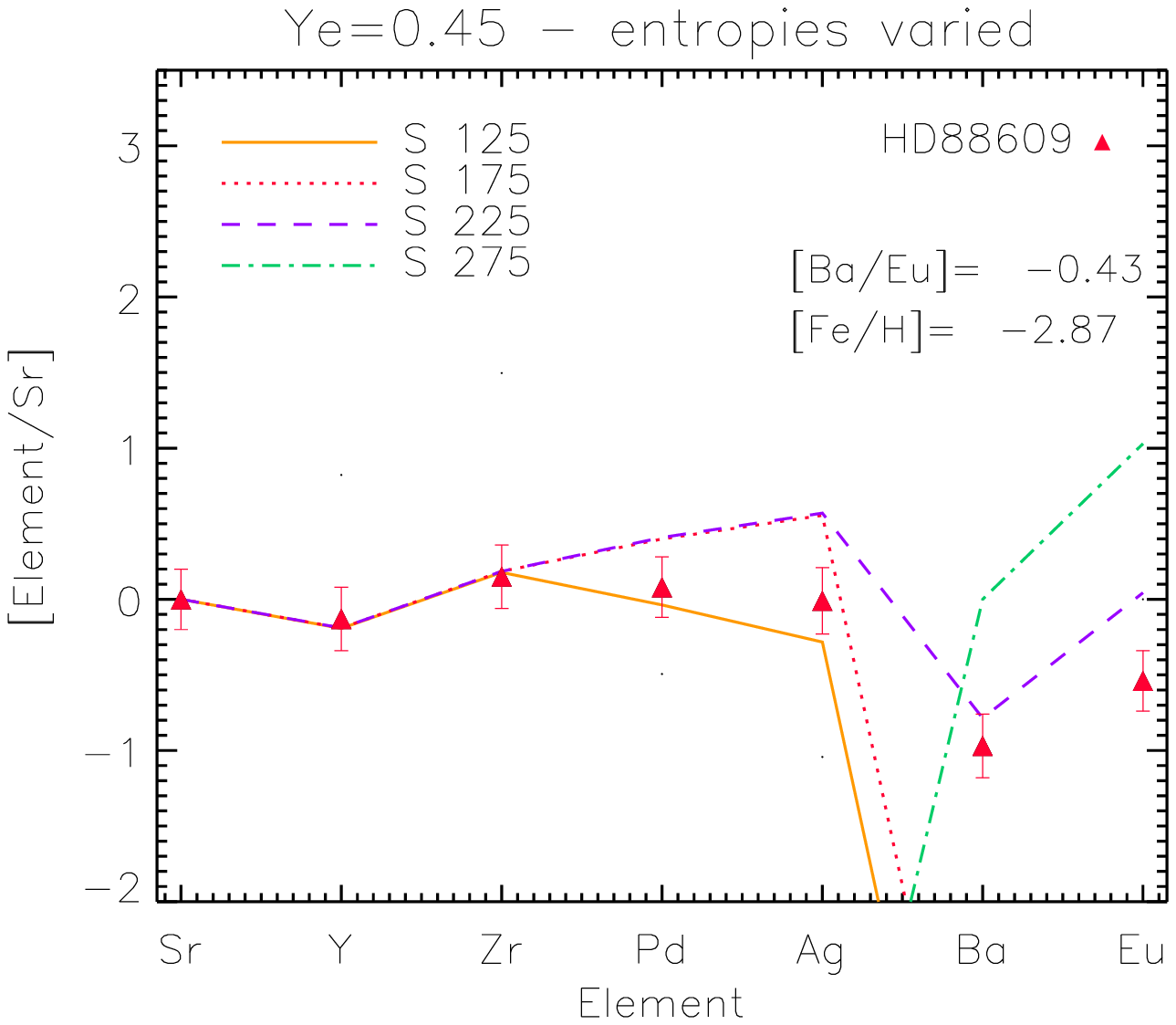}
\includegraphics[width=0.4\textwidth]{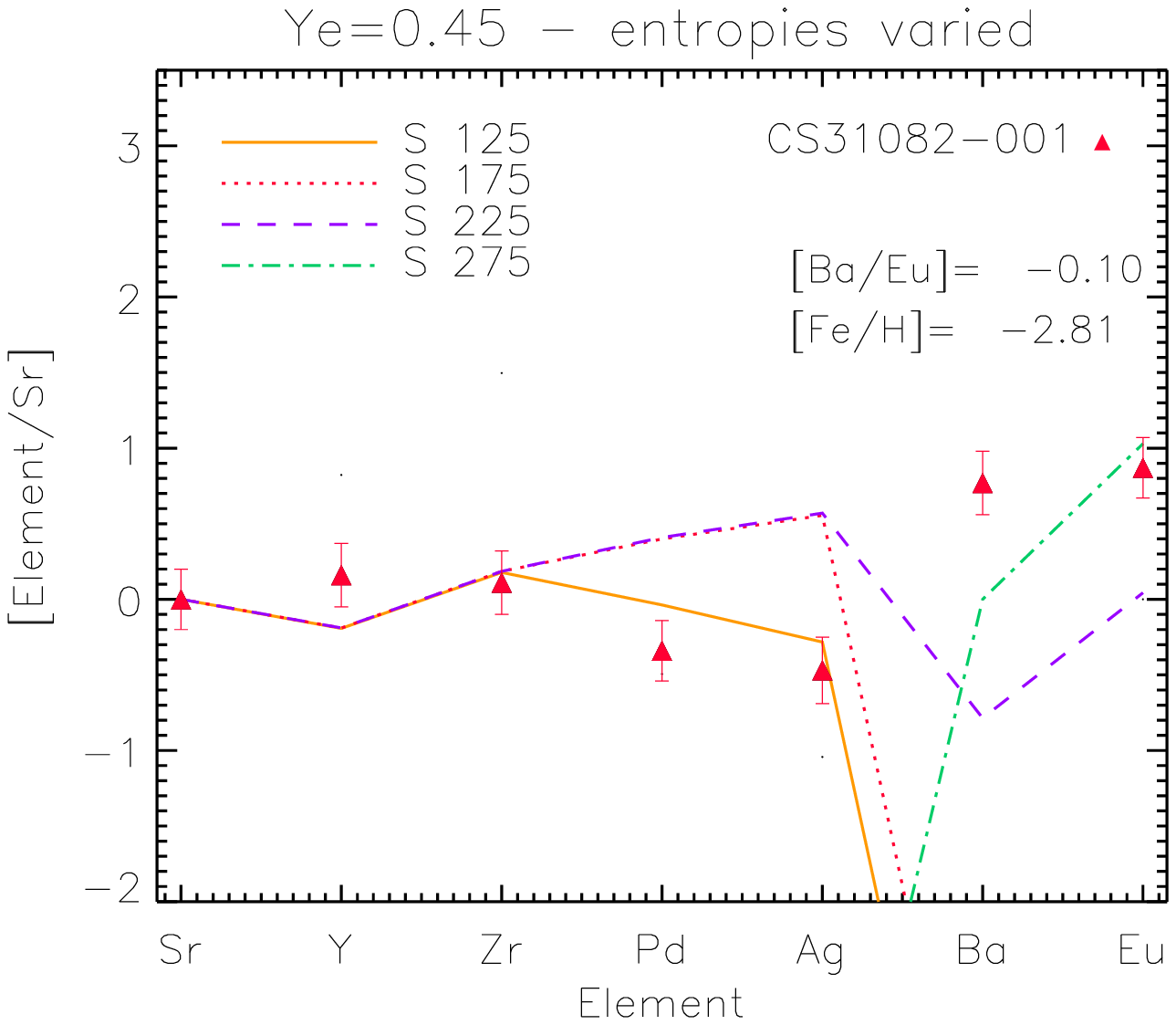}
\includegraphics[width=0.4\textwidth]{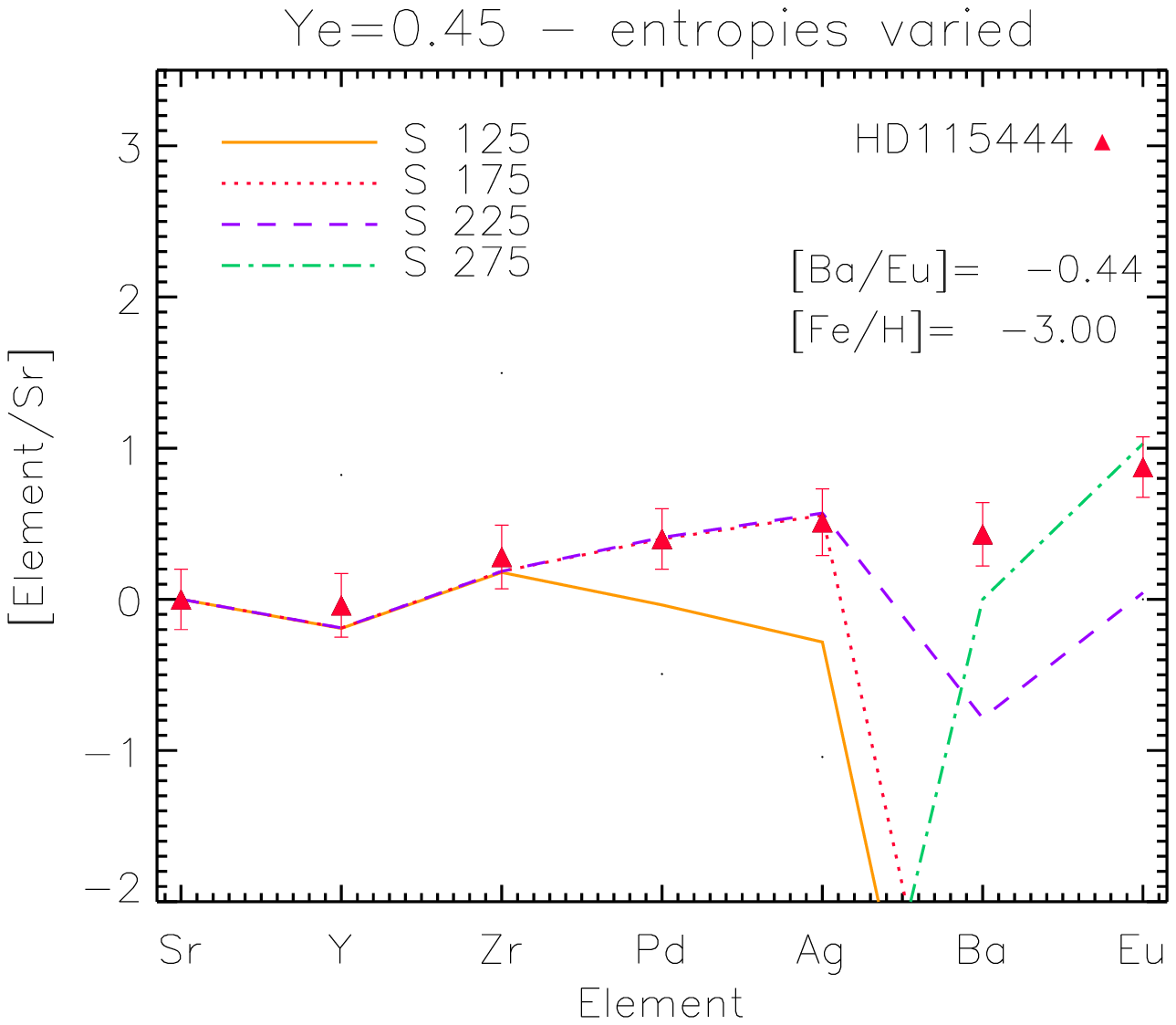}
\caption{HEW model yield predictions (computed assuming $Y_e = 0.45, V_{expan} = 7500$ km/s and four different entropies, $S$, see legend) compared to eight different stars (2 dwarfs, in blue; 6 giants, in red). 
}
\label{hewdw}
\end{center}
\end{figure*}
\clearpage
than Zr and Pd, which we see from the slowly increasing slope, with decreasing $Y_e$, in the main r-process plot. 
The heavy elements (Ba and Eu) need even more neutrons (lower $Y_e$'s) than Pd and Ag to be produced.
Comparing {\it Model 1b} yields to r-rich and r-poor stars (CS~31082--001 and HD122563, respectively), we see the increasing need for neutrons with growing atomic mass (see Fig. \ref{rpoorrich}). In the case of $Y_e = 0.49$, the lighter elements can be correctly reproduced by a charged particle freeze-out or a weak r-process, although, Ba and Eu require a main r-process entropy to be modelled in HD122563. The environment is overall too neutron-poor (or limited to medium entropy) to describe the abundances of a r-rich star (see Fig. \ref{rpoorrich}).
\begin{figure}[h!] 
\begin{center}
\includegraphics[width=0.5\textwidth]{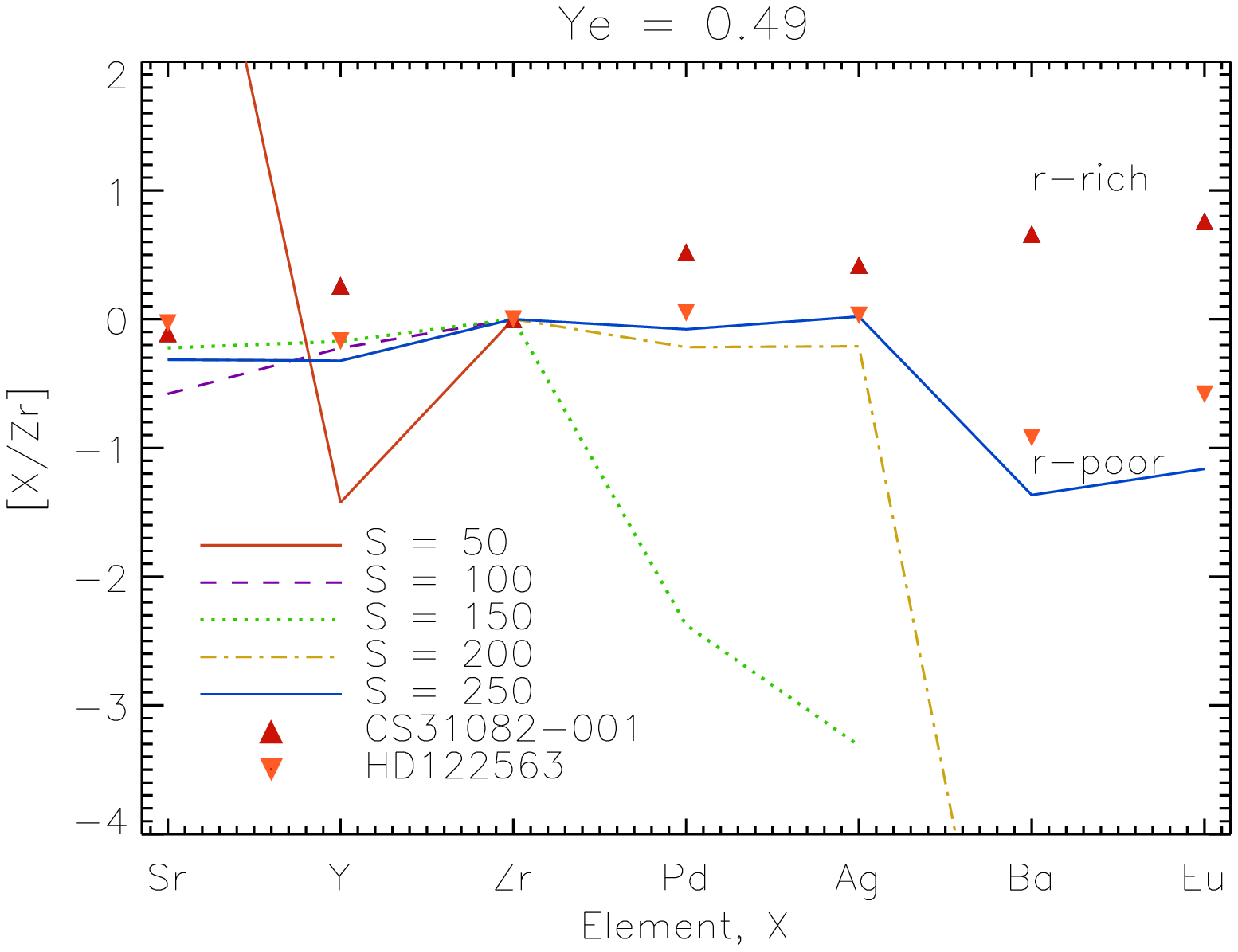}
\includegraphics[width=0.5\textwidth]{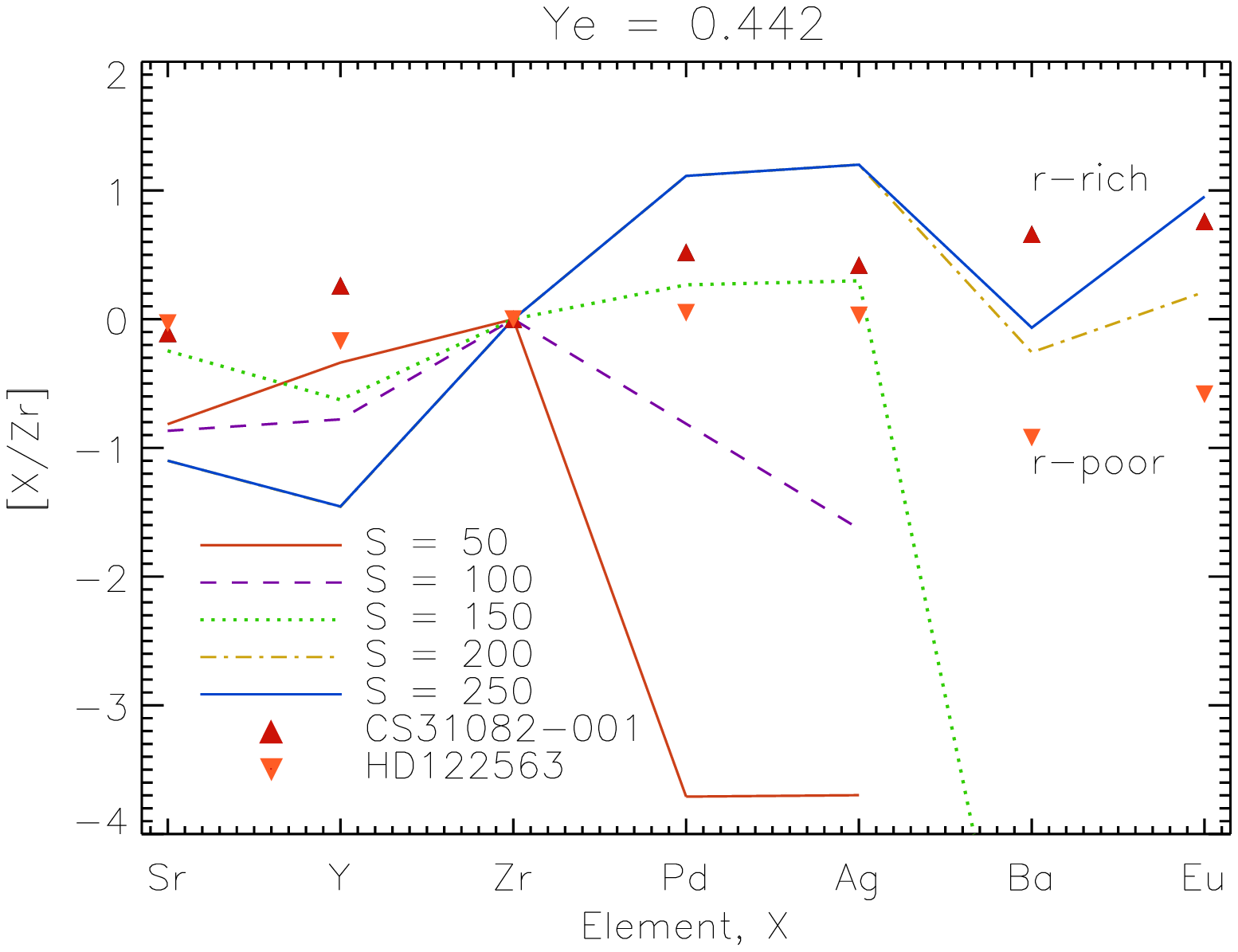}
\caption{HEW predictions with a neutron-poor environment ($Y_e=0.49$; upper plot) and a neutron-rich environment ($Y_e=0.442$; lower plot) compared to the r-poor star HD122563 and r-rich star CS~31082--001. }
\label{rpoorrich}
\end{center}
\end{figure}
In the $Y_e=0.442$ case, the lighter to intermediate mass elements are within the 0.2--0.25 dex uncertainties correctly reproduced by a weak r-process in both stars, whereas Ba and Eu are seen to need a main r-process and possibly an even larger neutron-to-seed ratio/lower $Y_e$ to be correctly reproduced. The need for two different processes at work is again expressed by these models and r-poor/r-rich stellar abundance patterns. The weak r-process cannot create Ba and Eu and the main r-process overproduces the intermediate elements (Pd and Ag). Moreover, it is also unable to correctly account for the lighter elements (Sr -- Zr), where a charged particle freeze-out is needed.

\subsection*{Model II}
The second type of model, which is connected to the low-mass SN of a collapsing O-Ne-Mg core are the two-dimensional (2D) models of \citet{wan11}. 
Neutrino interaction is included in the model, and the explosion is obtained self-consistently without any free parameters. The yields are calculated using post-processing networks, in which the output quantities from the supernova explosion, such as temperature, density, and $Y_e$ are applied as inputs without free parameters in contrast to the parametrised HEW models. Typical input values are entropy, $S \sim 10-20 k_B/\mathrm{baryon}$ (much lower than in the HEW predictions), and $Y_e \sim 0.4 $ to $ 0.5$. 
The scenario allows neutron-captures to take place in the neutron-rich clumps of matter, which are convectively transported to the outer layers.
\citet{wan11} found that there is
little production of elements heavier than Zr in these conditions but
suggest a possible reduction in the minimal electron fraction
$Y_\mathrm{e, min}$ below their original value of $Y_\mathrm{e, org} =
0.4$ (because of the limitation in the resolution and the two-dimensionality
of the model). The impact that lower $Y_\mathrm{e, min}$ values have
on the yields is explored in \citet{wan11}. 
To test
this, as in their work, the additional amount of neutron-capture
elements created in the neutron-rich clumps with an artificially
reduced $Y_\mathrm{e, min}$ value is added to their original yields

\begin{eqnarray*}
Y_Z(Y_{e,min}) = \frac{Y_Z (Y_{e,org}) M + \Sigma_i Y_Z(i) \Delta M_i}{M + \Sigma_i \Delta M_i}, 
\end{eqnarray*}

where $\Delta M_i$ is the relative mass of the elements to be ejected at an $Y_e$ lower than $Y_{\rm e,org} = 0.4$, and this mass is set to $2 \cdot 10^{-5}M_{\odot}$. The subscript $i$ runs over
$Y_\mathrm{e}$ from $Y_\mathrm{e, org}$ down to $Y_\mathrm{e, min}$ in
steps of $\Delta Y_\mathrm{e} = 0.005$. $M$ is the total mass of the ejecta calculated for the higher (original) $Y_{\rm e,org}$, and this mass is 1.136$\cdot 10^{-2}M_{\odot}$.  
$Y_Z$ is the predicted yield (mole fraction) of the element with atomic number Z. 
\begin{figure}[!hpb] 
\begin{center}
\includegraphics[width=0.48\textwidth]{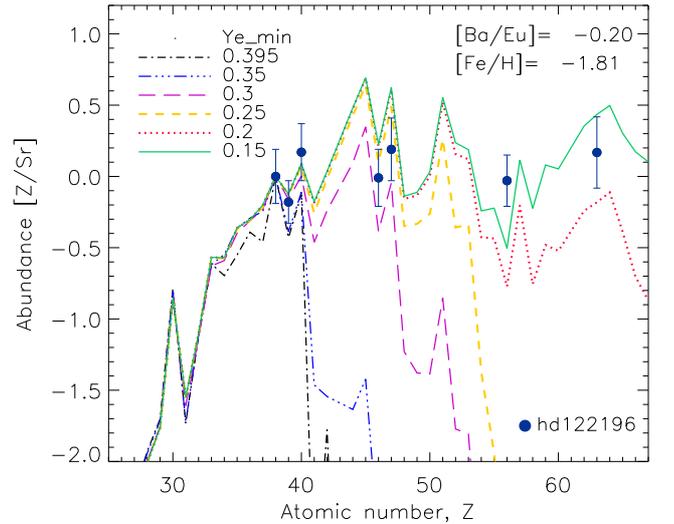}
\caption{HD122196 (dwarf star) compared to different yields calculated as a function of $Y_{\rm e,min}$. The highest electron fractions (0.395 and 0.35) are seen to have too few free neutrons to create the heavier elements (Ag -- Eu).}
\label{Yeselect}
\end{center}
\end{figure} 

\begin{figure*}[!hp] 
\begin{center}
\includegraphics[width=0.45\textwidth]{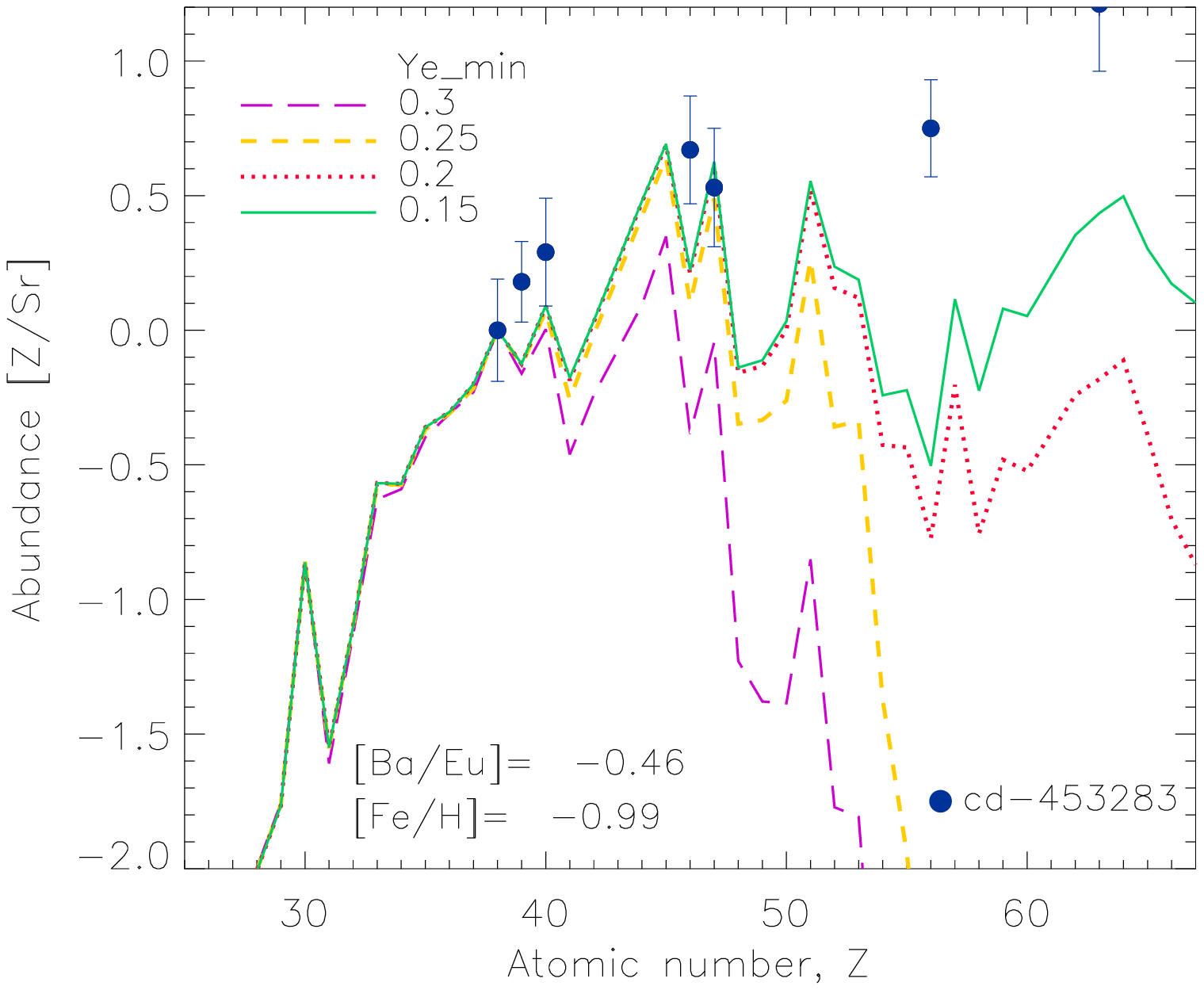}
\hspace{-0.5cm}
\vspace{-0.5cm}
\includegraphics[width=0.45\textwidth]{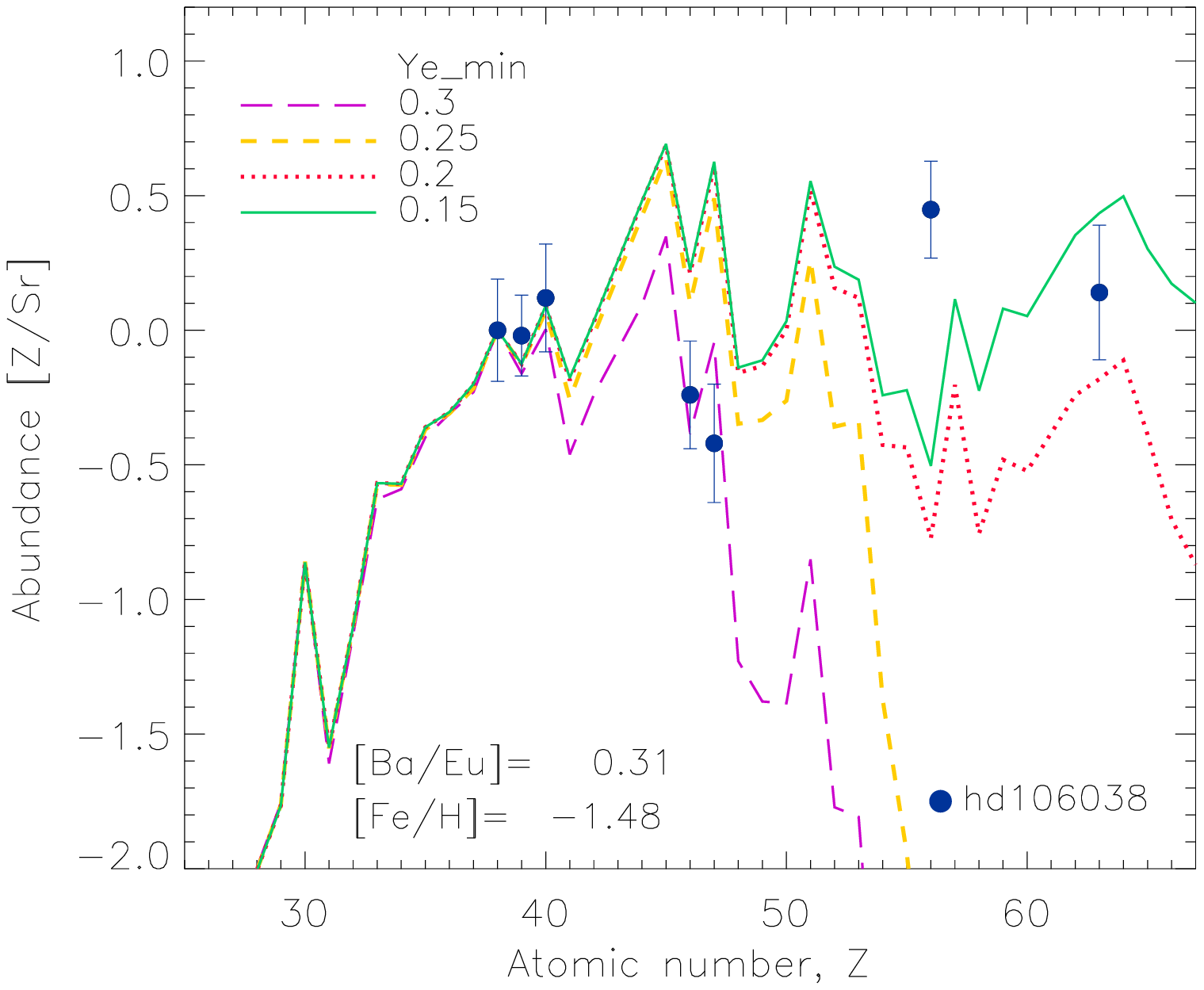}
\hspace{-0.5cm}
%\vspace{-0.5cm}
\includegraphics[width=0.45\textwidth]{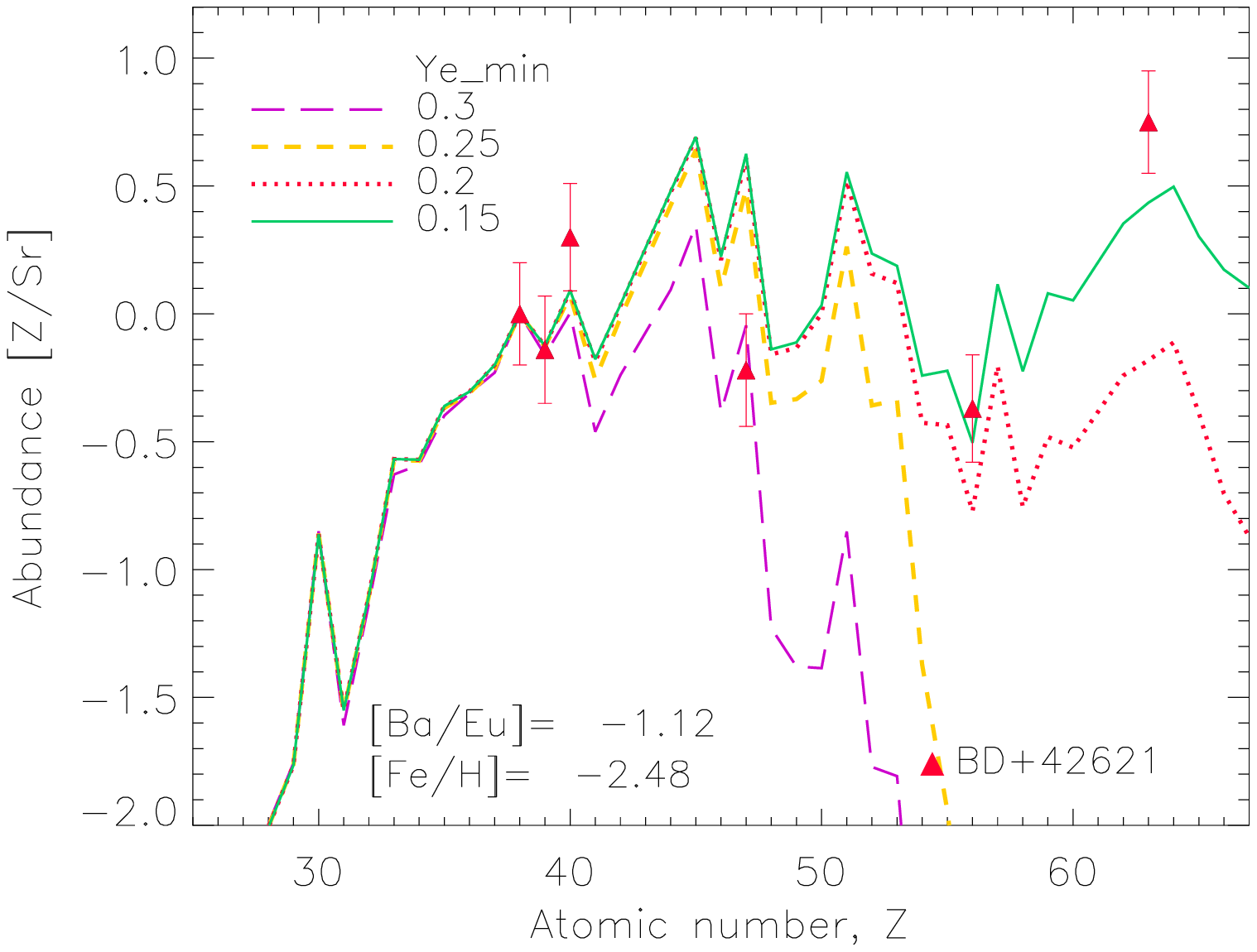}
\hspace{-0.5cm}
\vspace{-0.5cm}
\includegraphics[width=0.45\textwidth]{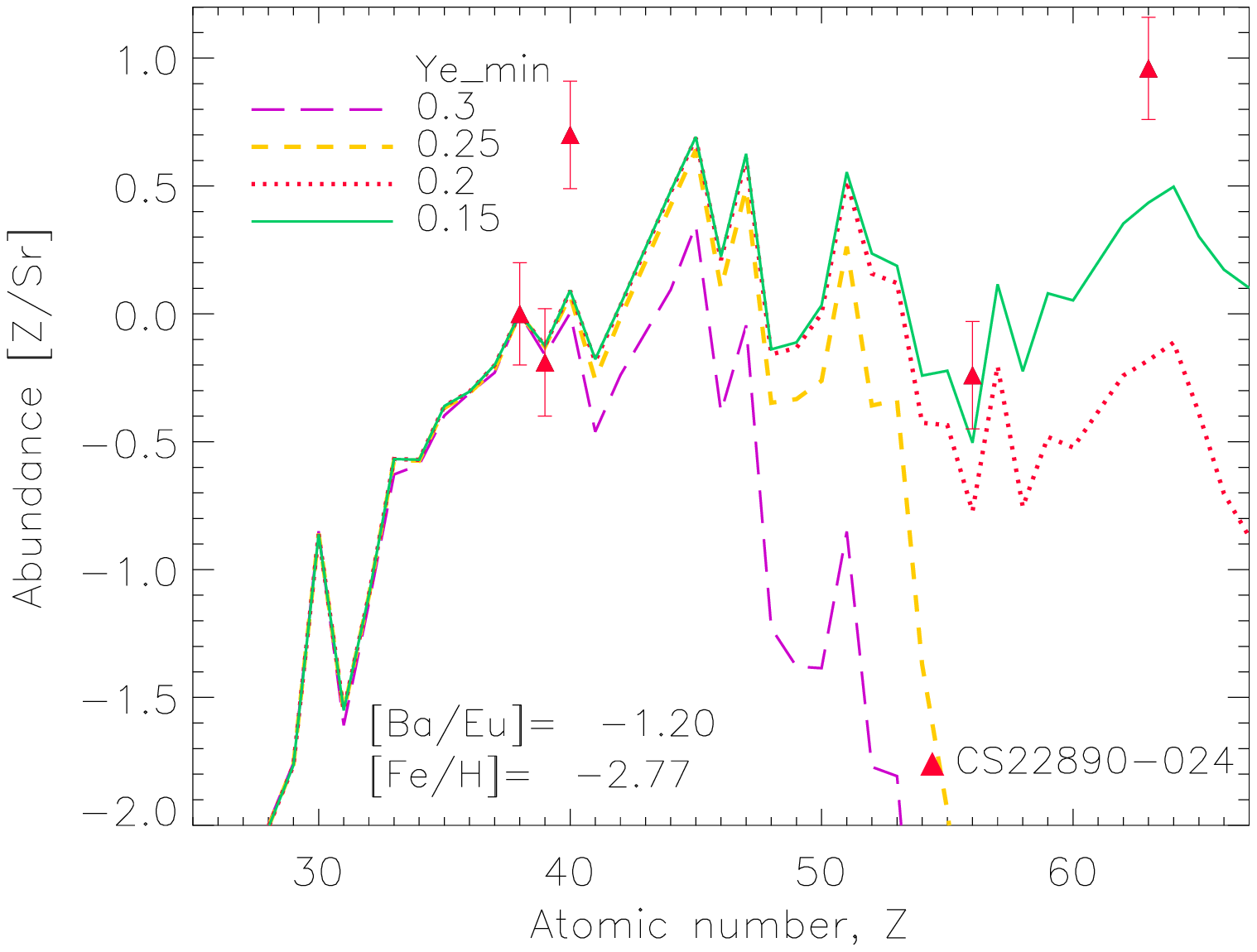}
\hspace{-0.5cm}
%\vspace{-0.5cm}
\includegraphics[width=0.45\textwidth]{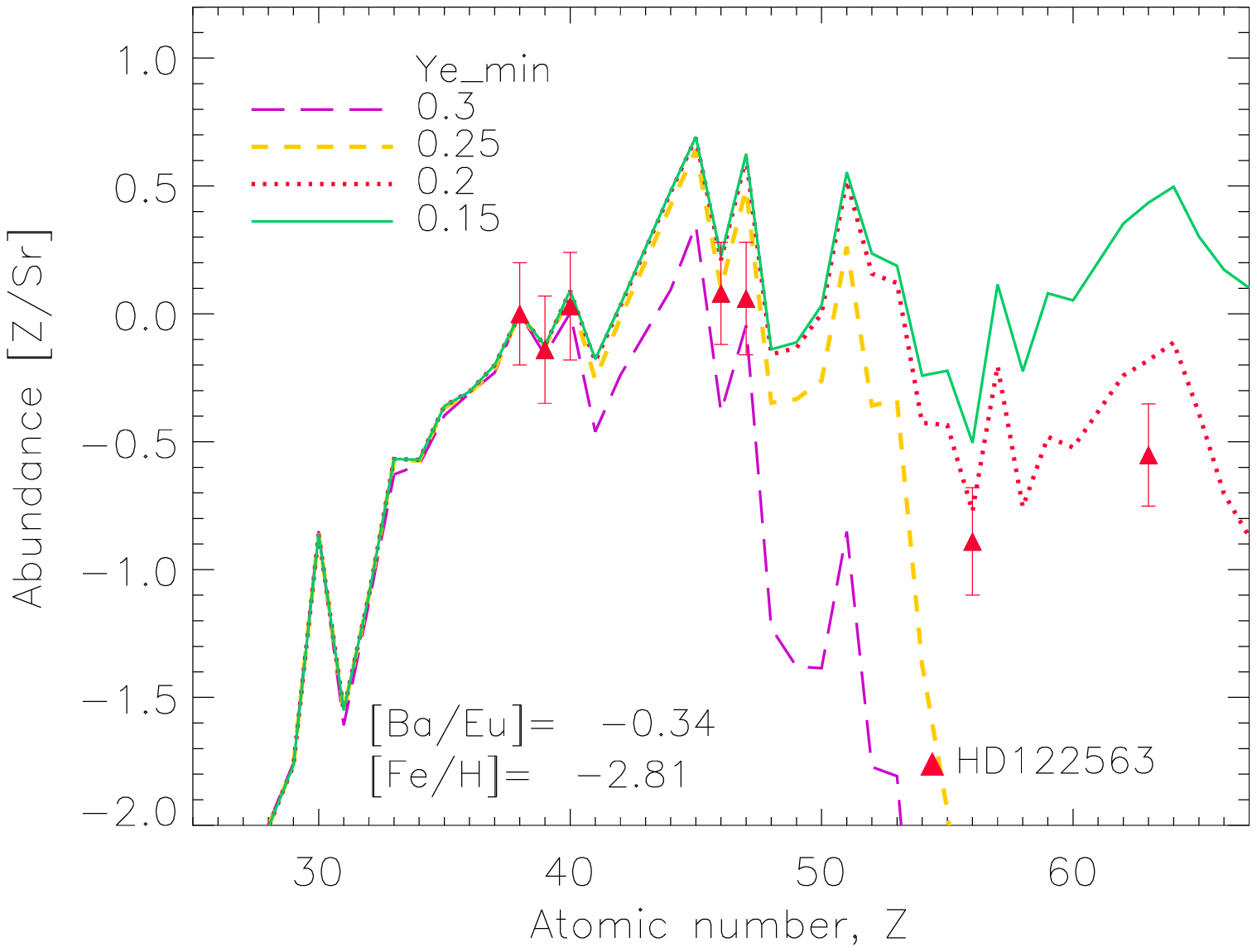}
\hspace{-0.5cm}
\vspace{-0.5cm}
\includegraphics[width=0.45\textwidth]{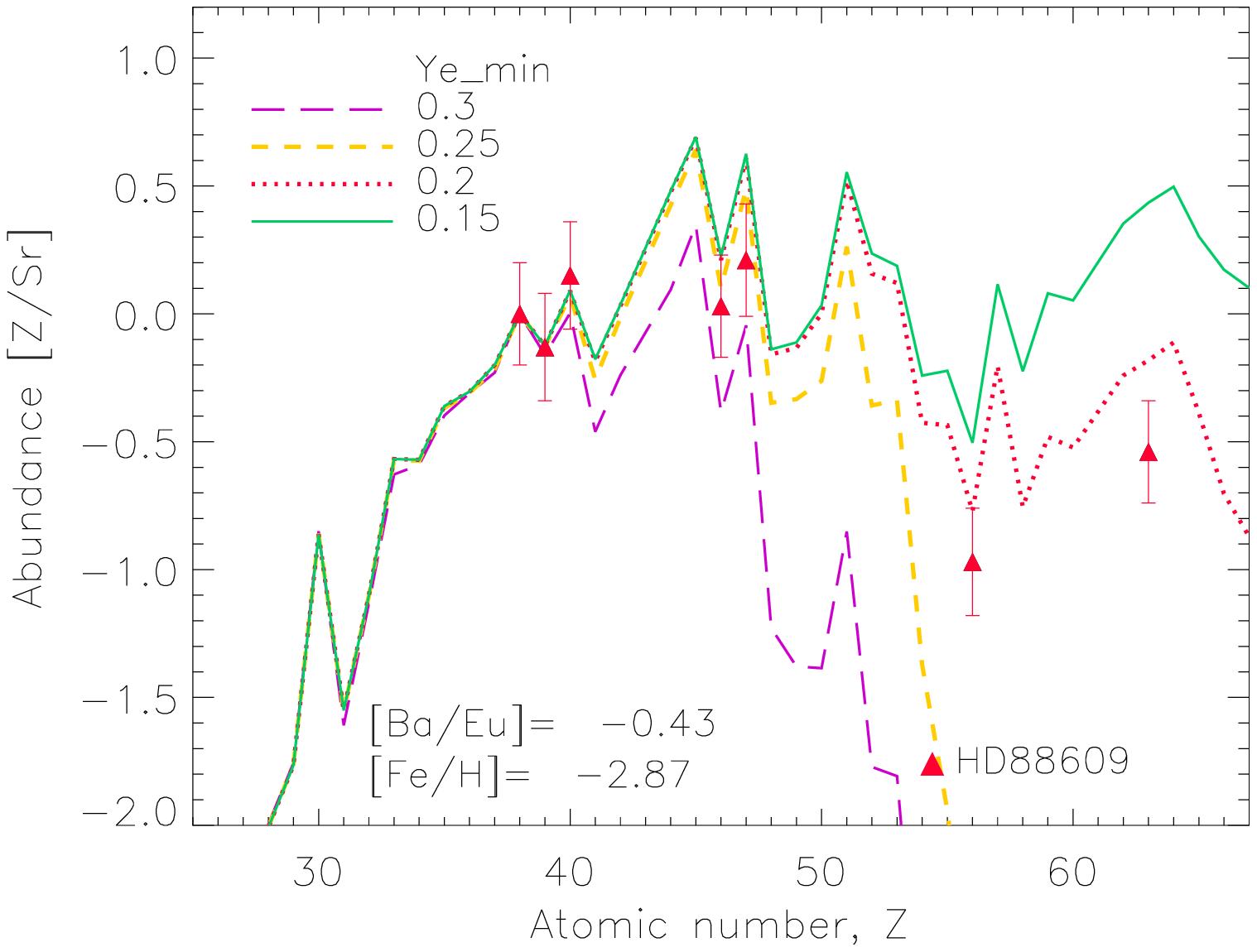}
\hspace{-0.5cm}
%\vspace{-0.5cm}
\includegraphics[width=0.45\textwidth]{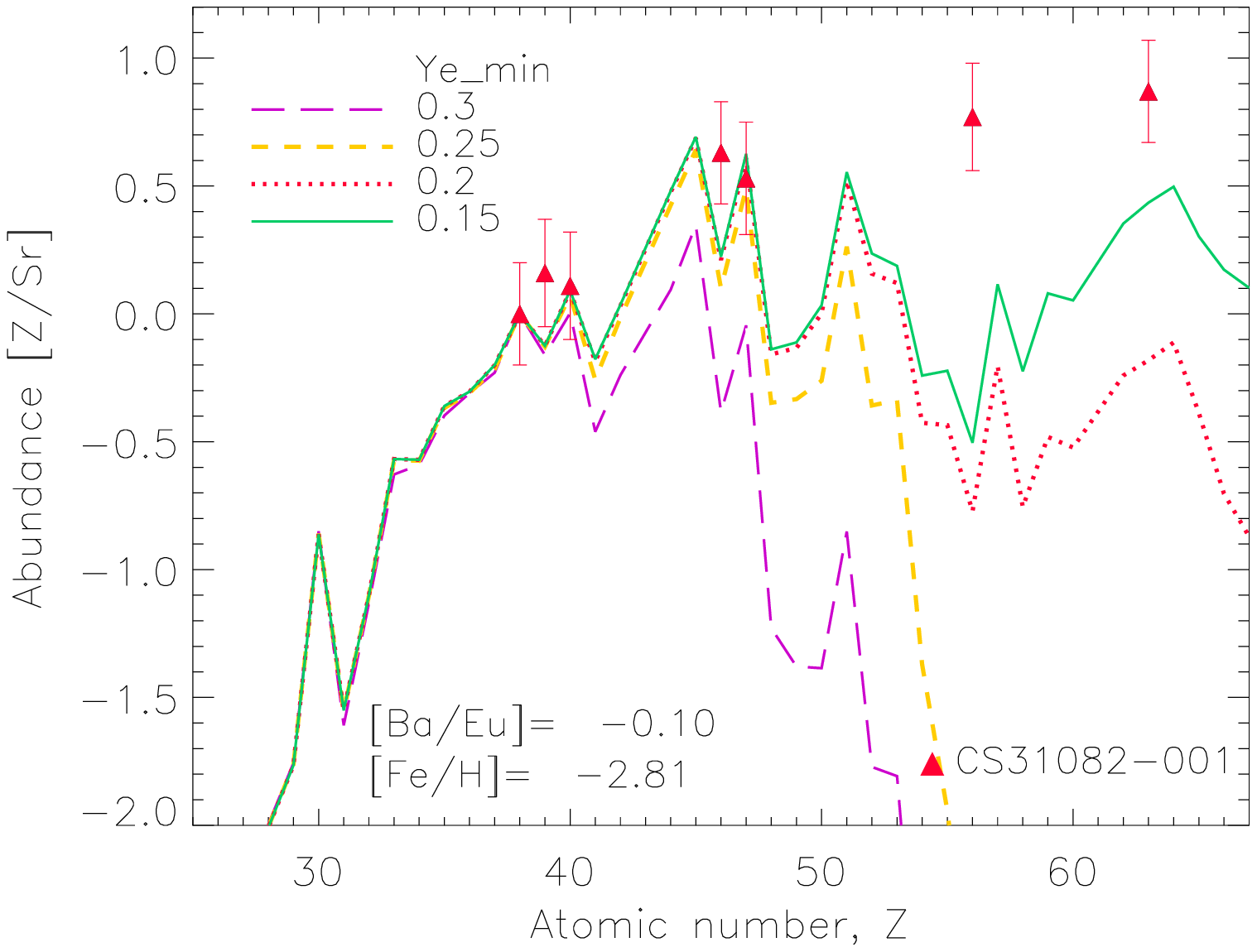}
\hspace{-0.5cm}
\vspace{-0.3cm}
\includegraphics[width=0.45\textwidth]{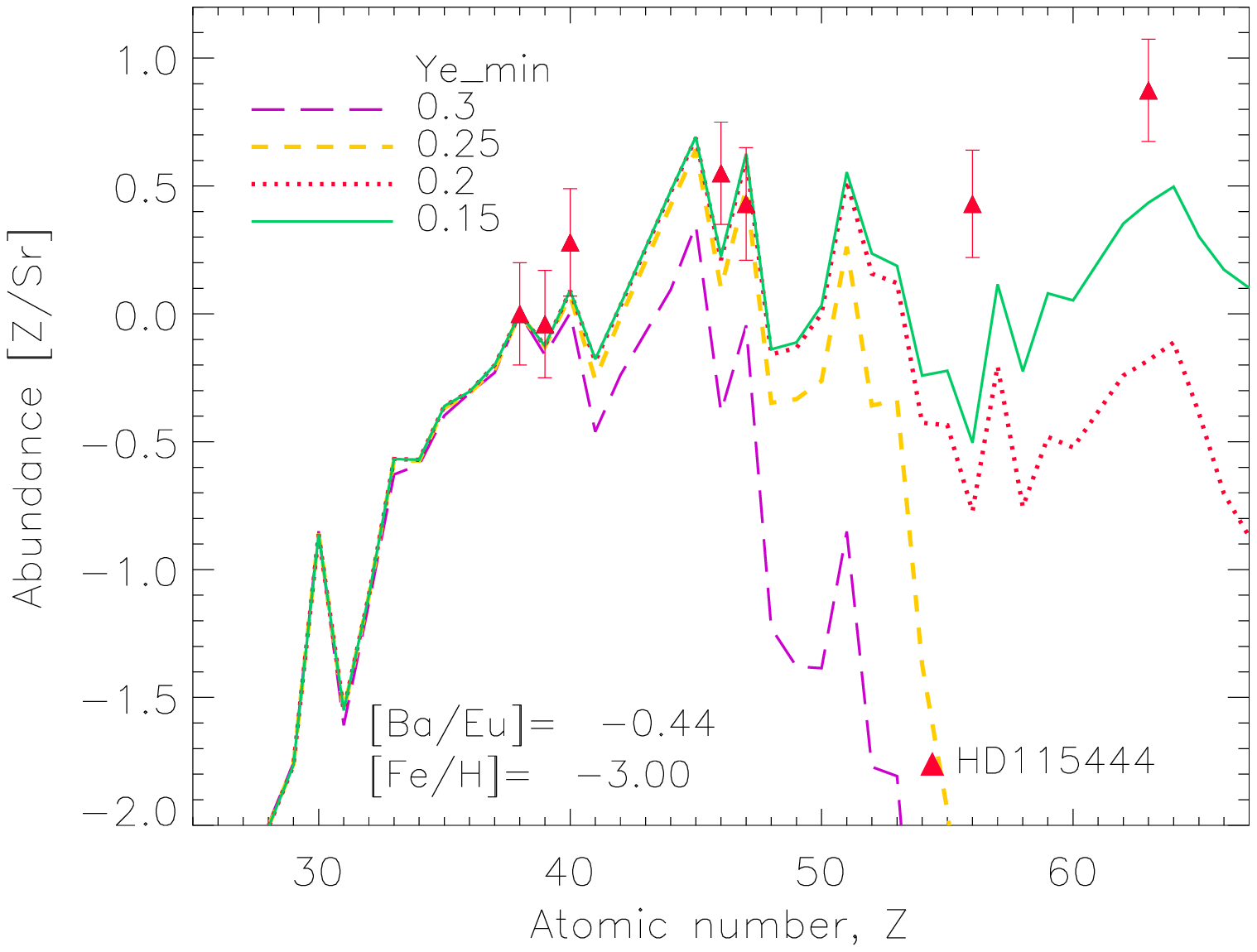}
\hspace{-0.5cm}
\caption{O-Ne-Mg SN model yields with 0.05 step sized decreasing $Y_{\rm e,min}$ starting from $\sim$ 0.3 to 0.15 compared to observationally derived abundances of dwarf stars (blue, top) and giants (red, bottom). The abundances were normalised to Sr. 
These model predictions fit the pattern of the r-poor star (HD122563) better than that of the r-rich star (CS~31082--001).}
\label{wanafig}
\end{center}
\end{figure*}
%\clearpage
Yields calculated with lower $Y_{\rm e,min}$ were necessary in order to obtain considerable amounts of Pd.

With electron fractions as small as $Y_{\rm e,min} =$ 0.2, \citet{wan11} found that the Ba and Eu abundances in HD122563 could be correctly reproduced.
These calculated yields are compared to the observationally derived abundances for one metal-poor giant in Fig. \ref{Yeselect}, in which we first constrain the $Y_{\rm e,min}$ values. 
For simplicity, calculated yields have been plotted in steps of 0.05. From this figure, it is clear that the predictions calculated with an electron fraction of 0.395 and 0.35 fail to produce sufficiently large amounts of any element heavier than zirconium. Hence, these models are not be considered any further.   

\begin{figure}[h] 
\begin{center}
\includegraphics[width=0.48\textwidth]{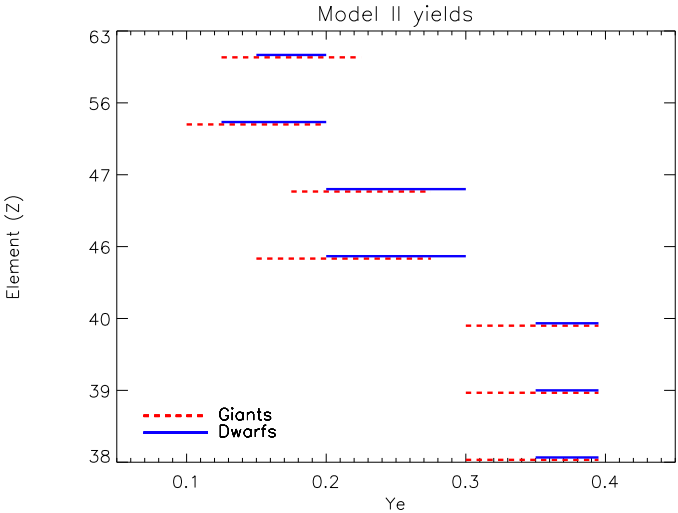}
\caption{Overview of the yields vs electron fraction. Blue represents the dwarfs and red the giants. The atomic number is shown on the y-axis.}
\label{histo}
\end{center}
\end{figure}

The usual seven heavy elements are seen to be created at different values of $Y_{\rm e,min}$, where the heavier ones require a lower $Y_{\rm e,min}$ because they need more neutrons to be produced.
For the Sr to Zr abundances, the model predictions calculated with a $Y_{\rm e,min}$ in the range from 0.35 to 0.395 show a good agreement (i.e. agree within the abundance uncertainty $\sim 0.15 - 0.3$dex) to several of the dwarf stars, Pd and Ag seem to start being 
produced in the proper amounts starting from values of 0.3 down to 0.2 (in $Y_{\rm e,min}$). 
Similar element--$Y_{\rm e,min}$ relations are found for the giants, though the $Y_{\rm e,min}$ values seem to be shifted slightly towards higher values (see Fig. \ref{wanafig} and \ref{histo}). Bins of the required $Y_{\rm e,min}$ values needed to create these seven heavy elements are for the sake of clarity shown in Fig. \ref{histo}.

This difference in the electron fraction needed to produce Eu in dwarf and giant stars, could be due to the different behaviour of Eu in dwarfs and giants, or that the giants need large NLTE abundance corrections. Since Eu is heavier than Ba, it seems unlikely that it would need fewer neutrons (larger $Y_{\rm e,min}$) to form than Ba does. 
To compare the more extreme cases, we compare stars with strong r-process enhancements and/or depletions (r-rich: CS 31082--001 and HD115444, r-poor: HD122563 and HD88609) to these model predictions as well as the HEW predictions.
From Fig. \ref{wanafig}, it becomes clear that the O-Ne-Mg core-collapse SNe may be the site for stars with weak enhancements as seen in both HD122563 and HD88609 (i.e. r-poor), although it is clear that the site is insufficiently neutron-rich to produce such large amounts of the heavy elements (Ba and Eu) and does not in general support the conditions needed for a main r-process.

\subsection{Discussion of yield predictions}

For the seven elements here scrutinised, both models provide  satisfactory agreement with the derived abundances within their associated uncertainties in more than 60\% of the stars. However, neither the HEW predictions nor the O-Ne-Mg SNe model alone can provide an agreement with these seven abundances applying only one set of input parameters. Since very different entropies or electron fractions are needed, two different processes seem necessary. The faint O-Ne-Mg SNe could very well be the formation site of these elements in stars such as HD122563 and HD88609, i.e. stars that are relatively speaking enhanced only in the lighter elements or generally speaking r-poor. However, these supernovae are not the formation site of abundance patterns of stars such as CS~31082--001, which is also enhanced in main r-process elements such as Eu (r-rich). 
The entropy or neutron-richness in the ejecta from O-Ne-Mg SNe are
too low to facilitate a main r-process similar to that in the parametrised HEW
winds explored here. Another possibility for creating the r-poor stars would be a HEW with multiple (medium+high) entropies and low electron fractions ($Y_e \leq 0.442$) --- if possible --- or a fairly high entropy and an electron fraction of 0.49 cf. Fig. \ref{rpoorrich}.

\citet{wan11} showed the effect that higher dimensional models have on the predicted yields, since 1D models could not create trans-iron
elements below Zr, whereas 2D models could (in their Fig. 3).
It is therefore important to consider 3D models before attempting to constrain the r-process site. The stellar atmospheres as well as synthesis codes may also need updates on the input physics, namely improvements in the 1D, LTE assumptions. With the current model predictions at hand, we may need to assume that there are several different sites and/or progenitors to explain the diverse abundances patterns we derive from stellar observations. 
On the basis of our model comparison, we cannot draw strong conclusions about the weak r-process site, but the O-Ne-Mg core-collapse SNe seem promising. However, from the abundance pattern of HD122563 and HD88609, the presence of Eu indicates that the ISM must have been pre-enriched by main r-process material, or that their gas was ejected from an object that allowed both the weak and main r-processes to coincide in the same object. For both processes the HEW winds are possible sites, since they offer the conditions needed\footnote{In their current state, the HEW model predictions allowing for a large range of parameters, especially the large span in entropies, can reproduce several different abundance patterns of all neutron-capture elements. Hence, assuming that these physical conditions are viable in one single site, they explain the patterns with one "continuous broad range r-process" \citep{klk08,Fara,roed2010}. }, but it remains unknown whether the high entropies are physically feasible. These statements depend very much on the efficiency of the mixing in the ISM at metallicities around and below about [Fe/H] $= -2.5$. Therefore, we need to investigate the abundance patterns of stars in the metallicity range $-3.3$ to $-5$ dex (i.e. below the metallicities of our sample), to determine how these patterns that reflect more pristine gases behave and compare to the model predictions, before we can address in detail the mixing efficiency in the early Galaxy. Unfortunately, these abundance patterns cannot be contrasted, since we would need very high-quality spectra (S/N $>$ 100 at 3200 \AA), which would in turn require observations of very long durations for these single faint stars, in order to derive the crucial abundances, such as those of Pd and Ag.

Here as well as in previous studies, it has become evident that knowing the precise values of $Y_e$ ($Y_{\rm e,min}$) is essential to accurately predicting the ejected abundance patterns. The work of \citet{winteler} showed that magneto-rotational core-collapse SN jets can reproduce the solar abundance pattern for $120 <$ A $< 210$. Moreover, \citet{arcones} argue that neutrino-driven winds, either proton- or neutron-rich, stemming from core-collapse SNe can create nuclei in the range 65 $<$ A $<$ 115. Despite the different sites suggested, both studies illustrated the importance of knowing $Y_e$.
This in turn translates into understanding the neutrino interactions and their effect on the electron fraction \citep{arcones}. 

From the comparison of our derived abundances to the HEW model predictions, we learn that we only weakly detect process contributions of the order of 10\% or less, but we need to assume that there has been more than a 10 -- 15\% contribution to see the effect of the process in our abundance ratios (cf. Table \ref{HEWpercent}). Alternatively, our abundance ratios might actually be sensitive down to and below a process contribution of 10\%, since some of the estimated fractional process contributions might have been slightly overestimated. This contribution would change drastically depending on the $Y_e$, which is affected by the other free input parameters as well as the estimated importance of neutrino interactions. 
  
\section{Summary and conclusion}
\subsection*{Summary}
Based on the correlations and anti-correlations of Sr, Y, Zr, Pd, Ag, Ba, and Eu, it has become evident with time that the formation of the heavy elements is not a straightforward process to model, and that the previously believed solar-scaled universal r-process pattern only provides good estimates for the heaviest elements, not the elements in the atomic mass range from 38 to 47. At low metallicities ([Fe/H] $< -$2.5), several studies combined with this have shown that at least four different neutron-capture processes (s- and r-processes with both weak and main components) are needed to explain the observationally determined abundances. Starting with the lighter element Sr, which can in part be produced via charged particle freeze-out, this element might also receive a considerable contribution from a weak s-process at higher metallicities. Zirconium has shown similarities to both Sr and Ag, which indicates that Zr is created by multiple processes, since the formations of Sr and Ag differ. Silver is created by a second/weak r-process. 
Its formation process clearly deviates from the main r-process responsible for forming Eu and to some extent Ba at low metallicity. The picture becomes even more differentiated when we try to understand the formation of Ba. To date, an inexplicably large star-to-star scatter is found for the Ba abundances both under LTE and NLTE assumptions. No single process can explain all of these results despite the possibility that some scatter may be caused by the process occurring at various sites, in the case of the extremely metal-poor stars. Two processes might be needed to fully explain the formation of Ba.

From the comparison to model predictions, we see that despite the different physics and parameter space investigated, the 2--3 dex star-to-star scatter in the stellar abundances cannot be explained by e.g. NLTE corrections, stellar parameter influence or sample biases, which confirms the need for at least two neutron-capture processes yielding heavy elements at very low metallicities (below $-2.5$ dex in [Fe/H]). Some scatter can arise from the different amount each site produces, and it seems necessary to have a combination of various sites to explain the individual abundance patterns that the different (r-poor vs r-rich) stars show. One possible explanation could be that massive supernovae facilitate high entropy winds, which create some amount of intermediate elements (in the atomic number interval range 40 -- 50) via a weak r-process, combined with yields of the heaviest main r-process elements. According to \citet{woosley}, different entropies can be found within one supernova, giving rise to various entropies and processes in the exploding winds. Another site contributing to the weak r-process elements Pd and Ag could be the O--Ne--Mg SNe, which are predicted to be very common. However, according to \citet{wan11}, this kind of SN cannot facilitate a main r-process, hence cannot produce Eu, and it seems unlikely to be the dominant r-process site, which was indicated in \citet{ishimaru}. However, 3D models might change this picture, though it seems unlikely that these supernovae would ever reach the proper conditions to create a full main r-process. Nevertheless, we still need several sites and r-processes to explain the abundances of Ag -- Eu.  

\subsection*{Conclusion}
We have found that in our observed sample of stars both dwarfs and giants show on average the same correlation/anti-correlation at all metallicities, thus we feel confident that the correlation trends combined with the large star-to-star scatter confirm the presence of two different r-processes. A second/weak r-process creating Zr -- Ag (generally elements in the atomic number range 40 -- 50), and a main r-process producing the very heavy elements. The second r-process seems to work in intermediate entropy, and neutron number density environments, and its path possibly lies closer to stability than that of the main r-process. However, on the basis of previous studies and the current state of the model predictions, we cannot disregard the possibility that our suggested second different r-process is in fact a lower end of a continuous broad-range main r-process. Many of the physical parameters differ between the weak and the main r-process, by many orders of magnitude, and we therefore need stronger constraints on what suffices and/or is necessary to define an individual process.  

In addition, it seems that several sites are needed to explain the diverse stellar abundance pattern coming from r-poor and r-rich stars. Possible formation sites are the high-entropy winds of SNe and neutron-rich ejecta of electron-capture (O--Ne--Mg) SNe.
The yields from these objects will be mixed in the ISM, which makes tracing the original site a challenging task; furthermore, these objects are unlikely to be the only sites, and we still do not know their frequency or mixing ratios.

\begin{acknowledgements}
This work was supported by Sonderforschungsbereich SFB 881 "The Milky Way System" (subproject A5) of the German Research Foundation (DFG). The authors are grateful to the referee for comments and criticism.
CJH thanks ESO for the support, L. Casagrande for providing temperature estimates and C. Sneden for spectra. Furthermore, CJH is very grateful to W. Walters and B. Nordstr\"om for help and discussions. FP acknowledges support from the Collaborative Research Project MASCHE, 
part of ESF EUROCORES programme EuroGENESIS. HH acknowledges support from the Swedish Research Council (VR) under contract 621-2011-4206. SW and CJH kindly thank H.-T. Janka and B. M\"uller for collaboration and fruitful discussions.
This research has made use of NASA's Astrophysics Data System, the SIMBAD database, operated at CDS, Strasbourg, France, and the Two Micron All Sky Survey, which is a joint project of the University of Massachusetts and the Infrared Processing and Analysis Center/California Institute of Technology, funded by the National Aeronautics and Space Administration and the National Science Foundation. 

\end{acknowledgements}

\bibliographystyle{aa}
\bibliography{Ag_CJH_18643}

\Online

\appendix
\section{Line lists}
Here we give the details of our line lists as well as our adopted solar abundances and additional useful information for all the heavy elements studied.
\begin{table}[h!]
\caption{Line parameters for the resonance 5s-5p in AgI. }
\label{table:nonisotopic}
\begin{tabular}{@{}cccc@{}cc}
\hline \hline
Isotope	& Lower level & Upper level & Flow-Fup & $\lambda_{\mathrm{air}}$ & $\log gf$\\
 & & & & [\AA] & \\
\hline					
107	& $^2$S$_{1/2}$	& $^2$P$_{1/2}$	& 0-1	& 3382.891	& -0.936 \\ 
107	& $^2$S$_{1/2}$	& $^2$P$_{1/2}$	& 1-0	& 3382.884	& -0.936 \\ 
107	& $^2$S$_{1/2}$	& $^2$P$_{1/2}$	& 1-1	& 3382.885	& -0.635 \\
109	& $^2$S$_{1/2}$	& $^2$P$_{1/2}$	& 0-1	& 3382.894	& -0.936 \\
109	& $^2$S$_{1/2}$	& $^2$P$_{1/2}$	& 1-0	& 3382.886	& -0.936 \\
109	& $^2$S$_{1/2}$	& $^2$P$_{1/2}$	& 1-1	& 3382.887	& -0.635 \\
&&&&&\\[-2mm]
107	& $^2$S$_{1/2}$	& $^2$P$_{3/2}$	& 0-1	& 3280.684	& -0.624 \\
107	& $^2$S$_{1/2}$	& $^2$P$_{3/2}$	& 1-1	& 3280.678	& -0.925 \\
107	& $^2$S$_{1/2}$	& $^2$P$_{3/2}$	& 1-2	& 3280.678	& -0.226 \\
109	& $^2$S$_{1/2}$	& $^2$P$_{3/2}$	& 0-1	& 3280.686	& -0.624 \\
109	& $^2$S$_{1/2}$	& $^2$P$_{3/2}$	& 1-1	& 3280.679	& -0.925 \\
109	& $^2$S$_{1/2}$	& $^2$P$_{3/2}$	& 1-2	& 3280.680	& -0.226 \\ \hline
\end{tabular}
\end{table}
Table \ref{table:nonisotopic} provides the atomic information for both of the silver lines. These calculations were made without any assumption about the natural isotopic ratio. If individual isotopic Ag abundances are needed, these $\log gf$ values should be applied instead of those listed in Table \ref{table:isotopic}.

Below we list the solar abundances that we used. These were adopted from \citet{asp09}. 
\begin{table}[!h]     
\begin{center}    
\caption{Element and adopted solar abundances.} 
\label{solval}    
\begin{tabular}{ c c }          
\hline 
Element & $\log \epsilon$ \\
\hline
Sr & 2.87  \\
Y  & 2.21  \\
Zr & 2.58   \\
Pd & 1.57   \\
Ag & 0.94   \\
Ba & 2.18   \\
Eu & 0.52   \\
\hline  
\end{tabular}  
\end{center}   
\end{table}  
     
We provide the atomic data used for the heavy elements in our line list. The values are taken from VALD \citep{VALD}. The molecular information in our line list was taken from Kurucz's home page and \citet{kurucz}. 
\begin{table}
\begin{center}
\caption{Atomic data for the strontium to europium: Wavelength, excitation potential, and $\log gf$. The 'T' indicates that the value is the total $\log gf$, which for Ba was split according to \citet{mcwil98} and for Eu according to \citet{ivans}.}
\label{heavylist}
\begin{tabular}{l c c }       
\hline
Sr II & $\chi$ & $\log gf$ \\
  & [eV] & [dex] \\
\hline
3464.45  &  3.04 &  0.49 \\  
4077.71  &  0.00 &  0.17 \\ 
4161.79  &  2.94 & -0.50 \\
4215.52  &  0.00 & -0.14 \\
\hline
Y II & $\chi$ & $\log gf$ \\
  & [eV] & [dex] \\
\hline
 3549.01  &  0.13 & -0.28  \\
 3600.74  &  0.18 &  0.28    \\
 3628.70  &  0.13 & -0.71   \\
 3774.34  &  0.13 &  0.21   \\
 3788.70  &  0.10 & -0.07  \\
 3950.36  &  0.10 & -0.49  \\
 4398.01  &  0.13 & -1.00  \\
 4854.87  &  0.99 & -0.38   \\
 4883.69  &  1.08 &  0.07   \\
 5087.42  &  1.08 & -0.17   \\
 5200.42  &  0.99 & -0.57   \\
\hline
Zr II & $\chi$  & $\log gf$ \\
  & [eV] & [dex] \\
\hline
 3356.09 &  0.09 & -0.51 \\
 3499.57 &  0.41 & -0.81  \\
 3551.96 &  0.09 & -0.31  \\
 3573.06 &  0.32 & -1.04  \\ 
 3607.38 &  1.24 & -0.64  \\
 3714.79 &  0.53 & -0.93  \\
 4050.33 &  0.71 & -1.00  \\
 4161.21 &  0.71 & -0.72  \\
 4208.98 &  0.71 & -0.46  \\
 4317.32 &  0.71 & -1.38 \\
 5112.28 &  1.66 & -0.59 \\
\hline
Pd I & $\chi$ & $\log gf$  \\
  & [eV] & [dex] \\
\hline
3404.58 & 0.814  &  0.320 \\  
\hline
Ba II & $\chi$ & $\log gf$ \\
  & [eV] & [dex] \\
\hline
4554.03  &  0.00 &  0.17$^T$  \\
4934.08  &  0.00 & -0.15$^T$   \\
5853.67  &  0.60 & -1.01$^T$  \\
\hline
Eu II & $\chi$  & $\log gf$ \\
  & [eV] & [dex] \\
\hline
 4129.73  &  0.00 &  0.22$^T$ \\
 4205.04  &  0.00 &  0.21$^T$ \\
 6645.06  &  1.38 &  0.12$^T$ \\    
\hline  
\end{tabular}  
\end{center}       
\end{table}    
\newpage

\section{Stellar parameters}
Below, we list the parameters needed to determine the temperature and gravity. Top: Giants, bottom: Dwarfs. The superscripts a,b, and c indicate the following: $^{(a)}$ Stars with T$_{eff}$ and log g derived from excitation potential and ionisation balance. $^{(b)}$ Stars with a special r-process pattern --- either r-poor or r-rich. $^{(c)}$ Stellar parameters were altered in accordance with (a) owing to uncertainties in colour, dereddening, and parallax.
\begin{small}
\begin{table*}[!h]
\begin{center}
\label{stelpar}
\begin{tabular}{l c c c c c c c c c c}        
\hline\hline
Star    &		V &	K &	   $\pi$ & $\sigma(\pi)$ &   E(B-V)  & Mass$^*$ & T	&  g	&  [Fe/H] & $\xi $  \\
\hline
BD-01 2916  	   &     9.31 &	 8.03	&     20.20& 16.60	     & 0.00	& \nod  & 4480$^a$  & 1.20$^a$   &  -1.99 &  2.4 \\
BD+8 2856	&	\nod&	\nod	&     \nod &  \nod	     & 0.00	& \nod  & 4600$^a$  & 0.80$^a$   &  -2.09 &  2.0\\
BD+30 2611	&	9.13&	 6.09	&      3.45&  1.31	&	0.02	& \nod  & 4238& 0.50$^a$   &  -1.20 &  1.7\\
BD+42 621	&	10.5&	 9.76	&     16.10& 30.50	     & 0.00	& \nod  & 4725$^a$  & 1.50$^a$    &  -2.48 &  1.7\\ 
BD+54 1323	&	9.34&	 7.37	&      1.22&  1.20	&	0.01	& \nod  & 5213  & 2.01$^c$ &  -1.64 &  1.5\\ 
CS22890-024	& 	13.41&	11.44	&      \nod&   \nod	&	0.05	& \nod  & 5400  & 2.65$^a$  &  -2.77 &  1.7\\ 
CS29512-073	&    	13.92&	12.51	&      \nod&   \nod	&	0.05	& \nod  & 5000$^a$  & 1.85$^a$  &  -2.67 &  1.1\\ 
CS30312-100	&    	13.05&	10.88	&      \nod&   \nod	&	0.08	& \nod  & 5200  & 2.35$^a$ &  -2.62 &  1.4\\ 
CS30312-059	&    	13.14&	10.70	&      \nod&   \nod	&	0.12	& \nod  & 5021  & 1.90$^a$  &  -3.06 &  1.5\\ 
CS31082-001$^b$	&	11.67&	 9.46	&      \nod&   \nod	     & 0.00	& \nod  & 4925  & 1.51$^a$  &  -2.81 &  1.4\\\
HD74462         &	8.69&	 6.05	&      1.55&  1.16	&	0.05	& \nod  & 4590  & 1.84$^c$  &  -1.48 &  1.1\\ 
HD83212         &     	8.33&	 5.61  &      1.96&  0.98	&	0.05	& \nod  & 4530  & 1.21$^c$  &  -1.25 &  1.8\\ 
HD88609$^b$	&	8.59&	 6.01	&      0.63&  1.14	&	0.01	& \nod  & 4568  & 1.01$^c$  &  -2.87 &  1.9\\
HD108317	&	8.03&	 6.15  &      4.53&  1.06	&	0.01	& \nod  & 5360  & 2.76 &  -2.11 &  1.2\\ 
HD110184	&	8.27& 	 5.35	&      1.00&  0.99	&	0.02	& \nod  & 4450$^a$  & 0.50$^c$   &  -2.40 &  2.1\\ 
HD115444$^b$	&	8.96&	 6.61	&      3.55&  1.12	&	0.01	& \nod  & 4785  & 1.50$^c$   &  -3.00 &  2.1\\
HD122563$^b$	&    	6.20&	 3.73	&      3.76&   \nod	&	0.025   & \nod  & 4560$^a$  & 0.90$^a$   &  -2.81 &  1.8\\ 
HD122956	&    	7.25&	 5.90	&      3.30&   \nod	&	0.083   & \nod  & 4700  & 1.51  &  -1.45 &  1.2\\
HD126238	&	7.66&	 5.34	&      3.81&  0.95	&	0.04	& \nod  & 4900  & 1.80   &  -1.92 &  1.5\\
HD126587	&	9.11&	 6.668	&      1.40&  1.44	&	0.09	& \nod  & 4700$^a$  & 1.05$^c$  &  -3.16 &  1.7\\ 
HD128279	&     	7.97&	 7.07	&      5.96&   \nod	&	0.10    & \nod  & 5200$^a$  & 2.20$^a$   &  -2.34 &  1.3\\ 
HD165195	&   	7.34&	 4.14	&      2.20&   \nod	&	0.195   & \nod  & 4200$^c$  & 0.90$^c$   &  -2.10 &  2.1\\ 
HD166161$^b$	&	8.12&	 5.34	&      3.25&  1.19	&	0.13	& \nod  & 5250$^a$ & 2.15$^c$ &  -1.25 &  1.9\\
HD175305	&	7.18&	 5.06	&      6.18&  0.56	&	0.03	& \nod  & 5100  & 2.70  &  -1.38 &  1.2\\
HD186478	&	9.14&	 6.44	&      1.34&  1.25	&	0.12	& \nod  & 4730  & 1.50$^c$  &  -2.42 &  1.8\\
HD204543	&	8.28&	 5.78  &  1.08 &  1.38	&	0.04	& \nod  & 4700  & 0.80$^a$   &  -1.84 &  2.0\\
HE0315+0000	&    	15.52&  13.20	&      \nod&   \nod	&	0.081	& \nod  & 5200  & 2.40$^a$  &  -2.59 &  1.6\\ 
HE0442-1234	&	12.91&   9.96  &      \nod&   \nod	&	0.133	& \nod  & 4530  & 0.65$^a$  &  -2.32 &  1.8\\
HE1219-0312	&	15.94&	13.89	&     \nod &  \nod	     & 0.00	& \nod  & 4600  & 1.05$^a$  &  -3.21 &  1.4\\
\hline
Star	&		V &	K &	 $\pi$ & $\sigma(\pi)$	&   E(B-V) &	  Mass  &   T	&  g	&  [Fe/H] & $\xi $ \\
\hline
BD+092190       &   11.15  & 9.91   &  1.04	 & 2.79    &  	   0.0281	& 0.8	& 6450$^a$  & 4.00  & -2.60 & 1.5  \\ 
BD-133442       &   10.29  & 9.02   & \nod	 & \nod    &  	   0.044	& 0.8	& 6450  & 4.20$^a$  & -2.56 & 1.5 \\  
CD-3018140      &    9.95  & 8.66   &  7.32	 & 1.56    &  	   0.030	& 0.75  & 6340  & 4.13  & -1.92 & 1.0 \\  
CD-33 3337      &     9.08 &  7.67  &   9.11	 & 1.01    &  	   -0.0155	& 0.8	& 5952  & 3.95  & -1.55 & 1.4 \\  
CD-45 3283      &    10.57 &  8.97  &  15.32	 & 1.38    &  	   0.0001	& 0.8	& 5657$^c$ & 4.97  & -0.99 & 0.8\\
CD-57 1633      &     9.53 &  8.09  &  10.68	 & 0.91    &  	   0.0  	& 0.8	& 5907  & 4.26  & -1.01 & 1.1 \\  
HD3567          &    9.26  & 7.89   &  9.57	 & 1.38    &  	   -0.0028	& 0.82  & 6035  & 4.08  & -1.33 & 1.5 \\  
HD19445         &    8.05  & 6.64   & 25.85	 & 1.14    &  	   -0.0014	& 0.70  & 5982  & 4.38  & -2.13 & 1.4 \\  
HD22879         &    6.69  & 5.18   & 41.07	 & 0.86    &  	   -0.0056	& 0.8	& 5792  & 4.29  & -0.95 & 1.2 \\  
HD25704         &    8.12  & 6.56   & 19.02	 & 0.87    &  	   -0.0211	& 0.8	& 5700  & 4.18  & -1.12 & 1.0 \\  
HD63077         &    5.36  & 3.75   & 65.79	 & 0.56    &  	   -0.0225	& 0.8	& 5629  & 4.15  & -1.05 & 0.9 \\  
HD63598         &    7.95  & 6.37   & 20.14	 & 1.09    &  	   0.0  	& 0.8	& 5680  & 4.16  & -0.99 & 0.9 \\  
HD76932         &    5.80  & 4.36   & 46.90	 & 0.97    &  	   -0.024	& 0.85  & 5905  & 4.08  & -0.97 & 1.3 \\  
HD103723        &   10.07  & 8.66   &  7.63	 & 1.62    &  	   0.038	& 0.88  & 6128  & 4.28  & -0.85 & 1.5 \\  
HD105004        &   10.31  & 8.87   &  2.68	 & 4.49    &  	   0.038	& 0.8	& 5900$^a$ & 4.30$^c$  & -0.84 & 1.1\\
HD106038$^b$      &   10.18  & 8.76   &  9.16	 & 1.50    &  	   -0.025	& 0.70  & 5950  & 4.33  & -1.48 & 1.1 \\
HD111980$^b$      &    8.37  & 6.77   & 12.48	 & 1.38    &  	   -0.0113	& 0.79  & 5653  & 3.90  & -1.31 & 1.2 \\ 
HD113679        &    9.70  & 8.11   &  6.82	 & 1.32    &  	   0.024	& 0.96  & 5759  & 4.04  & -0.63 & 0.9 \\ 
HD116064        &    8.81  & 7.31   & 15.54	 & 1.44    &  	   0.0352	& 0.8	& 5999  & 4.33  & -2.19 & 1.5 \\ 
HD120559        &    7.97  & 6.2   & 40.02	 & 1.00    &  	   0.0070	& 0.8	& 5411  & 4.75  & -1.33 & 0.7 \\ 
HD121004        &    9.03  & 7.43   & 16.73	 & 1.35    &  	   0.017	& 0.80  & 5711  & 4.46  & -0.73 & 0.7 \\ 
HD122196        &    8.73  & 7.28   &  9.77	 & 1.32    &  	   0.032	& 0.78  & 6048  & 3.89  & -1.81 & 1.2 \\ 
HD126681$^b$      &    9.30  & 7.63   & 19.16	 & 1.44    &  	   -0.0183	& 0.70  & 5532  & 4.58  & -1.28 & 0.6 \\ 
HD132475        &    8.56  & 6.91   & 10.85	 & 1.14    &  	   0.058	& 0.75  & 5838  & 3.90  & -1.52 & 1.5 \\ 
HD140283        &    7.21  & 5.59   & 17.44	 & 0.97    &  	   0.021	& 0.75  & 5738  & 3.73  & -2.58 & 1.3 \\ 
HD160617        &    8.73  & 7.31   &  8.66	 & 1.25    &  	   0.0155	& 0.82  & 6028  & 3.79  & -1.83 & 1.3 \\ 
HD166913        &    8.23  & 6.92   & 16.09	 & 1.04    &  	   -0.004	& 0.73  & 6155  & 4.07  & -1.30 & 1.5 \\ 
HD175179        &    9.07  & 7.54   & 11.85	 & 1.52    &  	   -0.0056	& 0.80  & 5758  & 4.16  & -0.72 & 0.9 \\ 
HD188510        &    8.83  & 7.13   & 25.32	 & 1.17    &  	   0.0141	& 0.68  & 5536  & 4.63  & -1.58 & 1.0 \\ 
HD189558        &    7.74  & 6.16   & 14.76	 & 1.10    &  	   0.0042	& 0.76  & 5712  & 3.79  & -1.18 & 1.2 \\ 
HD195633        &    8.55  & 7.10   &  8.63	 & 1.16    &  	   0.0253	& 1.10  & 6005  & 3.86  & -0.71 & 1.4 \\ 
HD205650        &    9.05  & 7.57   & 18.61	 & 1.23    &  	   -0.007	& 0.70  & 5842  & 4.49  & -1.19 & 0.9 \\ 
HD213657        &    9.66  & 8.35   &  5.68	 & 1.54    &  	   0.0099	& 0.77  & 6208  & 3.78  & -2.01 & 1.2 \\ 
HD298986        &   10.05  & 8.74   &  7.68	 & 1.43    &  	   0.000	& 0.76  & 6144  & 4.18  & -1.48 & 1.4 \\ 
G005-040        &   10.76  & 9.13   & \nod	 & \nod	   & 	   0.0366	& 0.8	& 5766  & 4.23$^a$ & -0.93 & 0.8 \\
G013-009        &    10.0  & 8.74   &  5.75	 & 1.55    &  	   0.027	& 0.76  & 6416  & 3.95  & -2.27 & 1.4 \\ 
G020-024        &   11.13  & 9.67   &  5.42	 & 2.32    &  	   0.118	& 0.78  & 6482  & 4.47  & -1.89 & 1.5 \\  
G064-012        &   11.46  &10.21   &  1.88	 & 2.90    &  	   0.042	& 0.8	& 6459  & 4.31$^c$  & -3.10 & 1.5 \\  
G064-037        &   11.14  & 9.92   &  2.88	 & 3.10    &  	   0.0127	& 0.8	& 6494  & 3.82$^c$  & -3.17 & 1.4 \\  
G088-032        &   10.78  & 9.54   &  3.07	 & 2.32    &  	   -0.0028	& 0.80  & 6327  & 3.65  & -2.50 & 1.5 \\ 
G088-040        &    8.93  & 7.51   & 12.15	 & 1.24    &  	   -0.0084	& 0.8	& 5929  & 4.14  & -0.90 & 1.4 \\  
G183-011        &    9.86  & 8.60   &  6.47	 & 7.85    &  	   0.0084	& 0.70  & 6309  & 3.97  & -2.12 & 1.0 \\  
\hline\hline
\end{tabular}
\end{center}
\tablefoot{
\tablefoottext{*}{For the giants we assume $ M = 1 M_{\odot}$}}
\end{table*}
\end{small}
\clearpage
\newpage

\section{Abundances}
Table \ref{DWallabun} provides an overview of all the abundances and associated uncertainties determined for our sample's dwarf stars. Table \ref{Gallabun} gives the abundances and uncertainties for the giant stars in our sample.
\begin{table*}[!hp]
\begin{center}
\caption{Stellar abundances of Fe, Sr, Y, Zr, Pd, Ag, Ba, and Eu for dwarfs. The $'<'$ indicates that the abundance is an upper limit.}
\label{DWallabun}
\begin{tabular}{l c c c c c c c c }       
\hline\hline
Star    &   [Fe/H] &  [Sr/Fe] &  [Y/Fe] &  [Zr/Fe] &  [Pd/Fe] &  [Ag/Fe] &   [Ba/Fe] &   [Eu/Fe] \\  
\hline
BD+09 2190&     -2.60  &       \nod       &     $-0.28\pm0.17$ &    $-0.02\pm0.14$ &    $0.72\pm0.27$  &          \nod     &          \nod     &      \nod\\   
BD-13 3442&     -2.56  &   $ 0.21\pm0.14$ &     $-0.02\pm0.13$ &    $ 0.44\pm0.18$ &       \nod        &          \nod     &          \nod     &      \nod\\   
CD-30 18140&     -1.92  &  $ 0.15\pm0.11$ &     $  0.1\pm0.12$ &    $ 0.47\pm0.14$ &       \nod        &          \nod     &   $ -0.10\pm0.25$ &      \nod\\  
CD-33 3337&     -1.55  &   $ 0.22\pm0.11$ &     $ 0.01\pm0.13$ &    $ 0.27\pm0.17$ &    $ 0.19\pm0.27$ &    $ 0.27\pm0.25$ &   $  0.18\pm0.14$ &      \nod\\   
CD-45 3283&    -0.99  &    $-0.15\pm0.30$ &     $ 0.03\pm0.15$ &    $ 0.14\pm0.17$ &    $ 0.52\pm0.19$ &    $ 0.38\pm0.25$ &   $  0.32\pm0.15$ &     $ 0.78\pm0.17$\\   
CD-57 1633&     -1.01  &   $ 0.00\pm0.10$ &     $-0.23\pm0.12$ &        \nod       &    $ 0.17\pm0.23$ &    $ 0.20\pm0.26$ &   $  0.13\pm0.14$ &     $ 0.55\pm0.22$\\   
HD3567&        -1.33  &    $ -0.1\pm0.20$ &     $-0.18\pm0.19$ &    $ 0.27\pm0.17$ &    $ 0.29\pm0.20$ &    $ 0.53\pm0.25$ &   $  0.26\pm0.17$ &     $ 0.70\pm0.18$\\    
HD19445&        -2.13  &   $ 0.13\pm0.11$ &     $-0.1 \pm0.12$ &    $ 0.29\pm0.15$ &    $ 0.00\pm0.27$ &          \nod     &   $ -0.02\pm0.14$ &     $ 0.37\pm0.23$\\    
HD22879&        -0.95  &   $ 0.33\pm0.14$ &     $-0.06\pm0.12$ &    $ 0.19\pm0.17$ &    $ 0.17\pm0.21$ &    $ 0.00\pm0.26$ &   $ -0.18\pm0.18$ &     $ 0.45\pm0.2 $ \\  
HD25704&        -1.12  &   $ 0.30\pm0.11$ &     $-0.05\pm0.12$ &    $-0.02\pm0.17$ &    $ 0.07\pm0.21$ &    $ 0.09\pm0.26$ &   $  0.21\pm0.15$ &     $ 0.48\pm0.17$\\    
HD63077&       -1.05  &    $ 0.36\pm0.25$ &     $ 0.06\pm0.13$ &    $ 0.06\pm0.17$ &    $ 0.00\pm0.22$ &    $ 0.07\pm0.26$ &   $  0.22\pm0.15$ &     $ 0.43\pm0.17$\\    
HD63598 &      -0.990  &   $ 0.41\pm0.14$ &     $ 0.09\pm0.12$ &    $ 0.20\pm0.22$ &    $ 0.09\pm0.19$ &    $-0.07\pm0.26$ &   $  0.19\pm0.15$ &     $ 0.65\pm0.27$\\   
HD76932&       -0.97  &      $<$0.27      &     $-0.07\pm0.13$ &    $ 0.19\pm0.17$ &    $ 0.20\pm0.20$ &    $ 0.25\pm0.25$ &   $  0.30\pm0.15$ &     $ 0.41\pm0.16$\\   
HD103723&      -0.85  &    $ 0.04\pm0.13$ &     $-0.27\pm0.15$ &    $ 0.05\pm0.17$ &    $ 0.29\pm0.19$ &    $ 0.21\pm0.25$ &   $  0.17\pm0.15$ &     $ 0.43\pm0.16$\\   
HD105004&      -0.84  &    $ 0.10\pm0.10$ &     $-0.19\pm0.12$ &    $ 0.02\pm0.17$ &    $ 0.12\pm0.19$ &    $ 0.01\pm0.27$ &   $  0.17\pm0.18$ &     $ 0.26\pm0.20$\\   
HD106038&      -1.48  &    $ 0.56\pm0.17$ &     $ 0.54\pm0.14$ &    $ 0.68\pm0.17$ &    $ 0.32\pm0.21$ &    $ 0.14\pm0.25$ &   $  0.76\pm0.17$ &     $ 0.45\pm0.16$\\   
HD113679&      -0.63  &        $<$0.44    &     $ 0.08\pm0.12$ &    $ 0.22\pm0.17$ &    $ 0.14\pm0.19$ &    $ 0.08\pm0.25$ &   $  0.25\pm0.15$ &     $ 0.08\pm0.21$\\   
HD111980&      -1.32  &    $ 0.45\pm0.22$ &     $ 0.23\pm0.13$ &    $ 0.44\pm0.17$ &    $ 0.17\pm0.21$ &    $ 0.10\pm0.25$ &   $  0.48\pm0.15$ &        $<$0.5\\   
HD116064&      -2.17  &         \nod      &     $ 0.00\pm0.13$ &    $ 0.33\pm0.14$ &    $ 0.37\pm0.21$ &          \nod     &   $ -0.36\pm0.17$ &         \nod\\   
HD120559&      -1.31  &    $ 0.22\pm0.12$ &     $ 0.25\pm0.15$ &    $ 0.34\pm0.17$ &    $ 0.44\pm0.21$ &    $ 0.48\pm0.25$ &   $  0.22\pm0.17$ &      $0.71\pm0.17$\\   
HD121004&      -0.73  &    $ 0.40\pm0.12$ &     $ 0.16\pm0.12$ &    $ 0.51\pm0.17$ &    $ 0.32\pm0.19$ &    $ 0.16\pm0.25$ &   $  0.36\pm0.18$ &      $0.50\pm0.19$\\   
HD122196&      -1.81  &    $ 0.03\pm0.14$ &     $-0.15\pm0.12$ &    $ 0.20\pm0.16$ &    $ 0.02\pm0.27$ &        $<$0.22    &   $  0.02\pm0.16$ &      $0.22\pm0.18$\\   
HD126681&      -1.28  &    $ 0.20\pm0.19$ &     $ 0.34\pm0.12$ &    $ 0.56\pm0.17$ &    $ 0.51\pm0.21$ &    $ 0.28\pm0.25$ &   $  0.55\pm0.15$ &      $0.47\pm0.17$\\   
HD132475&      -1.52  &    $ 0.34\pm0.11$ &     $ 0.17\pm0.15$ &    $ 0.46\pm0.17$ &    $ 0.34\pm0.19$ &    $ 0.20\pm0.26$ &   $  0.40\pm0.15$ &      $0.43\pm0.16$\\   
HD140283&      -2.58  &    $-0.27\pm0.10$ &     $-0.48\pm0.12$ &    $-0.20\pm0.14$ &          \nod     &          \nod     &      $<-$0.62     &          \nod\\   
HD160617&      -1.83  &    $ 0.04\pm0.14$ &     $-0.03\pm0.12$ &    $ 0.19\pm0.21$ &    $ 0.42\pm0.21$ &       $<$0.35     &    $ 0.41\pm0.16$ &          \nod\\   
HD166913&      -1.93  &    $ 0.47\pm0.15$ &     $ 0.39\pm0.14$ &    $ 0.65\pm0.18$ &    $ 0.42\pm0.23$ &       $<$0.63     &    $ 0.62\pm0.14$ &      $0.61\pm0.16$\\   
HD175179&      -0.72  &      $<$1.28      &       $<$0.95      &    $ 0.17\pm0.17$ &    $ 0.12\pm0.20$ &   $ 0.09\pm0.26$  &    $ 0.45\pm0.14$ &      $0.20\pm0.16$\\   
HD188510&      -1.58  &    $-0.04\pm0.14$ &     $-0.16\pm0.13$ &    $ 0.18\pm0.17$ &    $ 0.32\pm0.19$ &   $ 0.19\pm0.26$  &    $ 0.13\pm0.14$ &      $0.40\pm0.18$\\   
HD189558&      -1.18  &    $-0.70\pm0.14$ &     $ 0.15\pm0.13$ &    $ 0.44\pm0.17$ &    $ 0.30\pm0.21$ &   $ 0.16\pm0.25$  &    $ 0.32\pm0.17$ &      $0.39\pm0.16$\\   
HD195633&      -0.71  &       $<$1.11     &       $<$0.66      &    $-0.14\pm0.17$ &    $-0.13\pm0.20$ &   $-0.03\pm0.28$  &    $ 0.10\pm0.16$ &          $<$0.1  \\   
HD205650&      -1.19  &    $-0.02\pm0.25$ &     $ 0.05\pm0.13$ &    $ 0.19\pm0.17$ &    $ 0.22\pm0.21$ &   $ 0.14\pm0.25$  &    $ 0.20\pm0.14$ &      $0.45\pm0.16$\\   
HD213657&      -2.01  &    $ 0.04\pm0.11$ &     $-0.05\pm0.12$ &    $ 0.37\pm0.15$ &    $ 0.22\pm0.27$ &       $<$0.53     &    $-0.03\pm0.15$ &          \nod\\    
HD298986&      -1.48  &    $-0.03\pm0.17$ &     $-0.09\pm0.14$ &    $ 0.23\pm0.17$ &    $ 0.32\pm0.19$ &    $ 0.43\pm0.28$ &    $ 0.14\pm0.16$ &     $ 0.54\pm0.16$\\   
G 01-039&      -2.27  &    $ 0.16\pm0.10$ &     $-0.07\pm0.12$ &    $ 0.29\pm0.14$ &          \nod     &         \nod      &         \nod      &           \nod\\    
G 05-040&      -0.93  &       $<$1.29     &        $<$1.09     &    $ 0.37\pm0.17$ &    $0.16\pm0.21$  &     $0.17\pm0.26$ &    $ 0.35\pm0.14$ &         $<$0.5\\   
G 20-024&      -1.90  &    $ 0.22\pm0.11$ &     $ 0.17\pm0.12$ &    $ 0.56\pm0.15$ &    $0.77\pm0.27$  &         \nod      &    $ 0.32\pm0.14$ &          \nod\\   
G 64-012&      -3.10  &    $-0.05\pm0.10$ &     $ 0.02\pm0.13$ &          \nod     &         \nod      &         \nod      &    $-0.35\pm0.16$ &          \nod\\   
G 64-037&      -3.16  &    $-0.06\pm0.10$ &     $ 0.03\pm0.13$ &    $ 0.52\pm0.14$ &         \nod      &         \nod      &         \nod      &          \nod\\   
G 88-032&      -2.53  &       \nod        &     $-0.14\pm0.13$ &    $ 0.25\pm0.14$ &         \nod      &         \nod      &         \nod      &          \nod\\   
G 88-040&      -0.89  &    $ 0.04\pm0.10$ &     $-0.27\pm0.15$ &    $ 0.02\pm0.17$ &    $0.12\pm0.19$  &     $0.03\pm0.26$ &    $ 0.05\pm0.14$ &      $0.32\pm0.16$\\   
G183-011&      -2.12  &       \nod        &     $-0.24\pm0.12$ &    $ 0.18\pm0.17$ &         \nod      &         \nod      &         \nod      &          \nod\\    
\hline\hline
\end{tabular}
\end{center}
\end{table*}
\clearpage

\begin{table*}[!ht]
\begin{center}
\caption{Stellar abundances of Fe, Sr, Y, Zr, Pd, Ag, Ba, and Eu for giants. The $'<'$ indicates that the abundance is an upper limit. }
\label{Gallabun}
\begin{tabular}{l c c c c c c c c }       
\hline\hline
Star      &       [Fe/H] &  [Sr/Fe] &  [Y/Fe] &  [Zr/Fe] &  [Pd/Fe] &  [Ag/Fe] &   [Ba/Fe] &   [Eu/Fe] \\	
\hline
BD-01 2916  &   -1.99  &$    0.11\pm0.18 $&$     0.03\pm0.20 $&$     0.26\pm0.16 $&$     0.39\pm0.20 $&$     0.16\pm0.22 $&$        0.36\pm0.15 $&$     0.60\pm0.18$ \\ 	
BD+42 621  &    -2.48  &$   -0.18\pm0.15 $&$    -0.32\pm0.19 $&$     0.12\pm0.17 $&          \nod     &$    -0.40\pm0.32 $&$       -0.55\pm0.14 $&$     0.54\pm0.16$ \\ 	
BD+8 2856  &    -2.09  &$   -0.01\pm0.12 $&$     0.04\pm0.15 $&$     0.05\pm0.14 $&$     0.12\pm0.21 $&       $<$0.73     &$        0.26\pm0.14 $&$     0.41\pm0.16$ \\ 	
BD+30 2611  &   -1.20  &$   -0.09\pm0.26 $&$    -0.35\pm0.18 $&$     0.28\pm0.22 $&$    -0.01\pm0.21 $&$    -0.50\pm0.28 $&$        0.28\pm0.14 $&$     0.52\pm0.19$ \\ 	
BD+54 1323  &   -1.64  &$    0.05\pm0.12 $&$    -0.08\pm0.18 $&$     0.01\pm0.16 $&$    -0.33\pm0.20 $&       $<$-0.21    &$        0.35\pm0.14 $&$     0.18\pm0.16$ \\ 	
CS22890-024  &  -2.77  &$   -0.06\pm0.18 $&$    -0.25\pm0.16 $&      0.64$\pm0.16$&          \nod     &          \nod     &$       -0.30\pm0.16 $&          $<$0.9 \\ 	
CS29512-073  &  -2.67  &$    0.28\pm0.14 $&$     0.09\pm0.15 $&$    -0.07\pm0.16 $&          \nod     &          \nod     &             \nod     &$     0.08\pm0.22$ \\ 	
CS30312-059  &  -3.06  &$    0.09\pm0.15 $&$    -0.32\pm0.13 $&$     0.43\pm0.37 $&          \nod     &          \nod     &$        0.07\pm0.16 $&         \nod      \\ 	
CS30312-100  &  -2.62  &$   -0.34\pm0.20 $&$    -0.79\pm0.12 $&$    -0.17\pm0.17 $&          \nod     &          \nod     &             \nod     &$     0.12\pm0.21$ \\ 	
CS31082-001  &  -2.81  &$    0.66\pm0.11 $&      0.82$\pm0.16$&$     0.77\pm0.26 $&$     1.29\pm0.19 $&$     1.19\pm0.27 $&$        1.43\pm0.14 $&$     1.53\pm0.31$ \\ 	
HD74462  &      -1.48  &$    0.06\pm0.10 $&$     0.38\pm0.16 $&$     0.54\pm0.26 $&$     0.13\pm0.18 $&$    -0.33\pm0.25 $&$        0.37\pm0.14 $&$     0.52\pm0.17$ \\ 	
HD83212  &      -1.25  &$   -0.04\pm0.17 $&$     0.21\pm0.22 $&$     0.27\pm0.22 $&$    -0.20\pm0.19 $&$    -0.53\pm0.28 $&$        0.32\pm0.14 $&$     0.27\pm0.17$ \\ 	
HD88609  &      -2.87  &$    0.04\pm0.11 $&$    -0.09\pm0.23 $&$     0.19\pm0.14 $&         $<$0.07   &        $<$0.25    &$       -0.93\pm0.14 $&$    -0.50\pm0.16$ \\ 	
HD108317  &     -2.11  &$   -0.05\pm0.13 $&$    -0.22\pm0.12 $&$     0.07\pm0.14 $&$    -0.08\pm0.20 $&$     0.15\pm0.32 $&$        0.34\pm0.17 $&$     0.34\pm0.18$ \\ 	
HD110184  &     -2.40  &$   -0.05\pm0.22 $&$     0.15\pm0.14 $&$     0.47\pm0.18 $&$     0.22\pm0.23 $&$     0.15\pm0.22 $&$       -0.06\pm0.14 $&$     0.22\pm0.16$ \\ 	
HD115444  &     -3.00  &$   -0.08\pm0.16 $&$    -0.12\pm0.14 $&$     0.20\pm0.19 $&$     0.47\pm0.20 $&        $<$0.35    &$        0.35\pm0.14 $&$     0.79\pm0.16$ \\ 	
HD122563  &     -2.81  &$    0.04\pm0.13 $&$    -0.10\pm0.16 $&$     0.07\pm0.18 $&$     0.12\pm0.20 $&$     0.10\pm0.25 $&$       -0.85\pm0.14 $&$    -0.51\pm0.16$ \\ 	
HD122956  &     -1.45  &$   -0.04\pm0.14 $&$     0.00\pm0.28 $&$    -0.01\pm0.25 $&$     0.32\pm0.20 $&$    -0.28\pm0.22 $&$        0.34\pm0.14 $&$     0.24\pm0.16$ \\ 	
HD126238  &     -1.92  &$   -0.09\pm0.14 $&$    -0.27\pm0.12 $&$     0.06\pm0.14 $&$     0.24\pm0.19 $&$    -0.01\pm0.28 $&$        0.16\pm0.14 $&$     0.19\pm0.16$ \\ 	
HD126587  &     -3.16  &$   -0.01\pm0.10 $&$    -0.20\pm0.13 $&$     0.19\pm0.19 $&$     0.42\pm0.26 $&$     0.40\pm0.28 $&$        0.08\pm0.14 $&$     0.31\pm0.17$ \\ 	
HD128279  &     -2.34  &$   -0.36\pm0.11 $&$    -0.78\pm0.13 $&$    -0.35\pm0.14 $&$    -0.16\pm0.20 $&$    -0.24\pm0.22 $&$       -0.43\pm0.15 $&$    -0.28\pm0.16$ \\ 	
HD165195  &     -2.10  &$   -0.19\pm0.16 $&$    -0.01\pm0.20 $&$     0.09\pm0.20 $&$    -0.32\pm0.20 $&$    -0.60\pm0.25 $&$        0.58\pm0.14 $&$     0.89\pm0.17$ \\ 	
HD166161  &     -1.25  &$    0.29\pm0.14 $&$     0.29\pm0.18 $&$     0.26\pm0.17 $&$     0.00\pm0.19 $&$     0.05\pm0.25 $&$        0.55\pm0.15 $&$     0.07\pm0.16$ \\ 	
HD175305  &     -1.38  &$    0.01\pm0.19 $&$     0.12\pm0.16 $&$     0.29\pm0.18 $&$     0.04\pm0.18 $&$     0.04\pm0.28 $&$        0.35\pm0.15 $&$     0.56\pm0.16$ \\ 	
HD186478  &     -2.42  &$    0.08\pm0.11 $&$     0.01\pm0.12 $&$     0.31\pm0.14 $&$     0.26\pm0.18 $&$     0.23\pm0.25 $&$        0.25\pm0.14 $&$     0.50\pm0.16$ \\ 	
HD204543  &     -1.84  &$   -0.07\pm0.12 $&$    -0.14\pm0.15 $&$     0.07\pm0.18 $&$     0.04\pm0.18 $&$     0.04\pm0.25 $&$        0.23\pm0.15 $&$     0.10\pm0.17$ \\ 	
HE 0315+0000  &  -2.59 &$    0.13\pm0.12 $&$     0.08\pm0.15 $&$     0.30\pm0.16 $&          \nod     &          \nod     &             \nod     &$     0.70\pm0.21$ \\     
HE 0442-1234  &  -2.32 &$   -0.18\pm0.17 $&$    -0.24\pm0.16 $&$    -0.06\pm0.19 $&$     0.12\pm0.30 $&          \nod     &             \nod     &$     0.28\pm0.14$ \\     
HE 1219-0312  &  -3.21 &$    0.01\pm0.13 $&$    -0.19\pm0.21 $&$     0.19\pm0.14 $&          \nod     &          \nod     &     $    0.70\pm0.20 $&        \nod    \\     
\hline\hline
\end{tabular}
\end{center}
\end{table*}

\end{document}